\documentclass[11pt,a4paper]{article}
\usepackage{jheppub}
\usepackage{graphicx}
\usepackage{amsmath,amsthm,amssymb}
\usepackage{tikz}
\usepackage{tikz-cd}
\usepackage{mathtools}
\usepackage{caption}
\usepackage{subcaption}
\usepackage{pifont}
\usepackage{dsfont}
\usepackage{url}
\usepackage{stmaryrd}

\usepackage{footmisc}

\usepackage{sectsty}
\allsectionsfont{\boldmath}

\allowdisplaybreaks

\usepackage{customprelude}

\newcommand{\xrightarrowdbl}[2][]{\xrightarrow[#1]{#2}\mathrel{\mkern-14mu}\rightarrow}

\title{Anomaly-induced vanishing of brane partition functions}

\author[a]{Felix B. Christensen,}
\author[a]{Iñaki García Etxebarria,}
\author[b]{and Enoch Leung}
\affiliation[a]{Department of Mathematical Sciences, Durham University, Durham, DH1 3LE, United Kingdom}
\affiliation[b]{Max Planck Institute for Mathematics in the Sciences, Inselstraße 22, 04103 Leipzig, Germany}

\emailAdd{inaki.garcia-etxebarria@durham.ac.uk, felix.b.christensen@durham.ac.uk, enoch.leung@mis.mpg.de}

\abstract{In the presence of 't Hooft anomalies, backgrounds for the symmetries of a quantum field theory can lead to non-conservation of Noether currents, or more generally, to the presence of charged insertions in the path integral. When there is a net background charge, the partition function evaluated on closed manifolds will vanish. For anomalous symmetries, this statement can also be understood as the anomaly theory giving rise to a non-trivial anomalous phase for the partition function even for ``rigid'' transformations which leave all background fields unchanged. We use the generalisation of this second viewpoint to the setting of anomalous higher-form symmetries in order to show vanishing of the partition function for a number of examples, both with and without a Lagrangian description. In particular, we show how to derive from these considerations the analogue of the Freed-Witten anomaly cancellation condition for the M5-brane, and also that for the D3-brane in S-fold backgrounds.}


\begin{document}


\maketitle
\flushbottom

\addtocontents{toc}{\protect\setcounter{tocdepth}{2}}


\section{Introduction}\label{intro}

Consider a D-brane in weakly coupled Type II string theory, wrapping
some closed submanifold $D$ of ten dimensional spacetime. A very well
known result by Freed and Witten \cite{Witten:1998xy,Freed:1999vc}
states that the theory of open worldsheets ending on $D$ has a global
anomaly unless
\[
  \label{eq:FW-intro}
  [H]|_D = W_3(N_D)\, ,
\]
where $[H]|_D\in H^3(D;\bZ)$ is the characteristic class of the
NSNS 2-form field $B$ restricted to $D$, $N_D$ is the normal
bundle to $D$ in the ambient space, and $W_3$ its third integral
Stiefel-Whitney class.

\medskip

The main goal of this paper is to derive a generalisation of this
cancellation condition to situations where the original worldsheet
derivation does not apply due to the presence of strong string
coupling. Such generalisations have already been worked out in
multiple cases using diverse techniques (we will provide references
below), our contribution is to provide an anomaly-based argument that
allows us to rederive existing generalisations in a systematic manner,
and also allows us to study some cases that had not been understood
before. In contrast with the original approach of Freed and Witten,
our approach makes use of a (higher-form) anomaly on the worldvolume
of the brane itself, and it is applicable whenever the anomaly theory
for the Quantum Field Theory (QFT) on the brane is known. We emphasise that
this anomaly theory is often significantly easier to understand than
the worldvolume theory itself. Although in this paper we will be mostly interested in brane physics,
the argument we use is in fact purely field theoretical, and we will
also discuss how it applies to various QFTs of independent interest,
such as the 3d minimal TQFTs \cite{Hsin:2018vcg}, which we study in
Section~\ref{sec:3d-minimal-TQFTs}.

\medskip

In order to be able to treat some of the topological subtleties that
arise in the analysis we will work in the framework of differential
cohomology. The main ideas of the argument can be stated without
having to introduce this machinery, so we briefly summarise them here
to make the logic clearer. Our approach follows from two fundamental
observations. First, consider free Maxwell theory on $D$. This theory
has two $\U(1)$ 1-form symmetries \cite{Gaiotto:2014kfa}, which we
denote by $\U(1)^{[1]}_e$ and $\U(1)^{[1]}_m$. The partition function
of the theory coupled to a background $B_e$ for the electric symmetry
is given (slightly schematically, we will be more precise below) by
\[
  S = \frac{1}{2e^2} \int_D (F-B_e)\wedge \star (F-B_e)\, .
\]
This is a close cousin of the actual theory living on a single
D-brane, with the background field $B_e$ playing the role of the
restriction of the Type II NSNS field $B$ to the brane. In this
context, the Freed-Witten condition becomes the requirement that
$[dB_e]=0$ as an element of $H^3(D;\bZ)$. Hsieh, Tachikawa and
Yonekura argued convincingly in \cite{Hsieh:2020jpj} that whenever
$[dB_e]\neq 0$, the partition function of Maxwell theory should be
taken to be zero. We follow this point of view, and will interpret the
Freed-Witten anomaly cancellation condition as the statement that the
brane partition function vanishes whenever condition~\eqref{eq:FW-intro}
(or its generalisations) is not satisfied.

\medskip

Our second observation is that this vanishing of the partition
function can be understood as coming from the presence of a mixed
anomaly between $\U(1)_e^{[1]}$ and $\U(1)_m^{[1]}$.
Before explaining this point for Maxwell theory itself, let us
discuss the more familiar setting of a theory with an ordinary
anomaly-free global $\U(1)$ symmetry. Consider the
expectation value
\[
  \vev{\cO(x)} \df \frac{\int [d\Phi] \cO(x) e^{-S[\Phi]}}{\int
    [d\Phi] e^{-S[\Phi]}}\label{correlator}
\]
where $\cO(x)$ is an operator with charge $q$ under the $\U(1)$
symmetry, and $\Phi$ are the dynamical fields in this theory. There is
a well-known argument showing that for $q\neq 0$ this expectation
value vanishes on any closed\footnote{In order to have a simple form
  of the vanishing argument, we will always assume in this paper that
  $\cM^d$ is closed, otherwise the question of vanishing depends
  sensitively on the action of symmetries on the boundary conditions.}
manifold $\cM^d$: consider a small codimension-1 sphere $S$
surrounding the insertion point $x$, and insert a symmetry generator
for the $\U(1)$ symmetry on $S$. We denote this insertion by
$U_\alpha(S)$, with $\alpha\in \bR/\bZ$ parametrising the $\U(1)$
transformation generated by the operator.  Concretely, we have
\[
  \label{eq:U_alpha(S)}
  U_\alpha(S) \df \exp\left(2\pi i \alpha \int_S j\right)
\]
with $j$ the conserved $(d-1)$-form associated to the $\U(1)$ symmetry,
satisfying the Ward identity $dj=q \delta^d(x)$ inside the path
integral, with $\delta^d(x)$ a delta function localised at the point
$x\in \cM^d$, the location of the charged operator.

The sphere $S$ splits $\cM^d$ into two pieces, one of which contains
$x$, and one which does not. We will refer to these two pieces as the
interior $I_S$ of $S$, and the exterior $E_S$, respectively. Up to
orientation, both pieces have $S$ as their boundary. Using Stokes'
theorem, we can write
\[
  \label{eq:U_alpha-inout}
  U_\alpha(S) = \exp\left(2\pi i \alpha \int_{I_S} dj\right)
  = \exp\left(-2\pi i\alpha \int_{E_S} dj\right)\, .
\]
Since the exterior of $S$ contains no charged operators, the right
hand side of~\eqref{eq:U_alpha-inout} is the identity operator, and
therefore $\vev{\cO(x)} = \vev{U_\alpha(S)\cO(x)}$. But since $\cO(x)$
has charge $q$ under $\U(1)$, or equivalently due to the Ward identity
$dj=q\delta^d(x)$, we have
$\vev{U_\alpha(S)\cO(x)}=e^{2\pi iq\alpha}\vev{\cO(x)}$, so
$\vev{\cO(x)}$ vanishes unless $e^{2\pi i q\alpha}=1$ for all
$\alpha$, which is only possible if $q=0$.

The vanishing conditions that we find can be understood as a
generalisation of this argument to the case in which, due to 't Hooft
anomalies, backgrounds for symmetries are charged
themselves.\footnote{In the case of a mixed anomaly, the background
  for one symmetry is charged under the other.}  Consequently, the
insertion of a symmetry generator $U_\alpha(S)$ plays a role analogous
to the charged operator $\mathcal{O}(x)$ in the correlator
\eqref{correlator}, and we would like to examine its vanishing
properties. Since backgrounds for symmetries, particularly continuous
ones, are typically not localised anywhere on the manifold, we will
focus on the limiting case of the symmetry generators described above
when $I_S=\cM^d$, by introducing ``rigid'' symmetry
generators\footnote{Given that these operators have support over all
  $\cM^d$, it is perhaps also natural to think of them as $(-1)$-form
  symmetries, instead of a limiting configuration of ``filled-in''
  0-form generators. The $(-1)$-form characterisation is also natural
  in that the operators act on the whole partition function, and not
  on any one localised insertion.}
\[
  \label{eq:R-symmetry-operator}
  \sR_\alpha(\cM^d) \df \exp\left(2\pi i\alpha \int_{\cM^d}
    dj\right)\, .
\]
(We will motivate the terminology momentarily.) If the symmetry is conserved, $dj$ will vanish, due to the Ward identity, away from charged operator insertions, so \eqref{eq:R-symmetry-operator} will pick up phases precisely where any charged operators have been inserted. If the symmetry is anomalous, for example as in the case of the ABJ anomaly, $dj\propto F\wedge F$, so the background field itself makes $\sR_\alpha(\cM^d)$ potentially non-trivial. In this paper we view the $F\wedge F$ background as providing charged insertions in the path integral.

Ordinarily,
introducing a symmetry generator can be seen as a modification of the
background for the symmetry: since the background couples to the
current via a term $A\wedge j$ in the action, if we insert a symmetry
generator $U_\alpha(S)$ defined as above into the path integral, we
are effectively shifting the background $A$ for the $\U(1)$ symmetry
by $A\to A+2\pi \alpha \delta(S)$, with $\delta(S)$ a Dirac-delta
1-form localised on $S$. In terms of $I_S$, let us introduce a
generalised Heaviside function $\Theta(I_S)$ which is equal to 1
inside $I_S$, and 0 outside. We can then write
\[
  U_\alpha(S) = \exp\left(2\pi i\alpha \int_{\cM^d} dj\wedge \Theta(I_S)\right) = \exp\left(2\pi i\alpha (-1)^d \int_{\cM^d} j\wedge d\Theta(I_S)\right)\, .
\]
This is the same as~\eqref{eq:U_alpha(S)}, once we identify
$(-1)^dd\Theta(I_S)$ with $\delta(S)$, but it has a nice (and
well-known) interpretation: insertion of the symmetry generator acts
via a constant $e^{2\pi i\alpha}$ gauge transformation on $I_S$, and
the identity on $E_S$. Similarly, we can view the rigid symmetry
generator $\sR_\alpha$ as the limiting case in which we act with an
everywhere constant gauge transformation $e^{2\pi i\alpha}$, we will
refer to such gauge transformations as ``rigid''
transformations. Since $d\alpha=0$ such gauge transformations do not
change $A$; geometrically this is encoded in the fact that we are
integrating $dj$ over a manifold without boundary.

\medskip

All this, of course, is just reproducing well-known field theory
phenomena in a slightly different language, but we are now in a
position where we can generalise our discussion to higher symmetries,
and to theories without a Lagrangian description. Let us consider
first the generalisation to continuous higher-form symmetries, which
is immediate. Associated to every $\U(1)$ $p$-form symmetry there is a
conserved $(d-p-1)$ current $j$, satisfying
$dj = q\delta^{d-p}(\Sigma^p)$, where $\Sigma^p$ is the
$p$-dimensional locus where we have placed charged insertions. Rigid
operators are now parametrised by closed codimension $p$ submanifolds
$\cN^{d-p}$ of $\cM^d$:
\[
  \sR_\beta(\cN^{d-p}) = \exp\left(2\pi i \beta \int_{\cN^{d-p}} dj\right)\, .\label{rigid_operators}
\]
Inserting such an operator does not change the background for the
higher-form symmetry, and provides a higher-form generalisation of the
idea of a rigid (or constant) gauge transformation.

\medskip

As mentioned above, our goal in this paper is to construct a version
of the vanishing argument where the net charge comes from backgrounds
for the (anomalous) symmetries present in the system. We will do so in
the powerful language of \emph{anomaly theories}
\cite{Freed:2014iua,Monnier:2019ytc}: these are invertible field
theories in $d+1$ dimensions whose action reproduces the perturbative
anomalies one also gets from the traditional descent procedure
\cite{Zumino:1983rz}, but which can be defined much more generally. In
particular, anomaly theories enable us to treat non-perturbative anomalies and
anomalies for discrete symmetries on an equal footing with
perturbative anomalies for continuous symmetries. If we want to
reproduce the anomaly descent computation in the continuous case, we
can place the anomaly theory on the cylinder
$\cC \df [0,1]\times \cM^d$, and then perform a gauge transformation
of the background fields in the ($d+1$)-dimensional theory such that
the change in the action of the anomaly theory localises on one of the
endpoints of the cylinder. As a concrete example, if we are interested
in a rigid transformation $e^{i\alpha}$ for some $\U(1)$ symmetry with
background field $A$, we can perform a gauge transformation
$g(t)=e^{i\alpha t}$ where $t\in [0,1]$ is the coordinate along the
cylinder. This will induce a transformation $A\to A+ \alpha dt$ of the
background field on $\cC$. The actions for anomaly theories are gauge-invariant up to boundary terms, so the anomalous phase will come from
the boundary contributions.

\medskip

More generally, to see the effect of a rigid $p$-form symmetry
transformation $\sR_\alpha(\zeta)$ associated with some closed
codimension-$p$ submanifold $\zeta\subset\cM^d$, we modify the
background field of the symmetry by
$A\to A+d(t\delta^p(\zeta)) = A + dt\wedge\delta^p(\zeta)$, with as
usual $\delta^p(\zeta)$ a $p$-form (defined on $\cM^d$, and pulled
back to $\cC$) localised on the manifold $\zeta$. In the examples
below we will often see that the answer only depends on the homology
class of $\zeta$, so $\delta^p(\zeta)$ is a representative of the
class in cohomology Poincaré dual to $[\zeta]$. Unless otherwise
specified, we will choose arbitrary representatives of this cohomology
class.

\medskip

There is a more geometric way of understanding the
$A\to A + dt\wedge\delta^p(\zeta)$ configuration we are placing on
$\cC$ in order to show vanishing of the partition function.
Instead of the linear gauge parameter chosen above, choose a
more localised profile for the gauge transformation, of the form
$g=e^{i\alpha\, \vartheta(t-t_\sR)}$, with $\vartheta(x)$ the
Heaviside function:
\[
  \vartheta(x) = \begin{cases}
    1 & \text{if } x>0\,, \\
    0 & \text{otherwise}\, ,
  \end{cases}
\]
and $t_\sR\in (0,1)$. (After regularising $\vartheta(x)$ in the
standard way, this amounts to choosing a new parametrisation of
$[0,1]$.)  While the transformation is still pure gauge, the change in
$A$ is now localised at $t=t_\sR$: we have
$A\to A + \alpha \delta(t-t_\sR)$. This changes the partition function
of the anomaly theory precisely as an insertion of $\sR_\alpha(\cM^d)$
at $t=t_\sR$ would. In other words, one of the gauge transformations
that we could introduce on $\cC$ to compute the change of the phase in
the partition function due to the anomaly corresponds precisely to
pulling $\sR_\alpha(\cM^d)$ into the bulk of $\cC$.

\medskip

We now have all the ideas we need to discuss Maxwell theory from this
viewpoint. We can understand vanishing of the partition function by
considering a background on $\cC\df [0,1]\times \cM^4$ with $B_e$ a
pullback from $\cM^4$ (under the map which forgets the interval), and
$B_m=\lambda\wedge dt$, with $t$ a coordinate in $[0,1]$, and
$\lambda$ the pullback of a closed form on $\cM^4$. We assume that
$\cM^4$ has no torsion. (A proper understanding of cases with torsion
requires the use of differential cohomology, as in the main body of
the paper.) The closed 1-form $\lambda$ is the analogue of the
constant gauge parameter $\alpha$ in the case of the point operator,
and it is given by $\beta$ times the Poincaré dual of $\cN^{d-p}$
in~\eqref{rigid_operators}. Evaluating the anomaly theory
$2\pi i \int B_m\wedge dB_e$ on such a background leads to an
anomalous factor
\[
  \exp\left(2\pi i \int_{\cM^4} dB_e \wedge \lambda\right)\, .
\]
We get to choose $\lambda$ arbitrarily, as long as it is closed, so
this phase factor will only be the identity for all choices of
$\lambda$ if $[dB_e]=0\in H^3(\cM^4;\bR)$. Whenever this is not true,
the partition function vanishes. (The fact that the vanishing of the partition function of this theory follows from the mixed anomaly was already pointed out in \cite{Maeda:2025ycr}. See also \cite{Cordova:2019jqi} for applications of the same mapping torus sufficient condition for vanishing that we just reviewed to the study of gapless phases.) As we will see in the main text, a more careful analysis using differential cochains leads to the broader (in the presence of torsion) sufficient vanishing condition $[dB_e]\neq 0\in H^3(\cM^4;\bZ)$, where $[dB_e]$ denotes the characteristic class.

\medskip

Finally, let us provide an alternative viewpoint on the
vanishing result, which we will adopt in the rest of this paper. Take $\cC\df [0,1]\times \cM^4$ as before, and
consider as an example the theory of a free Dirac fermion in four
dimensions. We assume that the signature of $\cM^4$ vanishes, for
simplicity. As is well-known, this theory has a $\U(1)_V$ vector
symmetry and a $\U(1)_A$ axial symmetry, with a mixed anomaly encoded
in the presence of an $A_A\wedge F_V^2$ term in the anomaly theory
\cite{PhysRev.177.2426,Bell:1969}.\footnote{We are being a bit
  imprecise here: the anomaly theory is better described as the
  exponentiated $\eta$ invariant \cite{Dai:1994kq}, but for the level
  of precision we are aiming at in this introduction the description
  in the text is sufficient.} If we evaluate the anomaly for arbitrary
backgrounds $A_A$, $A_V$ on $\cC$ it will in general lead to a phase
factor different from one. This does not imply that the partition
function should vanish: if we slice $\cC$ into constant-$t$ slices, so
that we have families of backgrounds $A_A(t)$, $A_V(t)$ on $\cM^4$,
the partition function of the anomaly theory is telling us how the
phase of the partition function changes as we move along the families
of background connections. More precisely, we are studying how the
phase of the determinant line bundle changes under parallel transport
in the space of connections. Parallel transport on a generic bundle
does indeed induce non-trivial holonomies in general. For example, if
the bundle has curvature (which indicates the presence of perturbative
anomalies) small loops can generate non-trivial phases. These phases
do not imply that the partition function needs to vanish, they only
tell us that the connection of the determinant line bundle over the
space of background connections is not trivial.

From this point of view, the rigid connections that we will be
choosing are very special. For instance, in the ABJ case, in order to
probe vanishing we could choose a connection on $\cC$ of the form
$A_A = \alpha dt$ and $A_V$ a pullback of some connection on $\cM^4$
(that is, we assume $A_V$ to be constant on the $t$ direction, and to
have no $dt$ component).  If we evaluate $A_A\wedge F_V\wedge F_V$ on
such a background it leads to an anomalous phase given by
$\exp(\pi i\alpha\int_{M^4} F_V\wedge F_V)$, which implies vanishing
of the partition function whenever the index of the Dirac operator on
$\cM^4$ (which we recall we are assuming to have vanishing signature)
is non-zero. This of course agrees with the expectation from field
theory, since zero modes lead to the vanishing of the partition
function, and with the classical calculation by Fujikawa
\cite{10.1093/acprof:oso/9780198529132.001.0001}. When doing the
constant-$t$ slicing above, the connection on each constant-$t$ slice
is always the same, and in particular $A_A$ vanishes. From the point
of view of the determinant line bundle, we are finding that we can
modify the phase of the determinant while keeping the background
connection fixed.

There is an important subtlety in this last line of argument that we
would like to highlight. The subtlety is due to the fact that fixing a
one-parameter family of connections on $\cM^d$ does not uniquely
specify the actual gauge background to put on $\cC$. While we have
argued above that the relevant backgrounds for studying vanishing due
to anomaly-induced charges are those of the form $\xi\wedge dt$ on
$\cC$, with $\xi$ a closed form on $\cM^d$, it is not difficult to
construct backgrounds on $\cC$ which are not of this type, but which
restrict to constant configurations on the constant-$t$ slices, and
which do not imply vanishing. For instance, any $A_A = \alpha(x)dt$,
with $x$ a coordinate on $\cM^4$, will restrict to $A_A=0$ on every
constant-$t$ slice. But running the argument for such backgrounds with
generic choices of $\alpha(x)$ would lead to too strong results,
namely that the partition function vanishes unless $F_V^2=0$
\emph{locally} on $\cM^4$, which is certainly too strong. When
choosing backgrounds on $\cC$ to study vanishing of the partition
function below, we will always restrict ourselves to those of the type
$\xi\wedge dt$, with $\xi$ closed, so that it is clear that we are
actually computing a phase obtained by a rigid gauge
transformation. We will clarify this subtlety in
Section~\ref{basepoint_anomaly_section}, once we have introduced some
necessary geometric machinery.

\medskip

Let us finish this introduction by briefly mentioning that throughout
this paper we will be dealing only with (possibly higher-form) Abelian
symmetry groups. This leaves many possible directions for further work, for
example the generalisation of our arguments to more general
categorical symmetries, or alternatively to (possibly Abelian) subgroups of non-Abelian
transformations leaving the background fixed. A recent interesting work in this
last direction is \cite{Meynet:2025zem}.

\medskip

This paper is organised as follows. In
Section~\ref{basepoint_anomaly_section}, after a quick review of the
differential cohomology tools we need, we reformulate the criteria for
vanishing of the partition function we have just described in a
precise and general way, including in particular topologically
non-trivial backgrounds, both continuous and
discrete. Sections~\ref{sec:BF} to \ref{S-folds_section} are then
devoted to showing how our vanishing condition applies to a variety of
systems.\footnote{It is perhaps useful to emphasise at this stage that
  the conditions that we find are \emph{sufficient} conditions for
  vanishing. We are not claiming that every example where the
  partition function vanishes can be explained using our arguments.}
In the case where the resulting vanishing conditions were already known,
our arguments simply show how the vanishing can be understood in
terms of our streamlined general argument, but some of the results are
new. The appendices review some technical results
which are needed in the main text.



\section{Basepoint anomaly from the mapping torus}
\label{basepoint_anomaly_section}

In this section, we establish more precisely the relation between the vanishing of the partition function of a QFT and the evaluation of its anomaly on the mapping torus. As is well-known, a gauge field can carry non-trivial topological data, and an adequate language which captures all such data is {\it differential cohomology} \cite{10.1007/BFb0075216,Hopkins:2002rd}. The advantage of working with differential cocycles is that it makes the aforementioned relation manifest. For this purpose, we shall briefly review the notion of differential cohomology.

\subsection{A brief review of differential cohomology}\label{differential_cohomology_review}

Suppose we have a QFT on some closed $d$-dimensional spacetime manifold $X^d$, and it has a $p$-form $\U(1)$ global symmetry \cite{Gaiotto:2014kfa}. One can then couple the theory to a $(p+1)$-form background gauge field $A_{p+1}$. More precisely, let us work with {\it differential cochains} in the sense of \cite{Hopkins:2002rd}. The cochain complex is defined as
\begin{equation}
    \check{C}^{p+2}(X^d) = \{\check{A} = (\mathsf{C}_{p+2},A_{p+1},F_{p+2}) \in C^{p+2}(X^d;\mathbb{Z}) \times C^{p+1}(X^d;\mathbb{R}) \times \Omega^{p+2}(X^d)\} \, ,\label{differential_cochain_definition}
\end{equation}
with the differential given by
\begin{equation}
    \begin{gathered}
        d: \check{C}^{p+2}(X^d) \to \check{C}^{p+3}(X^d) \, ,\\
        (\mathsf{C}_{p+2},A_{p+1},F_{p+2}) \mapsto (\delta\mathsf{C}_{p+2},F_{p+2} - \mathsf{C}_{p+2} - \delta A_{p+1},dF_{p+2}) \, .
    \end{gathered}
\end{equation}
A gauge field is a {\it differential cocycle} which is, by definition, closed with respect to $d$, i.e.
\begin{equation}
    \check{Z}^{p+2}(X^d) = \{\check{A} = (\mathsf{C}_{p+2},A_{p+1},F_{p+2}) \in Z^{p+2}(X^d;\mathbb{Z}) \times C^{p+1}(X^d;\mathbb{R}) \times \Omega_\mathbb{Z}^{p+2}(X^d)\} \, ,
\end{equation}
where the triplet $(\mathsf{C}_{p+2},A_{p+1},F_{p+2})$, denoting respectively the {\it characteristic class}, {\it connection}, and {\it curvature} (or {\it field strength}), satisfies
\begin{equation}
    \delta\mathsf{C}_{p+2} = 0 \, , \qquad dF_{p+2} = 0 \, , \qquad \delta A_{p+1} = F_{p+2} - \mathsf{C}_{p+2} \, .
\end{equation}
On the RHS of the last relation, both $F_{p+2}$ and $\mathsf{C}_{p+2}$ are implicitly regarded as $\mathbb{R}$-valued cocycles using the suitable inclusion maps.

Among the space of differential cocycles $\check{Z}^{p+2}(X^d)$, those in which the integral cohomology class $[\mathsf{C}_{p+2}]_\mathbb{Z} \in H^{p+2}(X^d;\mathbb{Z})$ vanishes are {\it topologically trivial}. There is also a subspace $\check{Z}_\text{flat}^{p+2}(X^d)$ consisting of {\it flat} cocycles where the curvature $F_{p+2} \in \Omega_\mathbb{Z}^{p+2}(X^d)$ vanishes. A given element $\check{A} \in \check{Z}_\text{flat}^{p+2}(X^d)$ thus has its connection and characteristic class related by
\begin{equation}
    \delta A_{p+1} = -\mathsf{C}_{p+2} \, .
\end{equation}
Through the short exact sequence
\begin{equation}
    0 \to \mathbb{Z} \to \mathbb{R} \to \mathbb{R}/\mathbb{Z} \to 0 \, ,\label{short_exact_sequence}
\end{equation}
the flat cocycle $\check{A}$ determines and is determined uniquely (up to gauge transformations) by a cohomology class $[A_{p+1}]_{\mathbb{R}/\mathbb{Z}} \in H^{p+1}(X^d;\mathbb{R}/\mathbb{Z})$, such that
\begin{equation}
    \beta([A_{p+1}]_{\mathbb{R}/\mathbb{Z}}) = [\mathsf{C}_{p+2}]_\mathbb{Z} \, ,\label{flat_gauge_field_characteristic_class}
\end{equation}
where $\beta$ is the connecting homomorphism, known as the Bockstein homomorphism, associated with the long exact sequence in cohomology,
\begin{equation}
    \cdots \to H^i(-;\mathbb{Z}) \to H^i(-;\mathbb{R}) \to H^i(-;\mathbb{R}/\mathbb{Z}) \xrightarrow{\beta} H^{i+1}(-;\mathbb{Z}) \to \cdots \, ,
\end{equation}
induced by \eqref{short_exact_sequence}.

The product between two differential cocycles $\check{A} \in \check{C}^{p+2}(X^d)$ and $\check{A}' \in \check{C}^{q+2}(X^d)$ is the triplet,
\begin{equation}
    \check{A} \star \check{A}' = (\mathsf{C}_{p+2} \cup \mathsf{C}_{q+2}',A_{p+1} \cup F_{q+2}' + (-1)^{p+2} \mathsf{C}_{p+2} \cup A_{q+1}' + Q(F_{p+2},F_{q+2}'),F_{p+2} \wedge F_{q+2}') \, ,
\end{equation}
where $Q(\alpha,\beta) \in C^{|\alpha|+|\beta|-1}(X^d \times I;\mathbb{R})$ is any natural chain homotopy between the wedge product $\wedge$ for differential forms and the cup product $\cup$ for cochains \cite{10.1007/BFb0075216}, defined to be such that
\begin{equation}
    \alpha \wedge \beta - \alpha \cup \beta = Q(d\alpha,\beta) + (-1)^{|\alpha|} Q(\alpha,d\beta) + \delta Q(\alpha,\beta) \, .
\end{equation}
It follows that $\check{A} \star \check{A}'$ is topologically trivial if either $\check{A}$ or $\check{A}'$ is topologically trivial. Likewise, $\check{A} \star \check{A}'$ is flat if either $\check{A}$ or $\check{A}'$ is flat. One can also check that the product is associative up to chain homotopy.

A physically relevant quantity that one can build from the data of a differential cocycle is its holonomy $\chi(M^{p+1}) \coloneqq \exp(2\pi i \int_{M^{p+1}} A_{p+1})$ over some $(p+1)$-submanifold $M^{p+1} \subset X^d$. If $M^{p+1}$ is the boundary of some $N^{p+2}$, then we have
\begin{equation}
    \chi(M^{p+1}) = \exp\bigg(2\pi i \int_{M^{p+1}} A_{p+1}\bigg) = \exp\bigg(2\pi i \int_{N^{p+2}} F_{p+2}\bigg)
\end{equation}
by virtue of Stokes' theorem. One may attempt to define the cohomology of the complex $\check{C}^\ast(X^d)$ in the usual manner as
\begin{equation}
    \check{H}^{p+2}(X^d) \stackrel{?}{=} \frac{\text{ker}(d: \check{C}^{p+2}(X^d) \to \check{C}^{p+3}(X^d))}{\text{im}(d: \check{C}^{p+1}(X^d) \to \check{C}^{p+2}(X^d))} \, ,
\end{equation}
such that the corresponding equivalence relation is given by $\check{A} \sim \check{A} - d\check{a}$, i.e.
\begin{equation}
    (\mathsf{C}_{p+2},A_{p+1},F_{p+2}) \sim (\mathsf{C}_{p+2} - \delta\mathsf{c}_{p+1},A_{p+1} - f_{p+1} + \mathsf{c}_{p+1} + \delta a_p,F_{p+2} - df_{p+1}) \, ,\label{wrong_equivalence_relation}
\end{equation}
for any $\check{a} = (\mathsf{c}_{p+1},a_p,f_{p+1}) \in \check{C}^{p+1}(X^d)$. However, under $\check{A} \to \check{A} - d\check{a}$, the holonomy
\begin{equation}
    \chi(N^{p+2}) \mapsto \chi(N^{p+2}) \exp\bigg(\!-2\pi i \int_{M^{p+1}} f_{p+1}\bigg)
\end{equation}
is not invariant for generic $f \in \Omega^{p+1}(X^d)$.

To have a meaningful notion of a ``gauge field'' for a symmetry, we demand that physical observables, e.g.~the holonomy, are independent of gauge transformations. We therefore modify the equivalence relation above by imposing $\check{a}$ to be a ``flat'' differential cochain (not cocycle), i.e.~$f = 0$, so that the {\it differential cohomology group} $\check{H}^{p+2}(X^d) = \check{Z}^{p+2}(X^d)/\sim$ is defined by
\begin{equation}
    (\mathsf{C}_{p+2},A_{p+1},F_{p+2}) \sim (\mathsf{C}_{p+2} - \delta\Lambda_{p+1},A_{p+1} + \Lambda_{p+1} + \delta\lambda_p,F_{p+2})\label{differential_cohomology_equivalence_relation}
\end{equation}
where $\Lambda_{p+1} \in C^{p+1}(X^d;\mathbb{Z})$ and $\lambda_p \in C^p(X^d;\mathbb{R})$. There are two subclasses of this equivalence relation which are commonly known in the physics literature, namely, {\it small gauge transformations} where $\Lambda_{p+1}$ is exact and $\lambda_p \in C^p(X^d;\mathbb{R})$ is generic, as well as {\it large gauge transformations} where $\Lambda_{p+1} \in Z^{p+1}(X^d;\mathbb{Z})$ is closed with non-zero integral periods \cite{Diaconescu:2003bm,Moore:2004jv,Belov:2004ht,Belov:2006jd,Freed:2006yc}.

\subsection{Anomaly and the mapping cylinder}

By turning on a background gauge field for the global symmetry, the partition function $\mathcal{Z}[\check{A}]$ of the QFT becomes a functional of the differential cocycle $\check{A} \in \check{Z}^{p+2}(X^d)$. A priori, $\mathcal{Z}[\check{A}]$ is not necessarily well-defined under the equivalence relation \eqref{differential_cohomology_equivalence_relation}, i.e.~we may obtain a phase under a gauge transformation,
\begin{equation}
    \mathcal{Z}[\check{A} - d\check{\lambda}] = e^{2\pi i \mathcal{A}[\check{A},\check{\lambda}]} \mathcal{Z}[\check{A}] \, ,\label{anomalous_phase}
\end{equation}
where $\mathcal{A}[\check{A},\check{\lambda}]$ is known as the {\it 't Hooft anomaly} \cite{tHooft:1979rat}. For simplicity, we have assumed here that the partition function is a section of a line bundle $\mathcal{L}$ (i.e.~a vector bundle with rank $1$) over the moduli space of gauge fields, $\mathfrak{A} \simeq \check{Z}^{p+2}(X^d)$. Equivalently, $\mathcal{A}[\check{A},\check{\lambda}]$ can be realised in terms of an {\it invertible field theory} on a $(d+1)$-dimensional bulk $Y^{d+1}$ such that $\partial Y^{d+1} = X^d$. More generally, we can consider replacing the line bundle with a vector bundle of higher rank, but we will not study such cases in our work.

When $\mathcal{A}[\check{A},\check{\lambda}] \notin \mathbb{Z}$, one cannot uniquely define $\mathcal{Z}[\check{A}]$ for a given differential cohomology class $\check{A} \in \check{H}^{p+2}(X^d)$. This means that the partition function is not well-defined over $\mathfrak{A}/\mathfrak{G} \simeq \check{H}^{p+2}(X^d)$, i.e.~the space of gauge fields modulo gauge transformations, where $\mathfrak{G} \simeq \check{C}^{p+1}_\text{flat}(X^d)$, the space of degree-$(p+1)$ differential cochains with vanishing curvature component. Hence, the anomaly can be interpreted as an obstruction to {\it gauging} the global symmetry, i.e.~summing over all inequivalent classes of background gauge fields $\check{A} \in \check{H}^{p+2}(X^d)$ in the partition function. (If we insist in doing such a sum, the result will vanish, but this is a different kind of vanishing to the one in this paper, which does not involve gauging of the symmetry and happens only for specific background field configurations.) We stress that this notion of an anomaly applies not only to a $p$-form $\U(1)$ symmetry, but also to when the symmetry is any group, e.g.~non-Abelian or discrete \cite{Hopkins:2002rd,Hsieh:2020jpj}, and even when it is a higher group, as long as we define the suitable differential cocycles $\check{A}$ accordingly. In the latter case, $\mathcal{A}[\check{A},\check{\lambda}]$ is usually known as a {\it mixed anomaly} between the different levels of the associated Postnikov tower.

Geometrically, the relation \eqref{anomalous_phase} can be reinterpreted as a {\it parallel transport} of $\mathcal{Z}[\check{A}]$ over the mapping cylinder $X^d \times I$, where $I$ is the unit interval, which we parametrise with a coordinate $t\in [0,1]$. More explicitly, we formally extend the gauge field to $X^d \times I$ such that it interpolates between two different gauge fields $\check{A},\check{A}' \in \check{Z}^{p+2}(X^d)$ on the two ends (we will soon impose that $A$ and $A'$ are related by a gauge transformation), i.e.
\begin{equation}
    \underline{\check{A}} = \check{\mathbb{A}}(x,t) + \check{a}(x) \star \check{t} \, ,
\end{equation}
where $\check{\mathbb{A}} \in \check{C}^{p+2}(X^d \times I)$ is a ``differential cochain'' represented by the triplet,\footnote{The combination $(1-t)\mathsf{C}_{p+2} + t\mathsf{C}_{p+2}'$ is generally not an integral cochain in $C^{p+2}(X^d \times I;\mathbb{Z})$ for arbitrary $t \in [0,1]$, so $\check{\mathbb{A}}$, and hence the bulk gauge field $\underline{\check{A}}$, are not strictly differential cochains as we defined in \eqref{differential_cochain_definition}. However, we will ultimately be interested only in the case where $\mathsf{C}_{p+2} = \mathsf{C}_{p+2}'$, so the $t$ dependence in the characteristic class of $\underline{\check{A}}$ will drop off.}
\begin{equation}
    \check{\mathbb{A}} = ((1-t)\mathsf{C}_{p+2} + t\mathsf{C}_{p+2}',(1-t)A_{p+1} + tA_{p+1}',(1-t)F_{p+2} + tF_{p+2}') \, ,
\end{equation}
up to homotopy equivalence. Meanwhile, $\check{t} = (\delta t,0,dt) \in \check{Z}^1(I)$ is a differential 1-cocycle representing the ``volume form'' of the interval, such that the product of differential cochains is given by
\begin{equation}
    \check{a} \star \check{t} = (\mathsf{c}_{p+1} \cup \delta t,a_p \cup dt + Q(f_{p+1},dt),f_{p+1} \wedge dt) \in \check{C}^{p+2}(X^d \times I)
\end{equation}
for some $\check{a} = (\mathsf{c}_{p+1},a_p,f_{p+1}) \in \check{C}^{p+1}(X^d)$.\footnote{To be precise, we should send both $\check{a}$ and $\check{t}$ to differential cochains in $\check{C}^\ast(X^d \times I)$ via the inclusion maps $\iota_{X}: X^d \hookrightarrow X^d \times I$ and $\iota_I: I \hookrightarrow X^d \times I$, in order to make sense of their product. When one sets $\check{a}=0$, the differential cochain $\underline{\check{A}} = \check{\mathbb{A}} \in \check{C}^{p+2}(X^d)$ becomes ``pure-shift'' \cite{2008arXiv0810.4935S}.}

Acting on the bulk gauge field $\underline{\check{A}}$ with the differential $d$, we obtain
\begin{equation}
    d\underline{\check{A}} = \begin{cases} (-1)^p (\mathsf{C}_{p+2}' - \mathsf{C}_{p+2}) \cup \delta t + \delta\mathsf{c}_{p+1} \cup \delta t & \in Z^{p+3}(X^d \times I;\mathbb{Z}) \, ,\\[2ex] f_{p+1} \wedge dt - \mathsf{c}_{p+1} \cup \delta t - (-1)^{p+1} (A_{p+1}' - A_{p+1}) \cup \delta t\\ - \delta a_p \cup dt - \delta Q(f_{p+1},dt) & \in C^{p+2}(X^d \times I;\mathbb{R}) \, ,\\[2ex] (-1)^p (F_{p+2}' - F_{p+2}) \wedge dt + df_{p+1} \wedge dt & \in \Omega_\mathbb{Z}^{p+3}(X^d \times I) \, , \end{cases}
\end{equation}
so in order for $\underline{\check{A}}$ to be a differential cocycle, i.e.~$d\underline{\check{A}}=0$, we need
\begin{equation}
    \begin{gathered}
        \delta\mathsf{c}_{p+1} = (-1)^{p+1} (\mathsf{C}_{p+2}' - \mathsf{C}_{p+2}) \, , \qquad df_{p+1} = (-1)^{p+1} (F_{p+2}' - F_{p+2}) \, ,\\
        \delta a_p = f_{p+1} - \mathsf{c}_{p+1} - (-1)^{p+1} (A_{p+1}' - A_{p+1}) \, , \qquad Q(df_{p+1},dt) = 0 \, .
    \end{gathered}
\end{equation}
Therefore, $\check{a} \in \check{C}^{p+1}(X^d)$ can be regarded as a differential refinement of the relative {\it Chern-Simons form} associated with the pair $(\check{A},\check{A}')$, in the sense that it trivialises the difference between them, i.e.
\begin{equation}
    d\check{a} = (-1)^{p+1} (\check{A}' - \check{A}) \, .
\end{equation}
Note that with an abuse of notation, this $d$ denotes that for differential cochains but not differential forms.

We will hereafter be interested only in the case where $\check{A}$ and $\check{A}'$ are related by a gauge transformation, where
\begin{equation}
    (\mathsf{C}_{p+2}',A_{p+1}',F_{p+2}') = (\mathsf{C}_{p+2} - \delta\Lambda_{p+1},A_{p+1} + \Lambda_{p+1} + \delta\lambda_p,F_{p+2})
\end{equation}
for some $\check{\lambda} = (\Lambda_{p+1},\lambda_p,0) \in \check{C}_\text{flat}^{p+1}(X^d)$, then we find
\begin{equation}\delta\mathsf{c}_{p+1} = (-1)^p \delta\Lambda_{p+1} \, , \quad df_{p+1} = 0 \, , \quad \delta a_p = f_{p+1} - \mathsf{c}_{p+1} - (-1)^{p+1} (\delta\lambda_p + \Lambda_{p+1}) \, .
\end{equation}
The following combination,
\begin{equation}
    \check{a} + (-1)^{p+1} \check{\lambda} = (\mathsf{c}_{p+1} + (-1)^{p+1} \Lambda_{p+1}, a_p + (-1)^{p+1} \lambda_p, f_{p+1}) \, ,\label{auxiliary_field_redefinition}
\end{equation}
is closed under the differential $d$, so we may conveniently reparametrise
\begin{equation}
    \underline{\check{A}} = \check{\mathbb{A}} + (-1)^p \check{\lambda} \star \check{t} + \check{a} \star \check{t} = \check{A} - t d\check{\lambda} + (-1)^p \check{\lambda} \star \check{t} + \check{a} \star \check{t} \, ,\label{bulk_gauge_field_parametrization}
\end{equation}
such that $\check{a} \in \check{Z}^{p+1}(X^d)$ is now a differential cocycle which is independent of $\check{A}$ and $\check{\lambda}$.\footnote{As we will soon see, $\check{\lambda}$ and $\check{a}$ correspond respectively to {\it local} and {\it rigid} gauge transformations of $\check{A}$.}

In any given QFT, we shall define the anomaly $\mathcal{A}[\check{A},\check{\lambda},\check{a}] \in \mathbb{R}$ as in \eqref{anomalous_phase}, but with an auxiliary gauge field $\check{a}$ included, as the evaluation of some rational polynomial $\mathcal{P} \in \mathbb{Q}[\check{A},\check{\lambda},\check{a},\check{t}]$ over the mapping cylinder $X^d \times I$. We will discuss some explicit examples of the polynomial $\mathcal{P}$ in the sections that follow. Recall that the bulk gauge field $\underline{\check{A}} \in \check{Z}^{p+2}(X^d \times I)$ is defined solely by the constraints,
\begin{equation}
    \iota_{X \times \{0\}}^\ast \, \underline{\check{A}} = \check{A} \, , \qquad \iota_{X \times \{1\}}^\ast \, \underline{\check{A}} = \check{A}' = \check{A} - d\check{\lambda} \, ,
\end{equation}
where $\iota_{X \times \{t\}}: X^d \times \{t\} \hookrightarrow X^d \times I$ is an inclusion map for any $t \in [0,1]$. This gives rise to a space of possible candidates for what would appear naïvely as the ``identity element'' $\mathcal{A}[\check{A},0,\check{a}]$ in \eqref{anomalous_phase} with $\check{\lambda}$ trivial, corresponding to distinct choices of $\check{a} \in \check{Z}^{p+1}(X^d)$. For certain configurations of $\check{A}$, the quantity $\mathcal{A}[\check{A},0,\check{a}]$ does not depend on $\check{a}$ (as we will see in subsequent examples), but most generally, there is no reason to expect
\begin{equation}
    \mathcal{A}[\check{A},0,\check{a}] = \mathcal{A}[\check{A},0,\check{a}'] = \mathcal{A}[\check{A},0,0] \mod \mathbb{Z}
\end{equation}
for generic $\check{A},\check{a},\check{a}' \in \check{Z}^\ast(X^d)$. Hence, provided that $\check{a}$ is an arbitrary auxiliary gauge field, we have to understand the precise meaning of the relation,
\begin{equation}
    \mathcal{Z}[\check{A}] = e^{2\pi i \mathcal{A}[\check{A},0,\check{a}]} \mathcal{Z}[\check{A}] \, ,
\end{equation}
whose consistency implies the vanishing of $\mathcal{Z}[\check{A}]$ whenever $\mathcal{A}[\check{A},0,\check{a}] \notin \mathbb{Z}$.

\subsection{Vanishing theorem for the partition function}\label{vanishing_theorem_section}

To do so, let us analyze the quantity $\mathcal{A}[\check{A},0,\check{a}]$ more carefully. We generally have
\begin{equation}
    \mathcal{A}[\check{A},0,\check{a}] \neq \mathcal{A}[\check{A},0,\check{a}']
\end{equation}
if $[\mathsf{c}_{p+1}]_\mathbb{Z} \neq [\mathsf{c}_{p+1}']_\mathbb{Z}$. Moreover, from the ansatz for the bulk gauge field $\underline{\check{A}}$ in \eqref{bulk_gauge_field_parametrization}, we note that the roles of $\check{\lambda} \in \check{C}_\text{flat}^{p+1}(X^d)$ and $\check{a} \in \check{Z}^{p+1}(X^d)$ can be interchanged if we make $\check{a}$ flat and $\check{\lambda}$ closed. Under this condition, the fiber integration over the interval $I$ which yields the anomaly cannot distinguish between $\check{\lambda}$ and $\check{a}$. Therefore, it must be that
\begin{equation}
    \mathcal{A}[\check{A},\check{\lambda},\check{a}] = \mathcal{A}[\check{A},\check{a},\check{\lambda}]\label{local_global_correspondence}
\end{equation}
for any $\check{\lambda},\check{a} \in \check{Z}_\text{flat}^{p+1}(X^d)$.

By definition, the RHS of the relation above corresponds physically to the phase (after exponentiation) one obtains by performing the gauge transformation $\check{A} \to \check{A} - d\check{a}$, i.e.
\begin{equation}
    \mathcal{Z}[\check{A} - d\check{a}] = e^{2\pi i \mathcal{A}[\check{A},\check{a},\check{\lambda}]} \mathcal{Z}[\check{A}]\label{global_gauge_transformation_anomalous_phase}
\end{equation}
for some choice of $\check{\lambda} \in \check{Z}_\text{flat}^{p+1}(X^d)$ now regarded instead as an auxiliary gauge field. Since $\check{a} \in \check{Z}_\text{flat}^{p+1}(X^d)$ is closed under the differential $d$ by construction, we have $\check{A} - d\check{a} = \check{A}$. In other words, the space of flat differential cocycles $\check{Z}_\text{flat}^{p+1}(X^d)$ can be interpreted as the space of {\it rigid gauge transformations} that does not change the gauge field $\check{A} \in \check{Z}^{p+1}(X^d)$.\footnote{In the terminology of \cite{Diaconescu:2003bm,Moore:2004jv,Belov:2004ht,Belov:2006jd,Freed:2006yc}, these were referred to as {\it global gauge transformations}.} More explicitly, under such a transformation, we find
\begin{align}
    \check{A} = (\mathsf{C}_{p+2},A_{p+1},F_{p+2}) & \mapsto (\mathsf{C}_{p+2} - \delta\mathsf{c}_{p+1},A_{p+1} + \mathsf{c}_{p+1} + \delta a_p,F_{p+2})\nonumber\\
    & = (\mathsf{C}_{p+2},A_{p+1},F_{p+2}) \, ,
\end{align}
where $\mathsf{c}_{p+1} \in Z^{p+1}(X^d;\mathbb{Z})$ and $a_p \in C^p(X^d;\mathbb{R})$ satisfy $\delta a_p = -\mathsf{c}_{p+1}$. Isomorphism classes of flat differential cocycles $\check{a} \in \check{Z}_\text{flat}^{p+1}(X^d)$ form the cohomology group $H^p(X^d;\mathbb{R}/\mathbb{Z}) \cong \check{H}_\text{flat}^{p+1}(X^d)$, which can be identified as the automorphism group of $\check{A} \in \check{Z}^{p+2}(X^d)$.

Recycling a previous argument, there generally exist distinct choices of the ``identity element'' in \eqref{global_gauge_transformation_anomalous_phase}, i.e.
\begin{equation}
    \mathcal{A}[\check{A},0,\check{\lambda}] \neq \mathcal{A}[\check{A},0,\check{\lambda}'] \neq \mathcal{A}[\check{A},0,0] \, ,
\end{equation}
where $\check{\lambda},\check{\lambda}' \in \check{Z}_\text{flat}^{p+1}(X^d)$. This is compatible with the fact that a flat differential cocycle is determined uniquely (up to gauge equivalence) by an element $[\lambda_p]_{\mathbb{R}/\mathbb{Z}} \in H^p(X^d;\mathbb{R}/\mathbb{Z})$, whose characteristic class is given by $[\Lambda_{p+1}]_\mathbb{Z} = \beta([\lambda_p]_{\mathbb{R}/\mathbb{Z}}) \in H^{p+1}(X^d;\mathbb{Z})$ according to \eqref{flat_gauge_field_characteristic_class}. In the following, we will swap back the roles of $\check{\lambda}$ and $\check{a}$ to proceed with our analysis.

As mentioned earlier, the anomaly is the evaluation of a polynomial $\mathcal{P} \in \mathbb{Q}[\check{A},\check{\lambda},\check{a},\check{t}]$ over the mapping cylinder. In addition, the product of two differential cocycles is flat if either of them is flat. This implies that the quantity $\mathcal{A}[\check{A},0,\check{a}]$ for any $\check{a} \in \check{Z}_\text{flat}^{p+1}(X^d)$ is generically valued in $\mathbb{R}/\mathbb{Z}$ by virtue of Poincaré-Pontryagin duality, where the integral
\begin{equation}
    \int_{X^d}: H^{p+1}(X^d;\mathbb{R}/\mathbb{Z}) \times H^{d-p-1}(X^d;\mathbb{Z}) \to \mathbb{R}/\mathbb{Z}\label{Poincaré-Pontryagin_duality}
\end{equation}
is a perfect pairing, thus rendering
\begin{equation}
    \mathcal{A}[\check{A},0,\check{a}] = \mathcal{A}[\check{A},0,\check{a}'] \mod \mathbb{Z}
\end{equation}
if $[\mathsf{c}_{p+1}]_{\mathbb{R}/\mathbb{Z}} = [\mathsf{c}_{p+1}']_{\mathbb{R}/\mathbb{Z}} \in H^{p+1}(X^d;\mathbb{R}/\mathbb{Z})$. On the other hand, $\mathcal{A}[\check{A},0,\check{a}]$ is not well-defined as an element of $\mathbb{R}/\mathbb{Z}$ if $\check{a} \in \check{Z}^{p+1}(X^d)$ is not flat, i.e.~$f_{p+1} \neq 0$.

Let us revisit the relation,
\begin{equation}
    \mathcal{Z}[\check{A} - d\check{a}] = e^{2\pi i \mathcal{A}[\check{A},0,\check{a}]} \mathcal{Z}[\check{A}]\label{global_gauge_transformation_anomalous_phase_v2}
\end{equation}
for any $\check{a} \in \check{Z}_\text{flat}^{p+1}(X^d)$, where for conceptual clarity we have reinstated $d\check{a}$ in the argument on the LHS, even though it vanishes as a differential cochain. The non-trivial phase $e^{2\pi i \mathcal{A}[\check{A},0,\check{a}]}$ acquired from the parallel transport over the mapping cylinder $X^d \times I$ tells us how the partition function $\mathcal{Z}[\check{A}]$ transforms under the {\it rigid} gauge transformation $\check{A} \to \check{A} - d\check{a} = \check{A}$. More importantly, all flat differential cocycles $\check{a} \in \check{Z}_\text{flat}^{p+1}(X^d)$ are automorphisms of the background gauge field $\check{A} \in \check{Z}^{p+2}(X^d)$, so physical consistency between theories related by rigid gauge transformations implies that, whenever $\mathcal{A}[\check{A},0,\check{a}] \notin \mathbb{Z}$, there is a single solution to \eqref{global_gauge_transformation_anomalous_phase_v2}, albeit trivial, namely,
\begin{equation}
    \mathcal{Z}[\check{A}] = 0 \, .
\end{equation}

\begin{figure}[t!]
    \centering
    \includegraphics[width=\textwidth]{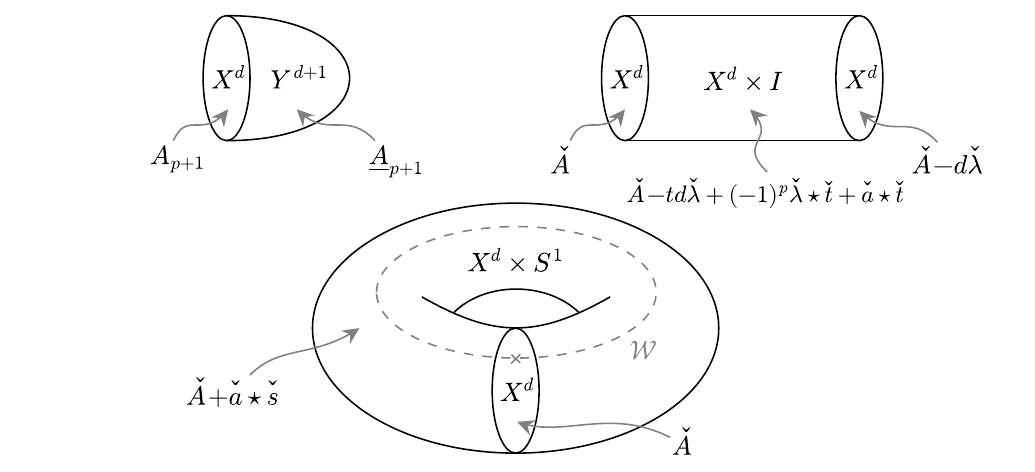}
    \caption{At the level of differential forms, the anomaly theory of a given QFT can be determined via inflow, and is supported on some $Y^{d+1}$ such that $\partial Y^{d+1}=X^d$. The connection $A_{p+1}$ is formally extended to $\underline{A}_{p+1}$ in the bulk. This statement can be promoted to the level of differential cochains by explicitly keeping track of the characteristic class of the gauge field, in which case the anomaly theory can be equivalently defined over the mapping cylinder $X^d \times I$. Here the bulk gauge field interpolates between two gauge-equivalent differential cocycles $\check{A},\check{A}-d\check{\lambda} \in \check{Z}^{p+2}(X^d)$. Crucially, there is an auxiliary degree of freedom $\check{a} \in \check{Z}_\text{flat}^{p+1}(X^d)$ corresponding to rigid gauge transformations of $\check{A}$. When $\check{\lambda}$ is trivial, we can glue the two ends of the cylinder to form the mapping torus $X^d \times S^1$. The anomaly $\mathcal{A}[\check{A},0,\check{a}]$ is then given by reducing the anomaly theory over $X^d \times S^1$. It is possible to ``cancel'' $\mathcal{A}[\check{A},0,\check{a}]$ by inserting a Wilson line $\mathcal{W}$ along $S^1$.}
    \label{mapping_torus_figure}
\end{figure}

As far as $\mathcal{A}[\check{A},0,\check{a}]$ is concerned, it will be convenient for us to carry out computations on the mapping torus $X^d \times S^1$ instead by gluing the two ends of the interval $I$, on which the bulk gauge field $\iota_{X \times \{0,1\}}^\ast \, \underline{\check{A}} = \check{A} \in \check{Z}^{p+2}(X^d)$ becomes identical when pulled back to $X^d$. In fact, $\mathcal{A}[\check{A},0,\check{a}]$ can now be interpreted as a holonomy of the polynomial $\mathcal{P}$ over $X^d \times S^1$. The practical advantage of doing so is that the mapping torus has no boundary, so if needed $\mathcal{P}$ can be further promoted to an invariant polynomial having support over a bounding manifold $Z^{d+2}$ such that $\partial Z^{d+2} = X^d \times S^1$.\footnote{In general, such an extension exists if and only if the pair $(X^d \times S^1,\mathsf{C}_{p+2})$ is trivial as an element of the bordism group $\Omega_{d+1}(K(\mathbb{Z},p+2))$, where $K(\mathbb{Z},p+2)$ is an Eilenberg-MacLane space whose only non-trivial homotopy group is $\pi_{p+2} = \mathbb{Z}$. The bordism group should be replaced by its suitable variants if we wish the extension to preserve some extra tangential structures. For instance, it is common to require the extension to preserve the Spin structure, in which case $S^1$ is null-bordant if and only if it has a Neveu-Schwarz, i.e.~bounding, Spin structure.} For example, we can take $Z^{d+2} = X^d \times D^2$, where $D^2$ is the solid 2-disk with $\partial D^2 = S^1$. More precisely, we would like to construct a differential cocycle $\check{\mathcal{P}} \in \check{H}^{d+2}(Z^{d+2})$ whose holonomy is given by $\chi(Z^{d+2}) = \exp(2\pi i \int_{X^d \times S^1} \mathcal{P}) = \mathcal{A}[\check{A},0,\check{a}]$. See Figure \ref{mapping_torus_figure} for an illustration.

For non-flat $\check{a} \in \check{Z}^{p+1}(X^d)$, there is no obvious interpretation in terms of gauge transformations of the background gauge field $\check{A} \in \check{Z}^{p+2}(X^d)$ in the $d$-dimensional QFT on $X^d$. One may naïvely try to use the correspondence \eqref{local_global_correspondence}, and regard $\check{a}$ as a {\it local} gauge transformation of $\check{A}$. However, this is not consistent with the fact that, by definition, a gauge equivalence is one that leaves the holonomy $\chi$ of a differential cocycle (in this case the anomaly polynomial $\mathcal{P}$) invariant, which corresponds to differential cochains, of one degree lower, that have vanishing curvature. This contradicts the assumption that $\check{a}$ is not flat. Henceforth, we restrict the auxiliary gauge field $\check{a} \in \check{Z}^{p+1}(X^d)$ to be flat.\footnote{See \cite{Diaconescu:2003bm,Belov:2004ht} for a related discussion of the more general case when $\check{a}$ is not flat.} As a side remark, the requirement of $\check{a}$ being flat formally amounts to imposing a {\it weight filtration} of a {\it differential function}, here taken to be the bulk gauge field $\underline{\check{A}}$, as defined by \cite{Hopkins:2002rd} (see also \cite{Diaconescu:2003bm} for a concise review). However, we will not make use of this language for the purpose of our work.

\begin{figure}[t!]
    \centering
    \includegraphics[width=\textwidth]{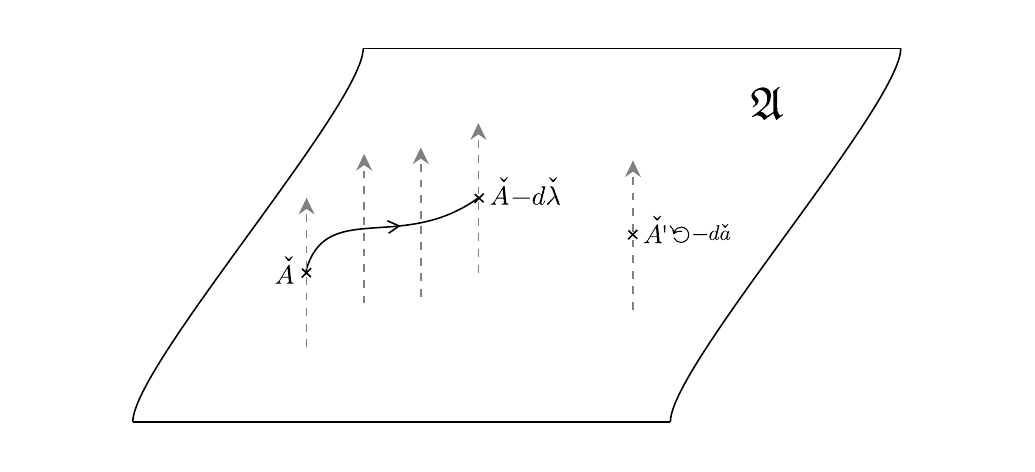}
    \caption{The partition function $\mathcal{Z}[\check{A}]$ is a section of a line bundle $\mathcal{L}$ over the space of gauge fields, $\mathfrak{A} \simeq \check{Z}^{p+2}(X^d)$. A 't Hooft anomaly is a phase $e^{2\pi i \mathcal{A}[\check{A},\check{\lambda},0]}$ acquired by $\mathcal{Z}[\check{A}]$ as we move along a non-trivial path $\check{A} \to \check{A} - d\check{\lambda}$ for some $\check{\lambda} \in \check{C}_\text{flat}^{p+1}(X^d)$. If the phase is trivial, then $\mathcal{L}$ can be lifted to a line bundle over $\mathfrak{A}/\mathfrak{G} \simeq \check{H}^{p+2}(X^d)$, where points along a given gauge orbit on $\mathfrak{A}$ are identified. On the other hand, a basepoint anomaly is a phase $e^{2\pi i \mathcal{A}[\check{A}',0,\check{a}]}$ acquired by $\mathcal{Z}[\check{A}']$ as we act on a fixed $\check{A}'$ with an automorphism $\check{A}' \to \check{A}'-d\check{a} = \check{A}'$ for some $\check{a} \in \check{Z}_\text{flat}^{p+1}(X^d)$. If such a phase is non-trivial, then $\mathcal{Z}[\check{A}']$ must vanish at this point on $\mathfrak{A}$.}
    \label{moduli_space_figure}
\end{figure}

It is important for us to remark on an important distinction between the anomalies $\mathcal{A}[\check{A},\check{\lambda},0]$ and $\mathcal{A}[\check{A},0,\check{a}]$, from the perspective of the space of gauge fields, $\mathfrak{A} \simeq \check{Z}^{p+2}(X^d)$. The former arises from the parallel transport of the partition function $\mathcal{Z}[\check{A}]$, as a section of the line bundle $\mathcal{L}$, when one moves along a path connecting $\check{A}$ and $\check{A} - d\check{\lambda}$ in $\mathfrak{A}$. There is nothing subtle about attaining a non-trivial phase $e^{2\pi i \mathcal{A}[\check{A},\check{\lambda},0]}$ from such a movement. The anomaly is merely the statement that the lift of $\cL$ from $\mathfrak{A}$ to $\mathfrak{A}/\mathfrak{G}$ is not single-valued.

In contrast, the latter arises when one simply sits at some fixed basepoint $\check{A} \in \mathfrak{A}$, so a non-trivial phase $e^{2\pi i \mathcal{A}[\check{A},0,\check{a}]}$ signifies that the section $\mathcal{Z}[\check{A}]$ itself has to vanish at this point, as depicted in Figure \ref{moduli_space_figure}. This happens because $\mathfrak{G}$ does not act freely, and its fixed points correspond precisely to the space of rigid gauge transformations, $\check{Z}_\text{flat}^{p+1}(X^d)$.\footnote{It is perhaps better to rephrase this picture in the language of {\it moduli stacks}. Loosely speaking, we should replace the moduli space with a lasagna-like object where each layer is a copy of $\mathfrak{A}$, then a rigid gauge transformation should correspond to a non-trivial path across layers while fixing the basepoint $\check{A}$ on each copy. This implies that the partition function $\mathcal{Z}[\check{A}]$, despite the notation, depends not only on the background gauge field, but also secretly on its automorphisms. A more detailed investigation on this perspective is left to future work. We also thank Victor Carmona for a useful discussion on this issue.} For this reason, we will hereafter call $\mathcal{A}[\check{A},0,\check{a}]$ the {\it basepoint anomaly}. See \cite{Seiberg:2018ntt,Meynet:2025zem} for a similar discussion in the context of anomalies of duality groups, in which case the ``gauge field'' is the coupling constant $\tau$, and $\mathfrak{A}/\mathfrak{G}$ becomes $\mathbb{H}/\text{SL}(2,\mathbb{Z})$, where $\mathbb{H}$ denotes the upper-half plane of $\mathbb{C}$.

Note that when $\mathcal{Z}[\check{A}] \neq 0$, the quantity $e^{2\pi i \mathcal{A}[\check{A},\check{\lambda},\check{a}]}$ satisfies a factorization property as follows. Compatibility with composition requires
\begin{align}
    \mathcal{Z}[\check{A} - d\check{a} - d\check{\lambda}] & = e^{2\pi i \mathcal{A}[\check{A} - d\check{a},\check{\lambda},0]} \mathcal{Z}[\check{A} - d\check{a}] = e^{2\pi i \mathcal{A}[\check{A},\check{\lambda},0]} \cdot e^{2\pi i \mathcal{A}[\check{A},0,\check{a}]} \mathcal{Z}[\check{A}]\nonumber\\
    & \stackrel{!}{=} e^{2\pi i \mathcal{A}[\check{A},\check{\lambda},\check{a}]} \mathcal{Z}[\check{A}]\label{anomaly_composition}
\end{align}
for any $\check{a} \in \check{Z}_\text{flat}^{p+1}(X^d)$ and $\check{\lambda} \in \check{C}_\text{flat}^{p+1}(X^d)$. Therefore, the anomaly decomposes as\footnote{Similarly, reversing the order of composition leads to $\mathcal{A}[\check{A},\check{\lambda},\check{a}] = \mathcal{A}[\check{A} - d\check{\lambda},0,\check{a}] + \mathcal{A}[\check{A},\check{\lambda},0] \ \text{mod} \ \mathbb{Z}$. Combining the two relations, one concludes that $\mathcal{A}[\check{A} - d\check{\lambda},0,\check{a}] = \mathcal{A}[\check{A},0,\check{a}] \ \text{mod} \ \mathbb{Z}$. This means that $\mathcal{A}[\check{A},0,\check{a}]$ depends on the background gauge field $\check{A}$ only through its class in $\check{H}^{p+2}(X^d)$.}
\begin{equation}
    \mathcal{A}[\check{A},\check{\lambda},\check{a}] = \mathcal{A}[\check{A},\check{\lambda},0] + \mathcal{A}[\check{A},0,\check{a}] \mod \mathbb{Z} \, .
\end{equation}
On the contrary, if $\mathcal{Z}[\check{A}] = 0$, then it follows from \eqref{anomaly_composition} that $\mathcal{Z}[\check{A} - d\check{\lambda}]$ also vanishes. Therefore, the vanishing of the partition function is a gauge-invariant statement.

\subsection{Relation to quantum Gauss law and SymTFT}

In the context where the symmetry in question is not a global symmetry but a {\it gauge symmetry} of the QFT, our analysis reproduces a previous result in the literature concerning the gauge invariance of physical wavefunctions \cite{Diaconescu:2003bm,Moore:2004jv,Belov:2004ht,Belov:2006jd,Freed:2006yc} (see also \cite{Hosseini:2025oka} for a recent study of anomaly cancellation in Type I string theory). Here, $\mathcal{L}$ is assumed to lift to a trivial line bundle over $\mathfrak{A}/\mathfrak{G}$, so the partition function is gauge-invariant, i.e.~$\mathcal{Z}[\check{A} - d\check{\lambda}] = \mathcal{Z}[\check{A}]$, and generally non-vanishing. Suppose we first neglect the auxiliary gauge field $\check{a} \in \check{Z}_\text{flat}^{p+1}(X^d)$, then we demand
\begin{equation}
	\mathcal{A}[\check{A},\check{\lambda},0] \in \mathbb{Z} \, ,
\end{equation}
which is known as the {\it classical Gauss law}.\footnote{At a semi-classical level, Gauss law constraints can be used to construct topological operators generating the respective global symmetries. See e.g.~\cite{Apruzzi:2022rei,Choi:2022fgx,Bah:2023ymy} for some recent applications to non-invertible symmetries.} On the contrary, one should also impose that the wavefunction is invariant under the action of non-trivial automorphisms $\check{a}$ of $\check{A}$. This further refines the constraint above, i.e.
\begin{equation}
	\mathcal{A}[\check{A},\check{\lambda},\check{a}] \in \mathbb{Z} \, ,
\end{equation}
referred to as the {\it quantum Gauss law}. In our language, such a refinement arises from the ambiguity of the basepoint anomaly $\mathcal{A}[\check{A},0,\check{a}]$ spanning over the space of rigid gauge transformations.

More precisely, despite giving similar conclusions, there is a subtle conceptual difference between \cite{Diaconescu:2003bm,Moore:2004jv,Belov:2004ht,Belov:2006jd,Freed:2006yc} and our approach. In the former, one is interested in wavefunctions $\Psi(\check{A}|_W)$ obtained after performing a Hamiltonian quantisation on the theory. This effectively treats $X^d$ as a manifold with boundary $\partial X^d = W^{d-1}$, such that the tangent bundle near the ``time-slice'' $W^{d-1}$ decomposes as $TX^d|_W \cong TW^{d-1} \oplus \mathbb{R}$. In our case, however, we are interested in the partition function $\mathcal{Z}[\check{A}]$ defined over the entire spacetime $X^d$.

Accordingly, $\Psi(\check{A}|_W)$ can be regarded as the partition function of the worldvolume theory associated with $W^{d-1}$, on which the gauge field $\check{A}|_W \in \check{Z}^{p+2}(W^{d-1})$ lives. The associated anomaly is given by the holonomy $\mathcal{A}[\check{A}|_W,\check{\lambda}|_W,\check{a}|_W] = \exp(2\pi i \int_{W^{d-1}} \mathcal{P}|_W)$ over the mapping torus $W^{d-1} \times S^1$. Note that such a holonomy is well-defined even for manifolds $X^d$ with boundary, because, by construction, we have defined the anomaly polynomial $\mathcal{P}$ directly over the mapping torus $X^d \times S^1$, without needing to invoke a bounding manifold $Y^{d+1}$ such that $\partial Y^{d+1} = X^d$. Lastly, demanding the wavefunction to be gauge-invariant then enforces the Gauss law $\mathcal{A}[\check{A}|_W,\check{\lambda}|_W,\check{a}|_W] \in \mathbb{Z}$ for all $\check{\lambda}|_W \in \check{C}_\text{flat}^{p+1}(W^{d-1})$ and $\check{a}|_W \in \check{Z}_\text{flat}^{p+1}(W^{d-1})$.

There is indeed a way to interpret our vanishing result for the partition function in terms of the quantum Gauss law. It has become a standard understanding in the modern literature on generalised symmetries that global symmetries in a $d$-dimensional QFT are associated with edge modes of gauge symmetries in a $(d+1)$-dimensional {\it Symmetry Topological Field Theory} (SymTFT) \cite{Apruzzi:2021nmk,Freed:2022qnc}. The anomaly theory is, loosely speaking, the SymTFT without the dynamics. For this particular discussion, it is more convenient to treat the latter (rather than the former) as a gauge theory in its own right, so that it makes more sense to talk about Gauss law constraints here. From the perspective of the SymTFT, the partition function $\mathcal{Z}[\check{A}]$ of the boundary theory supported on $X^d$ is identified as a wavefunction $\Psi(\underline{\check{A}}|_X)$ of the bulk theory, satisfying
\begin{equation}
    \Psi(\underline{\check{A}}|_X - d\underline{\check{a}}|_X - d\underline{\check{\lambda}}|_X) = e^{2\pi i \mathcal{A}[\underline{\check{A}}_X,\underline{\check{\lambda}}|_X,\underline{\check{a}}|_X]} \Psi(\underline{\check{A}}|_X) \, ,
\end{equation}
where the phase can be regarded as a Gauss law generator acting on charged states. We thus see that the wavefunction is gauge-invariant only if the bulk Gauss law
\begin{equation}
    \mathcal{A}[\underline{\check{A}}|_X,\underline{\check{\lambda}}|_X,\underline{\check{a}}|_X] \in \mathbb{Z}
\end{equation}
is satisfied, unless $\Psi(\underline{\check{A}}|_X)$ is itself vanishing. Importantly, even if $\check{\lambda}$ is trivial, i.e.~we are only acting with a rigid gauge transformation, one still demands the wavefunction to be annihilated by the Gauss law generator.\footnote{In this case, the Gauss law generator essentially corresponds to the rigid operator \eqref{rigid_operators}.} This is equivalent to our statement that the partition function vanishes whenever the basepoint anomaly is non-trivial.

\subsection{Insertion of sources}

The vanishing of the partition function $\mathcal{Z}[\check{A}]$ when $\mathcal{A}[\check{A},0,\check{a}]$ is non-trivial indicates that the correlation function (in the presence of a background gauge field) of general observables not charged under the symmetry in question,
\begin{equation}
    \langle \mathcal{O} \rangle_{\check{A}} = \mathcal{Z}_\mathcal{O}[\check{A}] \, ,
\end{equation}
evaluates to zero \cite{Witten:1999vg}. This is no longer true when $\mathcal{O} = \mathcal{O}[\check{A}]$ is an operator which also depends on the background gauge field $\check{A} \in \check{Z}^{p+2}(X^d)$, in a way such that its insertion enforces the modified basepoint anomaly
\begin{equation}
    \mathcal{A}_\mathcal{O}[\check{A},0,\check{a}] \stackrel{!}{=} 0
\end{equation}
to vanish identically as an element of $\mathbb{R}/\mathbb{Z}$, in the sense that $\mathcal{Z}_\mathcal{O}[\check{A} - d\check{a}] = \mathcal{Z}_\mathcal{O}[\check{A}]$ for any rigid gauge transformation $\check{a} \in \check{Z}^{p+2}_\text{flat}(X^d)$.

Heuristically, due to the Poincaré-Pontryagin duality \eqref{Poincaré-Pontryagin_duality}, such a constraint typically reduces to a vanishing condition on some linear combination of integral cohomology classes, in which case $\mathcal{O}$ is an (extended) operator sourced by the dual of $\check{A}$. From the perspective of the mapping torus $X^d \times S^1$, this amounts to the insertion of a ``Wilson line'' $\mathcal{W}$ extended along the circle $S^1$, such that $\mathcal{W}|_s = \mathcal{O}$ for any point $s \in S^1$, as illustrated in Figure \ref{mapping_torus_figure}. The statement will be made precise as we study explicit examples later.

It is important to note that the basepoint anomaly $\mathcal{A}[\check{A},0,\check{a}]$ is an ``anomaly'' not in the sense that the resultant theory is inconsistent, but simply that the corresponding partition function vanishes \cite{Hsieh:2020jpj,Yonekura:2024bvh}. As a result, it would not contribute to the path integral if one were to promote $\check{A}$ to a dynamical gauge field, e.g.~in a gravitational theory where all symmetries are expected to be gauged \cite{Harlow:2018tng}. Particularly, our formalism provides a unified approach to a variety of examples in string/M/F-theory, some of which were previously known, where certain ``tadpole constraints'' or ``consistency conditions'' are required to hold.


\section{BF theories}

\label{sec:BF}

The simplest class of examples in which there is a non-trivial basepoint anomaly is BF theory in generic spacetime dimensions. Any QFT whose anomaly is given by a BF-type theory can be analyzed similarly to the following discussion.

\subsection{Generalised Maxwell theory}

Concretely, let us consider generalised Maxwell theory in $d$ dimensions, with action
\begin{equation}
    S = -\frac{1}{4} \int_{X^d} F_{p+1} \wedge \ast F_{p+1} \, ,\label{Maxwell_action}
\end{equation}
where the differential form $F_{p+1}$ is the curvature of a dynamical differential cocycle $\check{A} \in \check{Z}^p(X^d)$, and $\ast$ is the Hodge star operator. This theory has a $p$-form $\U(1)_e$ electric symmetry and a $(d-p-2)$-form $\U(1)_m$ magnetic symmetry, arising respectively from the Bianchi identity and the equation of motion for $\check{A}$. We can therefore couple the theory to an electric background gauge field $\check{B} = (\mathsf{H}_{p+2},B_{p+1},H_{p+2}) \in \check{Z}^{p+2}(X^d)$ and a magnetic background gauge field $\check{C} = (\mathsf{G}_{d-p},C_{d-p-1},G_{d-p}) \in \check{Z}^{d-p}(X^d)$. At the level of differential forms, this amounts to replacing $F_{p+1}$ with $B_{p+1} + F_{p+1}$ in the action \eqref{Maxwell_action}, as well as adding the coupling $2\pi i \int_{X^d} C_{d-p-1} \wedge (B_{p+1} + F_{p+1})$.

These two global symmetries have a mixed anomaly given by the product, $(-1)^{d-p} \underline{\check{C}} \star \underline{\check{B}}$, of differential cocycles \cite{Hsieh:2020jpj}, or more specifically,
\begin{equation}
    \mathcal{A}[\underline{\check{B}},\underline{\check{C}}] = (-1)^{d-p} \int_{Y^{d+1}} \Big(\underline{C}_{d-p-1} \cup \underline{H}_{p+2} + (-1)^{d-p} \underline{\mathsf{G}}_{d-p} \cup \underline{B}_{p+1} + Q(\underline{G}_{d-p},\underline{H}_{p+2})\Big) \, .\label{Maxwell_anomaly}
\end{equation}
Via inflow, the anomaly theory above is a BF theory supported on a $(d+1)$-dimensional bulk $Y^{d+1}$ whose boundary is $\partial Y^{d+1} = X^d$, while $\underline{\check{B}},\underline{\check{C}} \in \check{Z}(Y^{d+1})$ are some extensions of $\check{B},\check{C} \in \check{Z}(X^d)$ onto $Y^{d+1}$.

Now we place the anomaly theory on the mapping cylinder $X^d \times I$. We would like to demonstrate how the partition function $\mathcal{Z}[\check{B},\check{C}]$ changes under the gauge transformations $\check{B} \to \check{B} - d\check{\lambda}^B$ and $\check{C} \to \check{C} - d\check{\lambda}^C$ for some $\check{\lambda}^B \in \check{C}_\text{flat}^{p+1}(X^d)$ and $\check{\lambda}^C \in \check{C}_\text{flat}^{d-p-1}(X^d)$.\footnote{Note the abuse of notation between $\check{C}$ as a differential cocycle and $\check{C}^\ast(X^d)$ as a cochain complex.} Following \eqref{bulk_gauge_field_parametrization}, the gauge fields on $X^d \times I$ can be parametrised as
\begin{equation}
    \begin{aligned}
        \underline{\check{B}} & = \check{B} - t d\check{\lambda}^B + (-1)^p \check{\lambda}^B \star \check{t} + \check{b} \ast \check{t} \, ,\\
        \underline{\check{C}} & = \check{C} - t d\check{\lambda}^C + (-1)^{d-p} \check{\lambda}^C \star \check{t} + \check{c} \ast \check{t} \, ,
    \end{aligned}
\end{equation}
for some $\check{b},\check{c} \in \check{Z}_\text{flat}^\ast(X^d)$. To evaluate the anomaly, we replace $Y^{d+1}$ with $X^d \times I$ in \eqref{Maxwell_anomaly} and further reduce over the interval $I$. The result is
\begin{align}
    \mathcal{A}[\check{B},\check{C},\check{\lambda}^B,\check{\lambda}^C,\check{b},\check{c}] & = \int_{X^d} \bigg((-1)^{d+1} (\Lambda^C_{d-p} + \delta\lambda^C_{d-p-2}) \cup B_{p+1}\nonumber\\
    & \phantom{=\ } + (-1)^p (\lambda^C_{d-p-2} + (-1)^{d-p} c_{d-p-2}) \cup \mathsf{H}_{p+2}\nonumber\\
    & \phantom{=\ } + (-1)^p \mathsf{G}_{d-p} \cup (\lambda^B_p + (-1)^p b_p)\nonumber\\
    & \phantom{=\ } + \frac{1}{2} \, (-1)^{d-p} (\Lambda^C_{d-p-1} + (-1)^{d-p} \mathsf{g}_{d-p-1}) \cup (\Lambda^B_{p+1} + \delta\lambda^B_p)\bigg) \, .
\end{align}
There are two special cases worth mentioning. The first of which is the limit where we turn off $\check{b}$ and $\check{c}$, and set $\Lambda^B_{p+1} = \Lambda^C_{d-p-1} = 0$, giving us
\begin{equation}
    \mathcal{A}[\check{B},\check{C},\check{\lambda}^B,\check{\lambda}^C,0,0] = \int_{X^d} \delta\lambda^C_{d-p-2} \cup B_{p+1} \, .
\end{equation}
Evidently, this is the standard perturbative anomaly of (generalised) Maxwell theory, which arises as the gauge variation of the topological term in the action. The partition function thus changes as
\begin{equation}
    \mathcal{Z}[\check{B} - d\check{\lambda}^B,\check{C} - d\check{\lambda}^C] = e^{2\pi i \mathcal{A}[\check{B},\check{C},\check{\lambda}^B,\check{\lambda}^C,0,0]} \mathcal{Z}[\check{B},\check{C}] \, ,
\end{equation}
confirming our prescription that the anomaly can be obtained from the mapping cylinder.

The second special case is when we set $\check{\lambda}^B = \check{\lambda}^C = 0$, and leave $\check{b},\check{c} \in \check{Z}_\text{flat}^\ast(X^d)$ generic. This yields the basepoint anomaly
\begin{equation}
    \mathcal{A}[\check{B},\check{C},0,0,\check{b},\check{c}] = \int_{X^d} \Big((-1)^d [c_{d-p-2}]_{\mathbb{R}/\mathbb{Z}} \cup [\mathsf{H}_{p+2}]_\mathbb{Z} + [\mathsf{G}_{d-p}]_\mathbb{Z} \cup [b_p]_{\mathbb{R}/\mathbb{Z}}\Big) \in \mathbb{R}/\mathbb{Z} \, .
\end{equation}
As discussed in Section \ref{differential_cohomology_review}, here we have made use of the fact that (isomorphism classes of) flat differential cocycles are equivalent to cohomology classes with $\mathbb{R}/\mathbb{Z}$ coefficients, so the terms above descend to a perfect pairing via the Poincaré-Pontryagin duality.

\subsubsection{Derivation with the mapping torus}

The basepoint anomaly can alternatively be derived by evaluating the anomaly theory on the mapping torus $X^d \times S^1$. In this case, the role of $\check{t} = (\delta t,0,dt) \in \check{Z}^1(I)$ should be replaced by a differential 1-cocycle
\begin{equation}
    \check{s} = (\phi,s,\text{vol}(S^1)) \in \check{Z}^1(S^1) \, ,
\end{equation}
where $\phi \in Z^1(S^1;\mathbb{Z})$ is a representative of the fundamental class of $S^1$, and $\text{vol}(S^1) \in \Omega_\mathbb{Z}^1(S^1)$ is the volume form of $S^1$, such that the connection $s \in C^0(S^1;\mathbb{R})$ satisfies $\delta s = \text{vol}(S^1) - \phi$.

The bulk gauge fields on $X^d \times S^1$ are parametrised as $\underline{\check{B}} = \check{B} + \check{b} \star \check{s}$ and $\underline{\check{C}} = \check{C} + \check{c} \star \check{s}$. More explicitly, the components of $\underline{\check{B}}$ are given by
\begin{equation}
    \begin{aligned}
	\underline{\mathsf{H}}_{p+2} & = \mathsf{H}_{p+2} + \mathsf{h}_{p+1} \cup \phi \, ,\\
	\underline{B}_{p+1} & = B_{p+1} + b_p \cup \text{vol}(S^1) + (-1)^{p+1} \mathsf{h}_{p+1} \cup s \, ,\\
	\underline{H}_{p+2} & = H_{p+2} + h_{p+1} \wedge \text{vol}(S^1) \, ,
    \end{aligned}
\end{equation}
and likewise for $\underline{\check{C}}$. It can then be shown that reducing the anomaly theory
\begin{equation}
    \mathcal{A}[\underline{\check{B}},\underline{\check{C}}] = (-1)^{d-p} \int_{X^d \times S^1} \Big(\underline{C}_{d-p-1} \cup \underline{H}_{p+2} + (-1)^{d-p} \underline{\mathsf{G}}_{d-p} \cup \underline{B}_{p+1} + Q(\underline{G}_{d-p},\underline{H}_{p+2})\Big)
\end{equation}
over $S^1$ gives rise to
\begin{equation}
    \mathcal{A}[\check{B},\check{C},\check{b},\check{c}] = \int_{X^d} \Big((-1)^d [c_{d-p-2}]_{\mathbb{R}/\mathbb{Z}} \cup [\mathsf{H}_{p+2}]_\mathbb{Z} + [\mathsf{G}_{d-p}]_\mathbb{Z} \cup [b_p]_{\mathbb{R}/\mathbb{Z}}\Big) \, ,\label{Maxwell_basepoint_anomaly}
\end{equation}
which is precisely the same basepoint anomaly we obtained earlier using the mapping cylinder construction.

Suppose the characteristic classes $[\mathsf{H}_{p+2}]_\mathbb{Z},[\mathsf{G}_{d-p}]_\mathbb{Z} \in H^\ast(X^d;\mathbb{Z})$ are non-trivial, then since the auxiliary gauge fields $\check{b},\check{c} \in \check{Z}_\text{flat}^\ast(X^d)$ are arbitrary, the basepoint anomaly \eqref{Maxwell_basepoint_anomaly} is generally a non-trivial element of $\mathbb{R}/\mathbb{Z}$ due to the perfectness of the pairing. At the same time, by definition, the partition function changes under the rigid gauge transformations $\check{B} \to \check{B} - d\check{b} = \check{B}$ and $\check{C} \to \check{C} - d\check{c} = \check{C}$ as
\begin{equation}
    \mathcal{Z}[\check{B},\check{C}] = e^{2\pi i \mathcal{A}[\check{B},\check{C},\check{b},\check{c}]} \mathcal{Z}[\check{B},\check{C}] \, .
\end{equation}
These two statements together imply that $\mathcal{Z}[\check{B},\check{C}]$ must be vanishing. In other words, the partition function is non-vanishing only if
\begin{equation}
    [\mathsf{H}_{p+2}]_\mathbb{Z} = [\mathsf{G}_{d-p}]_\mathbb{Z} = 0 \, ,
\end{equation}
i.e.~the gauge fields $\check{B},\check{C}$ are topologically trivial, hence proving the claim in \cite{Hsieh:2020jpj}.

\subsubsection{Flatness of auxiliary gauge fields}

In Section \ref{vanishing_theorem_section}, we already justified why the auxiliary gauge fields $\check{b}$ and $\check{c}$ should be flat, otherwise they would not qualify as rigid gauge transformations. Let us also provide a practical argument for why this has to be the case. Note that if the curvatures $h_{p+1}$ and $g_{d-p-1}$ were non-zero, then the terms $Q(h_{p+1},\text{vol}(S^1))$ and $Q(g_{d-p-1},\text{vol}(S^1))$ would contribute respectively in $\check{b} \star \check{s} \subset \underline{B}_{p+1}$ and $\check{c} \star \check{s} \subset \underline{C}_{d-p-1}$, such that the basepoint anomaly depends the choice of chain homotopy between the wedge product $\wedge$ for differential forms and the cup product $\cup$ for cochains.

In fact, we would have found extra terms in the basepoint anomaly,
\begin{align}
    \mathcal{A}'[\check{B},\check{C},\check{b},\check{c}] & \supset (-1)^{d-p} \int_{X^d \times S^1} \Big(Q(g_{d-p-1},\text{vol}(S^1)) \cup H_{p+2}\nonumber\\
    & \phantom{\supset\ } + (-1)^{d-p} \mathsf{G}_{d-p} \cup Q(h_{p+1},\text{vol}(S^1)) + Q(G_{d-p},h_{p+1} \wedge \text{vol}(S^1))\nonumber\\
    & \phantom{\supset\ } + Q(g_{d-p-1} \wedge \text{vol}(S^1),H_{p+2})\Big) \, .
\end{align}
The choice of the chain homotopy $Q(\alpha,\beta) \in C^{|\alpha|+|\beta|-1}(X^d \times S^1;\mathbb{R})$ is far from being unique. As an example, one can always pick a different definition of cup product $\cup': C^p(X^d \times S^1;\mathbb{R}) \times C^q(X^d \times S^1;\mathbb{R}) \to C^{p+q}(X^d \times S^1;\mathbb{R})$ at the level of cochains, and therefore a different chain homotopy $Q'(\alpha,\beta) \in C^{|\alpha|+|\beta|-1}(X^d \times S^1;\mathbb{R})$.

For the phase $e^{2\pi i \mathcal{A}[\check{B},\check{C},\check{b},\check{c}]}$ to be well-defined, we need $\mathcal{A}[\check{B},\check{C},\check{b},\check{c}] = \mathcal{A}'[\check{B},\check{C},\check{b},\check{c}]$ (mod $\mathbb{Z}$) for arbitrary choices of the cup product $\cup'$. This is satisfied if and only if at least one of the arguments of $Q(\alpha,\beta)$ is trivial. Since $G_{d-p}$ and $H_{p+2}$ are fixed input data that we seek to constrain, we are only left with the option to impose that $h_{p+1} = 0$ and $g_{d-p-1} = 0$, i.e.~the auxiliary gauge fields $\check{b}$ and $\check{c}$ have to be flat.

\subsubsection{Ordering (un)ambiguity}

As was noted in \cite{Hopkins:2002rd,Hsieh:2020jpj,GarciaEtxebarria:2024fuk}, unlike differential cohomology classes which are graded-commutative, the following two products of differential cocycles,
\begin{equation}
    \check{A} \star \check{A}' \, , \qquad (-1)^{(p+2)(q+2)} \check{A}' \star A \, ,
\end{equation}
for any $\check{A} \in \check{Z}^{p+2}(X^d)$ and $\check{A}' \in \check{Z}^{q+2}(X^d)$ are equivalent only up to an exact differential cocycle. One may therefore worry that there is an ordering ambiguity in the way we write down the anomaly theory \eqref{Maxwell_anomaly}.

Thanks to the fact that the difference between the two products is exact, one can explicitly check that when we evaluate
\begin{equation}
    \mathcal{A}[\underline{\check{C}},\underline{\check{B}}] = (-1)^{(d-p)(p+1)} \int_{X^d \times S^1} \Big(\underline{B}_{p+1} \cup \underline{G}_{d-p} + (-1)^p \underline{\mathsf{H}}_{p+2} \cup \underline{C}_{d-p-1} + Q(\underline{H}_{p+2},\underline{G}_{d-p})\Big) \, ,\label{Maxwell_anomaly_alternative_ordering}
\end{equation}
we will recover our previous result except for a total derivative, which actually vanishes since the mapping torus is closed. Therefore, the basepoint anomaly $\mathcal{A}[\check{B},\check{C},\check{b},\check{c}]$ is independent of the choice of ordering of gauge fields.

The same argument does not hold on the mapping cylinder $X^d \times I$, whose boundary is given by two copies of $X^d$. As a result, the full anomaly $\mathcal{A}[\check{B},\check{C},\check{\lambda}^B,\check{\lambda}^C,\check{b},\check{c}]$ does generally depend on the choice of ordering. This is not surprising, since the 't Hooft anomaly is defined only up to local counterterms in the action.

\subsubsection{Wilson and 't Hooft operators}

Consider the insertion of a Wilson operator
\begin{equation}
    \mathcal{W}(\Sigma^p) \coloneqq \exp\bigg(2\pi i (-1)^{(d-p)(p+1)} \int_{X^d} A_p \cup \delta(\Sigma^p)\bigg)
\end{equation}
along some $p$-cycle $\Sigma^p \in Z_p(X^d;\mathbb{Z})$.\footnote{The sign convention is chosen to better suit our definition of the anomaly theory.} The cocycle $\delta(\Sigma^p) \in \Omega_\mathbb{Z}^{d-p}(X^d)$ is a representative of the Poincaré dual of $[\Sigma^p] \in H_p(X^d;\mathbb{Z})$. For convenience, let us define a differential cocycle
\begin{equation}
    \check{\delta}(\Sigma^p) = (\varphi,\sigma,\delta(\Sigma^p)) \in \check{Z}^{d-p}(X^d) \, ,
\end{equation}
where $[\varphi] = \text{PD}([\Sigma^p])$ such that (with an abuse of notation) $\delta\sigma = \delta(\Sigma^p) - \varphi$. The anomaly theory \eqref{Maxwell_anomaly_alternative_ordering} then gets modified with $\check{C} \to \check{C} + \check{\delta}(\Sigma^p)$. In terms of the mapping torus, we can regard this as inserting a $(p+1)$-dimensional Wilson ``line'' with a leg extended along the circle $S^1$, i.e.~we have a copy of $\mathcal{W}(\Sigma^p)$ inserted on every ``time''-slice. If we reduce the effective anomaly theory over $S^1$, then the new conditions for the partition function (with $\mathcal{W}(\Sigma^p)$ inserted) to be non-vanishing become
\begin{equation}
    [\mathsf{H}_{p+2}]_\mathbb{Z} = 0 \, , \qquad [\mathsf{G}_{d-p}]_\mathbb{Z} + \text{PD}([\Sigma^p]) = 0 \, .
\end{equation}

By the same token, we can also insert a 't Hooft operator
\begin{equation}
    \mathcal{T}(\Sigma^{d-p-2}) \coloneqq \exp\bigg(2\pi i (-1)^{d-p} \int_{X^d} \tilde{A}_{d-p-2} \cup \delta(\Sigma^{d-p-2})\bigg)
\end{equation}
along some $(d-p-2)$-cycle $\Sigma^{d-p-2} \in Z_{d-p-2}(X^d;\mathbb{Z})$, where $\tilde{A}_{d-p-2}$ is the dual gauge field such that $\ast F_{p+1} = d\tilde{A}_{d-p-2}$ locally. This amounts to shifting $\check{B} \to \check{B} + \check{\delta}(\Sigma^{d-p-2})$ in the anomaly theory \eqref{Maxwell_anomaly}. In general, with both the Wilson and 't Hooft operators inserted, the conditions for the partition function to be non-vanishing are
\begin{equation}
    [\mathsf{H}_{p+2}]_\mathbb{Z} + \text{PD}([\Sigma^{d-p-2}]) = 0 \, , \qquad [\mathsf{G}_{d-p}]_\mathbb{Z} + \text{PD}([\Sigma^p]) = 0 \, .
\end{equation}

We discussed in Section \ref{intro} that the insertion of a Wilson or 't Hooft operator (i.e.~the charged operators of Maxwell theory) generally leads to a vanishing correlator, but it becomes non-zero precisely when the conditions above are satisfied, which reflects the fact that Maxwell theory equipped with non-trivial cohomology classes $[\mathsf{H}_{p+2}]_\mathbb{Z},[\mathsf{G}_{d-p}]_\mathbb{Z}$ is a perfectly well-defined theory. Despite having a vanishing partition function, one should not interpret such a theory as being inconsistent, as was pointed out by \cite{Yonekura:2024bvh}.

\subsection{Freed-Witten anomaly}\label{FW_anomaly_section}

There is a straightforward adaptation of the previous analysis, where we now take the dynamical $\U(1)$ gauge field in 4d Maxwell theory to rather be a $\text{Spin}^c \coloneqq \text{Spin} \times_{\mathbb{Z}_2} \U(1)$ gauge field (also known as all-fermion electrodynamics). In this case, one defines a flat differential cocycle
\begin{equation}
    \check{w} = (\mathsf{W}_3,w_2,0) \in \check{Z}_\text{flat}^3(X^4) \, ,
\end{equation}
where $w_2 \in C^2(X^4;\mathbb{R})$ is an uplift of the $\mathbb{Z}_2$-valued {\it second Stiefel-Whitney class} of $X^4$ to a real cochain \cite{Hsieh:2020jpj}, via the following diagram.
\begin{equation}
    \begin{tikzcd}
        0 \arrow[r] & \mathbb{Z} \arrow[r,"\times 2"] \arrow[d,"="] & \mathbb{Z} \arrow[r,"\text{mod} \, 2"] \arrow[d] & \mathbb{Z}_2 \arrow[r] \arrow[d] & 0 \\
        0 \arrow[r] & \mathbb{Z} \arrow[r] & \mathbb{R} \arrow[r] & \mathbb{R}/\mathbb{Z} \arrow[r] & 0
    \end{tikzcd}\label{Z2_embedding_diagram}
\end{equation}
By construction, $\delta w_2 = -\mathsf{W}_3 \in Z^3(X^4;\mathbb{Z})$ is equal to the third integral Stiefel-Whitney class of $X^4$, or equivalently,
\begin{equation}
    \beta([w_2]_{\mathbb{R}/\mathbb{Z}}) =  [\mathsf{W}_3]_\mathbb{Z} \, ,
\end{equation}
where $\beta: H^2(X^4;\mathbb{R}/\mathbb{Z}) \to H^3(X^4;\mathbb{Z})$ is the Bockstein homomorphism induced from the short exact sequence $0 \to \mathbb{Z} \to \mathbb{R} \to \mathbb{R}/\mathbb{Z} \to 0$.

As argued by \cite{Kapustin:2014tfa,Thorngren:2014pza,Wang:2018qoy,Hsieh:2020jpj,Yonekura:2024bvh}, the anomaly theory of 4d Maxwell theory with a $\text{Spin}^c$ gauge field is obtained effectively by replacing $\underline{\check{B}} \to \underline{\check{B}} + \underline{\check{w}}$ and $\underline{\check{C}} \to \underline{\check{C}} + \underline{\check{w}}$ in \eqref{Maxwell_anomaly}. More explicitly, we have
\begin{align}
    \mathcal{A}[\underline{\check{B}},\underline{\check{C}},\underline{\check{w}}] & = \int_{Y^5} \Big(\underline{B}_2 \cup \underline{G}_3 - \underline{\mathsf{H}}_3 \cup \underline{C}_2 + Q(\underline{H}_3,\underline{G}_3) + \underline{w}_2 \cup \underline{G}_3 - \underline{\mathsf{W}}_3 \cup \underline{C}_2\nonumber\\
    & \phantom{=\ } - \underline{\mathsf{H}}_3 \cup \underline{w}_2 - \underline{\mathsf{W}}_3 \cup \underline{w}_2\Big) \, .
\end{align}
One can then compute the basepoint anomaly as before on the mapping torus $X^4 \times S^1$, and deduce that the partition function is non-vanishing only if
\begin{equation}
    [\mathsf{H}_3]_\mathbb{Z} + [\mathsf{W}_3]_\mathbb{Z} = [\mathsf{G}_3]_\mathbb{Z} + [\mathsf{W}_3]_\mathbb{Z} = 0 \, .\label{Freed-Witten_anomaly_cancellation_condition}
\end{equation}
Moreover, since $[\mathsf{W}_3]_\mathbb{Z} = \beta([w_2]_{\mathbb{R}/\mathbb{Z}})$ with $[w_2]_{\mathbb{R}/\mathbb{Z}}$ being 2-torsion, by exactness it must be that $2[\mathsf{H}_3]_\mathbb{Z} = 2[\mathsf{G}_3]_\mathbb{Z} = 0$ as well.\footnote{See a similar consistency condition recently derived by \cite{Kim:2025fpz} on 5d $\mathcal{N}=1$ $\text{SU}(2)$ super Yang-Mills theory on $X^4 \times S^1$, wherein its relation to K-theoretic Donaldson invariants was discussed.}

This recovers the celebrated {\it Freed-Witten anomaly cancellation condition} for D-branes in Type IIB string theory \cite{Freed:1999vc}, as well as its S-dual counterpart \cite{Diaconescu:2000wy,Yonekura:2024bvh}. Note that all orientable 4-manifolds admit $\text{Spin}^c$ structures \cite{Teichner:1994abc}, in which case $[\mathsf{W}_3]_\mathbb{Z} = 0$. Again we stress that a violation of \eqref{Freed-Witten_anomaly_cancellation_condition} does not imply that placing the D-brane on such a background is prohibited, but rather that its partition function is necessarily vanishing.

As we saw in the Maxwell theory example, one can insert Wilson and 't Hooft operators to trivialise the basepoint anomaly. In the context of the D3-brane, the Wilson operator is an F1-string ending on a 1-submanifold of $X^4$, whereas the 't Hooft operator is a D1-string also ending on a 1-submanifold of $X^4$. With these sources inserted, the non-vanishing conditions for the D3-brane partition function are schematically
\begin{equation}
    [\mathsf{H}_3]_\mathbb{Z} + [\mathsf{W}_3]_\mathbb{Z} + \text{PD}([\text{D1}]) = 0 \, , \qquad [\mathsf{G}_3]_\mathbb{Z} + [\mathsf{W}_3]_\mathbb{Z} + \text{PD}([\text{F1}]) = 0 \, .\label{D3-brane_FW_anomaly_with_sources}
\end{equation}
Consequently, even if the manifold is $\text{spin}^c$, one may still be able to have a non-vanishing D3-brane partition function in a background with non-trivial $[\mathsf{H}_3]_\mathbb{Z}$ or $[\mathsf{G}_3]_\mathbb{Z}$, as long as the appropriate sources are inserted \cite{Witten:1998xy,Maldacena_2001}.

In the derivation of the basepoint anomaly above, the shifts $\underline{\check{B}} \to \underline{\check{B}} + \underline{\check{w}}$ and $\underline{\check{C}} \to \underline{\check{C}} + \underline{\check{w}}$ may seem somewhat ad hoc. However, in Section \ref{S-folds_section} we will see how they arise from a top-down perspective by studying the dimensional reduction of an M5-brane. We will see how the M5-brane analysis also constrains other terms in the D3-brane action which have not been discussed here.

\subsection{Dijkgraaf-Witten theory}

With some minor modifications, we can apply our earlier results to theories whose anomaly is given by a Dijkgraaf-Witten theory \cite{Dijkgraaf:1989pz} with $\mathbb{Z}_N$ gauge fields. One way to model such discrete gauge fields with differential cocycles is to uplift them to real cochains, analogously to what we did in the diagram \eqref{Z2_embedding_diagram}. Alternatively, we can directly define such a gauge field as a doublet, i.e.
\begin{equation}
    \check{Z}^{p+2}(X^d) = \{\check{A} = (\mathsf{C}_{p+2},A_{p+1}) \in Z^{p+2}(X^d;\mathbb{Z}) \times C^{p+1}(X^d;\mathbb{Z}_N)\} \, .
\end{equation}
By construction, a differential cocycle under this definition is automatically flat, i.e.~it has a vanishing field strength. Its connection and characteristic class are related by $\delta A_{p+1} = -\mathsf{C}_{p+2}$ mod $N$, or equivalently,
\begin{equation}
    \beta([A_{p+1}]_{\mathbb{Z}_N}) = [\mathsf{C}_{p+2}]_\mathbb{Z} \, ,
\end{equation}
where the Bockstein homomorphism is that induced from the short exact sequence $0 \to \mathbb{Z} \to \mathbb{Z} \to \mathbb{Z}_N \to 0$.\footnote{There are slightly different versions of Bockstein homomorphisms used throughout this work, but it should be clear from context which particular one is being invoked.}

The product between two differential cocycles $\check{A} \in \check{Z}^{p+2}(X^d)$ and $\check{A}' \in \check{Z}^{q+2}(X^d)$ is the doublet,
\begin{equation}
    \check{A} \star \check{A}' = (\mathsf{C}_{p+2} \cup \mathsf{C}_{q+2}',(-1)^{p+2} \mathsf{C}_{p+2} \cup A_{q+1}') \, .
\end{equation}
One can readily check that this product is closed under the differential $d$. Similarly to the ordinary case, the differential cohomology group $\check{H}^{p+2}(X^d) = \check{Z}^{p+2}(X^d)/\sim$ is defined via the gauge equivalence,
\begin{equation}
    (\mathsf{C}_{p+2},A_{p+1}) \sim (\mathsf{C}_{p+2} - \delta\Lambda_{p+1},A_{p+1} + \delta\lambda_p + \Lambda_{p+1} \, \text{mod} \, N) \, ,
\end{equation}
where $\Lambda_{p+1} \in C^{p+1}(X^d;\mathbb{Z})$ and $\lambda_p \in C^p(X^d;\mathbb{Z}_N)$.

At the level of differential cocycles, the Dijkgraaf-Witten theory can be expressed, up to a prefactor, as the product, $\underline{\check{C}} \star \underline{\check{B}}$, of the magnetic and electric background gauge fields. Explicitly, the anomaly theory is
\begin{equation}
    \mathcal{A}[\underline{\check{B}},\underline{\check{C}}] = \frac{1}{N} \int_{Y^{d+1}} \underline{\mathsf{G}}_{d-p} \cup \underline{B}_{p+1} \, .\label{DW_anomaly}
\end{equation}
We can place this theory on the mapping torus $X^d \times S^1$, and parametrise the bulk gauge fields as $\underline{\check{B}} = \check{B} + \check{s} \star \check{b}$ and $\underline{\check{C}} = \check{C} + \check{s} \star \check{c}$, where $\check{s} = (\phi,0) \in \check{Z}^1(S^1)$. Reducing over $S^1$ then yields the basepoint anomaly as an integral over $X^d$,
\begin{equation}
    \mathcal{A}[\check{B},\check{C},\check{b},\check{c}] = \frac{(-1)^{p+1}}{N} \int_{X^d} \Big([c_{d-p-2}]_{\mathbb{Z}_N} \cup [\mathsf{H}_{p+2}]_\mathbb{Z} + [\mathsf{G}_{d-p}]_\mathbb{Z} \cup [b_p]_{\mathbb{Z}_N}\Big) \, ,
\end{equation}
which is essentially the same as \eqref{Maxwell_basepoint_anomaly}, except for the change in coefficients.

Since $[b_p]_{\mathbb{Z}_N}, [c_{d-p-2}]_{\mathbb{Z}_N} \in H^\ast(X^d;\mathbb{Z}_k)$ are arbitrary, the perfect pairing
\begin{equation}
	H^n(X^d;\mathbb{Z}_N) \times H^{d-n}(X^d;\mathbb{Z}) \to \mathbb{Z}_N
\end{equation}
leads to the conclusion that the phase $e^{2\pi i \mathcal{A}[\check{B},\check{C}]}$ is non-trivial. Hence, the partition function is non-vanishing only if
\begin{equation}
    [\mathsf{H}_{p+2}]_\mathbb{Z} = [\mathsf{G}_{d-p}]_\mathbb{Z} = 0 \, .
\end{equation}
By exactness, this is equivalent to the condition that $[B_{p+1}]_{\mathbb{Z}_N}$ is the mod $N$ reduction of some element in $H^p(X^d;\mathbb{Z})$, and similarly for $[C_{d-p-1}]_{\mathbb{Z}_N}$.

\subsection{4d $\mathfrak{u}(N)$ gauge theory}

Let us review the construction of a 4d $\mathfrak{u}(N)$ gauge theory, making use of the precriptions by \cite{Kapustin:2014gua,Gaiotto:2014kfa,Gaiotto:2017yup}. We start with a 4d Yang-Mills theory where the global form of its gauge group is $G = \text{SU}(N) \times_{\mathbb{Z}_k} \U(1)$ for some generic divisor $k$ of $N$, whose connection can be parametrised as
\begin{equation}
    A_1^G = A_1^{\mathfrak{su}(N)} + \frac{1}{k} \, A_1^{\mathfrak{u}(1)} \mathds{1}_N \, ,
\end{equation}
such that $\text{Tr}\big(F_2^G\big) = F_2^{\mathfrak{u}(1)}$. Note that in the special case of $k=N$, we recover $G = \text{SU}(N) \times_{\mathbb{Z}_N} \U(1) \cong \U(N)$. The quotienting of the $\U(1)$ gauge field by $\mathbb{Z}_k$ is implemented by imposing the gauge equivalence,
\begin{equation}
    A_1^{\mathfrak{u}(1)} \sim A_1^{\mathfrak{u}(1)} - k \lambda_1 \, ,\label{U(1)/Zk_gauge_symmetry}
\end{equation}
where $\lambda_1$ is a $\mathfrak{u}(1)$-valued $1$-form.

The action of the gauge theory is given by
\begin{equation}
    S_\text{YM} = S_\text{kin} + S_\theta = -\frac{4\pi^2}{g^2} \int_{X^4} \text{Tr}\big(F_2^G \wedge \ast F_2^G\big) + \frac{i\theta}{2} \int_{X^4} \text{Tr}\big(F_2^G \wedge F_2^G\big) \, ,
\end{equation}
where the theta term can be further expanded into
\begin{equation}
    S_\theta = S_\theta^{\mathfrak{su}(N)} + S_\theta^{\mathfrak{u}(1)} = \frac{i\theta}{2} \int_{X^4} \text{Tr}\big(F_2^{\mathfrak{su}(N)} \wedge F_2^{\mathfrak{su}(N)}\big) + \frac{iN\theta}{2k^2} \int_{X^4} F_2^{\mathfrak{u}(1)} \wedge F_2^{\mathfrak{u}(1)} \, .
\end{equation}
Similarly to Maxwell theory, we may couple the system to a pair of $\mathfrak{u}(1)$-valued background gauge fields $B_2,C_2$, i.e.
\begin{gather}
    S_\text{kin}[B_2] = -\frac{4\pi^2}{g^2} \int_{X^4} \text{Tr}\Big(\big(B_2 \mathds{1}_N + F_2^G\big) \wedge \ast \big(B_2 \mathds{1}_N + F_2^G\big)\Big) \, ,\\
    S_\theta^{\mathfrak{u}(1)}[B_2] = \frac{iN\theta}{2k^2} \int_{X^4} \big(kB_2 + F_2^{\mathfrak{u}(1)}\big) \wedge \big(kB_2 + F_2^{\mathfrak{u}(1)}\big) \, ,\\
    S_\text{mixed}[B_2,C_2] = \frac{iN}{k} \int_{X^4} C_2 \wedge \big(kB_2 + F_2^{\mathfrak{u}(1)}\big) \, ,\label{S_mixed}\\
    S_\text{YM}[B_2,C_2] = S_\text{kin}[B_2] + S_\theta^{\mathfrak{su}(N)} + S_\theta^{\mathfrak{u}(1)}[B_2] + S_\text{mixed}[B_2,C_2] \, .
\end{gather}
Loosely speaking, $B_2$ is the background field associated with $\ast F_2^{\mathfrak{u}(1)}$, while $C_2$ is the background field associated with $F_2^{\mathfrak{u}(1)}$. Note that $kB_2 + F_2^{\mathfrak{u}(1)}$ is a gauge-invariant combination under the equivalence \eqref{U(1)/Zk_gauge_symmetry}.

In general, we can add an additional counterterm proportional to $S_\theta^{\mathfrak{u}(1)}[B_2]$, but we neglect it here for simplicity. Moreover, if one wants to obtain an $\text{SU}(N)/\mathbb{Z}_k$ gauge theory, then $A_1^{\mathfrak{u}(1)}$ should be integrated out and $B_2$ should be promoted to a dynamical field. This is not the route we wish to pursue here though. We would like to work with an $\text{SU}(N) \times_{\mathbb{Z}_k} \U(1)$ gauge theory, such that $B_2,C_2$ are regarded as background fields.

If we were to promote $B_2,C_2$ to dynamical fields, then $S_\text{mixed}[B_2,C_2]$ would no longer be gauge-invariant. This tells us that the mixed anomaly between the two $\U(1)$ symmetries of the $\text{SU}(N) \times_{\mathbb{Z}_k} \U(1)$ gauge theory, at the level of differential forms, is captured by the 5d anomaly theory,\footnote{There is an ambiguity coming from the $N/k$ prefactor in \eqref{S_mixed}, provided that $B_2$ and $C_2$ should be $\mathbb{Z}_k$ gauge fields. This is similar to the ambiguity between the $\mathcal{A}^{N,p}$ and $\mathcal{A}^{Np,1}$ minimal TQFTs \cite{Hsin:2018vcg}.}
\begin{equation}
    S_\text{anom}[\underline{B}_2,\underline{C}_2] = 2\pi i k \int_{Y^5} d\underline{C}_2 \wedge \underline{B}_2 \, .\label{u(N)_gauge_theory_continuum_anomaly}
\end{equation}
This is the continuum version of a 5d $\mathbb{Z}_k$ Dijkgraaf-Witten theory, where the equations of motion imply that $B_2,C_2$ are both $k$-torsion. Its differential-cocycle formulation is precisely \eqref{DW_anomaly}, repeated below for convenience,
\begin{equation}
    \mathcal{A}[\underline{\check{B}},\underline{\check{C}}] = \frac{1}{k} \int_{Y^5} \underline{\mathsf{G}}_3 \cup \underline{B}_2 \, .
\end{equation}
We may therefore quote our previous result, i.e.~the partition function of the $\text{SU}(N) \times_{\mathbb{Z}_k} \U(1)$ gauge theory is non-vanishing only if
\begin{equation}
    [\mathsf{H}_3]_\mathbb{Z} = [\mathsf{G}_3]_\mathbb{Z} = 0 \, .\label{u(N)_gauge_theory_FW_anomaly}
\end{equation}
By exactness, this implies that $[B_2]_{\mathbb{Z}_k}$ and $[C_2]_{\mathbb{Z}_k}$ must be mod $k$ reductions of some classes in $H^2(X^4;\mathbb{Z})$, so we can equivalently interpret \eqref{u(N)_gauge_theory_FW_anomaly} as saying that the partition function is non-vanishing only if the continuum limit \eqref{u(N)_gauge_theory_continuum_anomaly} is well-defined.

It is important to note that this is not a statement for the $\text{SU}(N)/\mathbb{Z}_k$ gauge theory. The $\U(1)$ factor in the gauge group actually matters. In fact, the cohomology classes $[B_2]_{\mathbb{Z}_k} \in H^2(X^4;\mathbb{Z}_k)$ and $[\mathsf{H}_3]_\mathbb{Z} \in H^3(X^4;\mathbb{Z})$ play rather different roles depending on whether we are working with an $\text{SU}(N)/\mathbb{Z}_k$ gauge theory or an $\text{SU}(N) \times_{\mathbb{Z}_k} \U(1)$ gauge theory. Particularly, in the former case, a non-trivial $[B_2]_{\mathbb{Z}_k}$ is an obstruction to lift to an $\text{SU}(N)$ gauge theory, while a non-trivial $[\mathsf{H}_3]_\mathbb{Z}$ is an obstruction to lift to an $\text{SU}(N) \times_{\mathbb{Z}_k} \U(1)$ gauge theory.\footnote{For this reason, $[B_2]_{\mathbb{Z}_k}$ is sometimes referred to in the physics literature as the second Stiefel-Whitney class of the $\text{SU}(N)/\mathbb{Z}_k$ gauge bundle \cite{Gaiotto:2017yup,Hsin:2018vcg,Brennan:2023vsa}.} See Table \ref{non-trivial_gauge_field_role} for a summary.

\begin{table}[t!]
    \begin{center}
        \begin{tabular}{|c|c|c|}
            \hline
            Gauge group & $[B_2]_{\mathbb{Z}_k}$ & $[\mathsf{H}_3]_\mathbb{Z}$ \\
            \hline
            \hline
            $\text{SU}(N)/\mathbb{Z}_k$ & Obstruction to lift to $\text{SU}(N)$ & Obstruction to lift to $G$ \\
            \hline
            $G$ & Background connection & Vanishing partition function\\
            \hline
        \end{tabular}
        \caption{\label{non-trivial_gauge_field_role} Implications of non-trivial $[B_2]_{\mathbb{Z}_k}$ and $[\mathsf{H}_3]_\mathbb{Z}$ in $\text{SU}(N)/\mathbb{Z}_k$ gauge theory vs.~$G = \text{SU}(N) \times_{\mathbb{Z}_k} \U(1)$ gauge theory. When $k=N$, we have $G \cong \U(N)$.}
    \end{center}
\end{table}

\subsection{Coincident D-branes}\label{coincident_branes_section}

It was observed by Kapustin in \cite{Kapustin:1999di} that the Freed-Witten anomaly cancellation condition \eqref{Freed-Witten_anomaly_cancellation_condition} receives a correction when dealing with a stack of coincident D-branes. In the case of D3-branes, this result can indeed be recast in our language as follows.

Suppose we have a stack of $N$ coincident D3-branes. The endpoints of F1-strings can live on any of the branes in this stack, thus enhancing the structure group of the Chan-Paton bundle from $\U(1)^N$ to $\text{SU}(N) \times_{\mathbb{Z}_k} \U(1)$ for some divisor $k$ of $N$. The bosonic part of the corresponding topological action includes the following terms,
\begin{align}
    S_\text{D3} & \supset 2\pi i \int_{X^4} \bigg(C_4 + C_2 \wedge \text{Tr}\big(B_2 \mathds{1}_N + F_2^G\big) + \frac{1}{2} \, C_0 \wedge \text{Tr}\big(B_2 \mathds{1}_N + F_2^G\big)^2\bigg)\nonumber\\
    & \supset 2\pi i \int_{X^4} \bigg(C_4 + \frac{N}{k} \, C_2 \wedge \big(kB_2 + F_2^{\mathfrak{u}(1)}\big) + \frac{N}{2k^2} \, C_0 \wedge \big(kB_2 + F_2^{\mathfrak{u}(1)}\big)^2\bigg) \, .\label{coincident_Dbrane_action}
\end{align}
We see that the second and third terms together can be modeled by an $\text{SU}(N) \times_{\mathbb{Z}_k} \U(1)$ gauge theory, with the identification $C_0 \sim \theta$. For the moment, let us focus on these two terms and neglect the first term.

In our previous analysis of the 4d $\text{SU}(N) \times_{\mathbb{Z}_k} \U(1)$ gauge theory, the interpretations of $B_2$ and $C_2$ are clear. They are respectively the background gauge fields associated with the 1-form $\U(1)$ electric and magnetic symmetries, which arise from the action of the dynamical field $F_2^G$. However, here the D3-branes are placed in a 10d string theory background, which has its own Kalb-Ramond field $B_2$ and Ramond-Ramond field $C_2$. In other words, there should really be two contributions to what we call $B_2$ in \eqref{coincident_Dbrane_action}, i.e.~the first is the pullback of the ``actual'' $B_2$ onto the brane worldvolume $X^4$, whereas the second is a localised mode $\zeta_2^B$ regarded as a background gauge field coupling to the $\U(1)$ center-of-mass mode of the Chan-Paton gauge field on the D3-branes. At the level of differential cocycles, the former can be modeled as $\check{B} = (\mathsf{H}_3,B_2,H_3) \in Z^3(X^4;\mathbb{Z}) \times C^2(X^4;\mathbb{R}) \times \Omega_\mathbb{Z}^3(X^4)$, and the latter can be modeled as $\check{\zeta}^B = (\beta([\zeta_2^B]),\zeta_2^B,0) \in Z^3(X^4;\mathbb{Z}) \times C^2(X^4;\mathbb{Z}_k) \times \ast$. The same applies to their duals $\check{C}$ and $\check{\zeta}^C$.

Furthermore, as discussed in Section \ref{FW_anomaly_section}, the $\U(1)$ gauge field admits a shifted quantisation since it receives a contribution from the worldvolume fermions. We implement all these contributions by effectively shifting $\check{B} \to \check{B} + \check{\zeta}^B + \check{w}$, and similarly $\check{C} \to \check{C} + \check{\zeta}^C + \check{w}$. More precisely, we can combine these differential cocycles together by promoting them to elements of $\check{H}^3(X^4) \subset Z^3(X^4;\mathbb{Z}) \times C^2(X^4;\mathbb{R}) \times \Omega_\mathbb{Z}^3(X^4)$ via the diagram below.
\begin{equation}
    \begin{tikzcd}
        0 \arrow[r] & \mathbb{Z} \arrow[r,"\times k"] \arrow[d,"="] & \mathbb{Z} \arrow[r,"\text{mod} \, k"] \arrow[d] & \mathbb{Z}_k \arrow[r] \arrow[d] & 0 \\
        0 \arrow[r] & \mathbb{Z} \arrow[r] & \mathbb{R} \arrow[r] & \mathbb{R}/\mathbb{Z} \arrow[r] & 0 \\
        0 \arrow[r] & \mathbb{Z} \arrow[r,"\times 2"] \arrow[u,"="] & \mathbb{Z} \arrow[r,"\text{mod} \, 2"] \arrow[u] & \mathbb{Z}_2 \arrow[r] \arrow[u] & 0
    \end{tikzcd}
\end{equation}
Running through the same analysis as before then leads to the conclusion that the partition function of the stack of D3-branes is non-vanishing only if
\begin{equation}
    [\mathsf{H}_3]_\mathbb{Z} + \beta_k([\zeta_2^B]) + [\mathsf{W}_3]_\mathbb{Z} = 0 \, , \qquad [\mathsf{G}_3]_\mathbb{Z} + \beta_k([\zeta_2^C]) + [\mathsf{W}_3]_\mathbb{Z} = 0 \, .\label{Freed-Kapustin-Witten_anomaly}
\end{equation}
To distinguish between the various Bockstein homomorphisms involved here, we denote the ``standard'' version simply as $\beta: H^2(X^4;\mathbb{R}/\mathbb{Z}) \to H^3(X^4;\mathbb{Z})$, whereas $\beta_n: H^2(X^4;\mathbb{Z}_n) \to H^3(X^4;\mathbb{Z})$ for any $n \in \mathbb{Z}^+$, such that $[\mathsf{H}_3]_\mathbb{Z} = \beta([B_2]_{\mathbb{R}/\mathbb{Z}})$, $[\mathsf{G}_3]_\mathbb{Z} = \beta([C_2]_{\mathbb{R}/\mathbb{Z}})$, and $[\mathsf{W}_3]_\mathbb{Z} = \beta_2([w_2]_{\mathbb{Z}_2})$.

When we take $k=N$, the first relation in \eqref{Freed-Kapustin-Witten_anomaly} matches with the original result in \cite{Kapustin:1999di}. By S-duality, we expect the second relation to hold as well. Nevertheless, our interpretations of these conditions are slightly different from \cite{Kapustin:1999di}. The concern therein was whether the Chan-Paton structure group can be lifted from $\text{SU}(N)/\mathbb{Z}_N$ to $\U(N)$, but here we interpret the result as whether the partition function of the coincident D3-branes is vanishing or not. For example, suppose $[\mathsf{W}_3]_\mathbb{Z} = 0$, then the partition function is non-vanishing only if the characteristic class $\beta_k([\zeta_2^B])$ of the Chan-Paton bundle cancels out (the pullback of) $[\mathsf{H}_3]_\mathbb{Z}$, and likewise for $\beta_k([\zeta_2^C])$ and $[\mathsf{G}_3]_\mathbb{Z}$.


\section{Other finite symmetry examples}\label{finite_symmetry_sec}

Let us discuss below a couple of other examples involving anomalies of finite symmetries.

\subsection{3d minimal TQFT}\label{sec:3d-minimal-TQFTs}

It was shown in \cite{Hsin:2018vcg} that any 3d topological quantum field theory (TQFT) with a 1-form $\mathbb{Z}_N$ global symmetry admits a factorisation, assuming $\gcd(N,p)=1$ for simplicity,
\begin{equation}
    \mathcal{T} \cong \mathcal{A}^{N,p} \otimes \mathcal{T}' \, ,
\end{equation}
where $\mathcal{A}^{N,p}$ is called a {\it minimal Abelian TQFT}, and $\mathcal{T}'$ is a decoupled sector. In particular, the former consists of line operators charged under the $\mathbb{Z}_N$ symmetry, and have non-trivial braiding statistics labeled by an integer $p$, which takes values in $\mathbb{Z}_N$ on spin manifolds.\footnote{For manifolds which are not spin (but orientable), $p \in \mathbb{Z}_{2N-1}$ for $N$ even and $p \in \mathbb{Z}_{2N-2}$ for $N$ odd. To be concrete, we focus on spin manifolds in this discussion.} As an aside, the minimal TQFT plays a crucial role in the construction of non-invertible symmetry defects in 4d QFTs (see, e.g.~\cite{Kaidi:2021xfk,Choi:2022zal}).

For $N$ odd, the self-anomaly of $\mathcal{A}^{N,p}$, and hence $\mathcal{T}$, is described by a 4d anomaly theory
\begin{equation}
    S_\text{anom}[\underline{B}_2] = \frac{2\pi i p}{N} \int_{Y^4} \frac{1}{2} \, \underline{B}_2 \cup \underline{B}_2 \, ,\label{odd_N_minimal_TQFT_anomaly}
\end{equation}
where $\underline{B}_2 \in C^2(Y^4;\mathbb{Z}_N)$. Note that the integral is well-defined as an element of $\mathbb{Z}_N$, because $\gcd(N,2)=1$, so $2^{-1}$ mod $N$ exists. Such a background gauge field can be modeled as a differential cocycle,
\begin{equation}
    \check{B} = (\mathsf{H}_3,B_2) \in Z^3(X^3;\mathbb{Z}) \times C^2(X^3;\mathbb{Z}_N) \, ,
\end{equation}
where $\delta B_2 = -\mathsf{H}_3$ mod $N$. The action \eqref{odd_N_minimal_TQFT_anomaly} does not really take the form of a BF theory, but some of our previously developed techniques will be applicable.

As usual, we replace $Y^4$ in \eqref{odd_N_minimal_TQFT_anomaly} with the mapping torus $X^3 \times S^1$. Employing the parametrisation $\check{\underline{B}} = \check{B} + \check{s} \star \check{b}$, the anomaly theory can be reduced over $S^1$ to yield the basepoint anomaly,
\begin{equation}
    \mathcal{A}[\check{B},\check{b}] = \frac{2\pi i p}{N} \int_{X^3} [b_1]_{\mathbb{Z}_N} \cup [B_2]_{\mathbb{Z}_N} \, .\label{minimal_TQFT_basepoint_anomaly}
\end{equation}
Since $[b_1]_{\mathbb{Z}_N}$ corresponds to an arbitrary rigid gauge transformation, we conclude that the partition function of the 3d TQFT $\mathcal{T}$ is non-vanishing only if
\begin{equation}
    [B_2]_{\mathbb{Z}_N} = 0 \, .
\end{equation}
This is qualitatively different from all of our previous examples where the characteristic classes, rather than the connections, are required to be topologically trivial. What we have here is a stronger statement, i.e.~$B_2$ as a $\mathbb{Z}_N$-valued connection needs to be cohomologically trivial. The triviality of the characteristic class $[\mathsf{H}_3]_\mathbb{Z} = \beta([B_2]_{\mathbb{Z}_N})$ follows automatically.

By exactness, $[B_2]_{\mathbb{Z}_N}=0$ being in the kernel of $H^2(X^3;\mathbb{Z}) \xrightarrow{\text{mod} \, N} H^2(X^3;\mathbb{Z}_N)$ implies that its integral lift must be in the image of $H^2(X^3;\mathbb{Z}) \xrightarrow{\times N} H^2(X^3;\mathbb{Z})$. In other words, a continuum version of \eqref{odd_N_minimal_TQFT_anomaly} can be written as
\begin{equation}
    S_\text{anom}[\widehat{\underline{B}}_2] = 2\pi i pN \int_{Y^4} \frac{1}{2} \, \widehat{\underline{B}}_2 \cup \widehat{\underline{B}}_2 \, ,
\end{equation}
where $\widehat{\underline{B}}_2 \in H^2(Y^4;\mathbb{Z})$. Equivalently, as was pointed out by \cite{Karasik:2022kkq}, the minimal TQFT $\mathcal{A}^{N,p}$ has a non-vanishing partition function only if the flux of the background connection through every 2-cycle of $X^3$ is a multiple of $N$.\footnote{A similar observation was made in Appendix A of \cite{Choi:2022zal}.}

For $N=2m$ even with $m$ odd, the 4d anomaly theory is given instead by
\begin{equation}
    S_\text{anom}[\underline{B}_2] = \frac{2\pi i p}{N} \int_{Y^4} \frac{1}{2} \, \mathfrak{B}_2([\underline{B}_2]_{\mathbb{Z}_N}) \, ,\label{even_N_minimal_TQFT_anomaly}
\end{equation}
where the cohomology operation $\mathfrak{B}_p: H^i(-;\mathbb{Z}_{pm}) \to H^{pi}(-;\mathbb{Z}_{p^2m})$ for any prime $p$ and any $m \in \mathbb{Z}^+$ is the {\it Pontryagin $p$-th power} \cite{b8effd3d-9c0a-396c-a024-ee4cd0bee32e}. The case of $p=2$ and $m=1$ corresponds to the standard Pontryagin square operation \cite{Whitehead1949}. One of its axioms is that
\begin{equation}
    (\text{mod} \ pm) \circ \mathfrak{B}_p(x) = x^p \, .
\end{equation}
In addition, the short exact sequence $0 \to \mathbb{Z}_{2m} \xrightarrow{\times 2} \mathbb{Z}_{4m} \xrightarrow{\text{mod} \, 2} \mathbb{Z}_2 \to 0$ induces a long exact sequence in cohomology,
\begin{equation}
    \cdots \to H^4(-;\mathbb{Z}_{2m}) \xrightarrow{\times 2} H^4(-;\mathbb{Z}_{4m}) \xrightarrow{\text{mod} \, 2} H^4(-;\mathbb{Z}_2) \to \cdots \, .
\end{equation}
Given $\gcd(m,2)=1$, we can decompose $x \in H^2(Y^4;\mathbb{Z}_{2m})$ into $y \cup z$ with $y \in H^2(Y^4;\mathbb{Z}_2)$ and $z \in H^0(Y^4;\mathbb{Z}_m)$, such that $(\text{mod} \ 2m) \circ \mathfrak{B}_2(x) = y^2 \cup z^2$. On spin 4-manifolds, $y^2$ is always a trivial element in $H^4(Y^4;\mathbb{Z}_2)$.\footnote{This follows from the properties of the second Wu class on 4-manifolds. We will take up this issue in greater generality in Section \ref{Chern-Simons_section} as we discuss quadratic refinements.} By exactness, this implies $\mathfrak{B}_2(x)$ is divisible by 2, and so the integral in \eqref{even_N_minimal_TQFT_anomaly} is well-defined as an element of $\mathbb{Z}_{2m}$. The case for $m$ even is discussed in, e.g.~\cite{Aharony:2013hda,Benini:2018reh}.

An explicit formula for the Pontryagin $p$-th power is
\begin{equation}
    \mathfrak{B}_p(x) = x^p + x^{p-1} \cup_1 \delta x \, ,
\end{equation}
where $\cup_i: C^{|x|}(-;\mathbb{Z}_{pm}) \times C^{|y|}(-;\mathbb{Z}_{pm}) \to C^{|x|+|y|-i}(-;\mathbb{Z}_{p^2m})$ denotes the cup-$i$ product \cite{4fc46540-015c-3832-bc1c-ffd13cef2752}, satisfying
\begin{equation}
    \delta(x \cup_i y) = (-1)^{|x|+|y|-i} x \cup_{i-1} y + (-1)^{|x||y| + |x| + |y|} y \cup_{i-1} x + \delta x \cup_i y + (-1)^{|x|} x \cup_i \delta y \, .
\end{equation}
Nevertheless, at the level of cohomology classes, it can be shown that such higher-order corrections do not contribute to the basepoint anomaly, so \eqref{minimal_TQFT_basepoint_anomaly} still holds.

\subsection{Vanishing of the RR partition function in 2d}

Consider a free Dirac fermion in $2$
dimensions. Like the $4d$
system, this has vector and axial $\U(1)$ symmetries with a mixed
anomaly, leading to a vanishing condition similar to the one in
the case of the ABJ anomaly discussed in the introduction. However, in
this case, there is an additional structure that we can exploit, as
discussed in \cite{Karch_2019}.

To define a fermion theory on a Riemann surface $\Sigma^2$, we need to specify a Spin structure. One can think of this as specifying periodic (R) or antiperiodic (NS) boundary conditions to each cycle on $\Sigma^2$. A classical result, derived e.g.~in \cite{Hori2003MirrorSymmetry}, is that the partition function of a free fermion on the torus with RR Spin structure vanishes. When quantising the theory, one finds that there is a fermionic zero mode, which causes the path integral to vanish. On general $\Sigma^2$, the appearance of such zero modes, and therefore the vanishing, is controlled by the mod $2$ index, i.e.~the number of zero modes modulo $2$. In $2d$, this index can be non-zero and is a topological invariant. A theorem due to Atiyah \cite{Atiyah1971} states that
\begin{equation}\label{eq:mod2indexthm}
    \mathcal{I}(\rho_{\Sigma^2})=\text{Arf}(q_{\rho_{\Sigma^2}})\,,
\end{equation}
where $\mathcal{I}(\rho_{\Sigma^2})$ is the mod $2$ index associated to the Spin structure $\rho_{\Sigma^2}$. To understand the RHS, recall that the first homology group with $\mathbb{Z}_2$ coefficients of the genus $g$ Riemann surface is given by
\begin{equation}
    H_1(\Sigma^2;\mathbb{Z}_2)=(\mathbb{Z}_2)^{2g}\,.
\end{equation}
We may define the intersection form for two cycles represented by closed curves $\gamma$, $\delta$ in $\Sigma^2$:\footnote{We are making the additional technical assumption that $\gamma$ and $\delta$ only intersect transversely. Intuitively, we can always deform them slightly to satisfy this assumption without changing the homology classes they represent. A proof can be found in standard references for differential topology, such as \cite{guillemin2010differential}.}
\begin{equation}
    \gamma\cdot\delta=\begin{cases}
        0 \text{ }|\text{ if }\gamma\text{ and }\delta\text{ intersect an even number of times}\\
        1\text{ }|\text{ if }\gamma\text{ and }\delta\text{ intersect an odd number of times}
    \end{cases}
\end{equation}
It can be shown that this definition is well-defined and descends to a symmetric bilinear form on homology. Given the intersection form, we may then choose a symplectic basis for $H_1(\Sigma^2;\mathbb{Z}_2)$. The generators are denoted $\{\alpha_i,\beta_i\}$ for $1\leq i\leq g$ and the intersection form is given by $\alpha_i\cdot\alpha_j=\beta_i\cdot\beta_j=0$ and $\alpha_i\cdot\beta_j=\delta_{ij}$. As mentioned before, a Spin structure $\rho_{\Sigma^2}$ assigns periodic or anti-periodic boundary conditions to each generator of $H_1(\Sigma^2;\mathbb{Z}_2)$. We define a function $q_{\rho_{\Sigma^2}}: H_1(\Sigma^2;\mathbb{Z}_2)\rightarrow\mathbb{Z}_2$ by $q_{\rho_{\Sigma^2}}(\alpha_i)=1$ if $\alpha_i$ is periodic, $q_{\rho_{\Sigma^2}}(\alpha_i)=0$ if $\alpha_i$ is anti-periodic, and similarly for $\beta_i$. To extend the definition to all of $H_1(\Sigma^2;\mathbb{Z}_2)$, we then demand that $q_{\rho_{\Sigma^2}}$ is a \emph{quadratic refinement} of the intersection pairing, that is, it satisfies 
\begin{equation}
    q_{\rho_{\Sigma^2}}(a+b)=q_{\rho_{\Sigma^2}}(a)+q_{\rho_{\Sigma^2}}(b)+a\cdot b
\end{equation}
for any $a,b\in H_1(\Sigma^2;\mathbb{Z}_2)$. It can be shown that
there is a bijection between quadratic refinements of the intersection
form and spin structures on $\Sigma^2$ \cite{johnson1980spin}. Once we
have such a quadratic refinement of the intersection form, we can
define its associated $\mathbb{Z}_2$-valued {\it Arf invariant},
\begin{equation}
    \text{Arf}(q_{\rho_{\Sigma^2}})=\sum_{i=1}^g q_{\rho_{\Sigma^2}}(\alpha_i)\,q_{\rho_{\Sigma^2}}(\beta_i) \, .
\end{equation}
It can be shown that the Arf invariant is well-defined, i.e.~it does not depend on our choice of symplectic basis. This discussion was somewhat abstract, so it may be helpful to consider the example of the torus again. In this case, the basis is given by $\{\alpha_1,\beta_1\}$, so the sum consists of a single term and we see that the Arf invariant is non-zero only if $q_{\rho_{\Sigma^2}}(\alpha_1)=q_{\rho_{\Sigma^2}}(\beta_1)=1$, i.e.~for the RR Spin structure, as we anticipated. On a general genus $g$ Riemann surface, there are $2^{2g}$ possible Spin structures, and by (\ref{eq:mod2indexthm}), we have a sufficient condition for the partition function to vanish given a choice of Spin structure: $\text{Arf}(q_{\rho_{\Sigma^2}})=1$. This is the case for exactly $2^{g-1}(2^g-1)$ choices of Spin structure.

We would now like to understand this condition in terms of a 't Hooft
anomaly. Recall that a Dirac fermion may be written as the sum of two
Majorana fermions in 2d. We will therefore analyse a single Majorana
fermion $\chi$ with the understanding that the same symmetries and
anomalies will be present for a Dirac fermion as well, albeit embedded
in a much larger symmetry group. The crucial ingredient is the
symmetry generated by $(-1)^{F_L}$, which acts as
\begin{equation}
    \mathbb{Z}_2^{F_L}: \chi\mapsto \gamma^3\chi\,,
\end{equation}
where $\gamma^3$ is the chirality matrix\cite{Karch_2019,Delmastro:2021xox}. By general arguments \cite{witten2020anomalyinflowetainvariant}, the
anomaly theory of this system is given by the $\eta$ invariant of a
real Dirac operator in $3d$ coupled to this
$\mathbb{Z}_2$-background. Potential anomalies of this symmetry are
classified by
$\Omega^{\text{Spin}}_3(B\mathbb{Z}_2)=\mathbb{Z}_8$. This group is
generated by $\mathbb{RP}^3$ with non-trivial $\mathbb{Z}_2$
bundle. As pointed out in \cite{Tachikawa:2018njr}, the relevant $\eta$-invariant evaluates to
$\pm1/8$ on this generator, so our $(-1)^{F_L}$-symmetry is indeed
anomalous. To see the connection between this $\eta$-invariant and the
Arf invariant discussed above we need to take one additional
mathematical detour.

A \emph{quadratic enhancement} of the intersection form is a function $\tilde{q}: H_1(\Sigma^2;\mathbb{Z}_2)\rightarrow\mathbb{Z}_4$ such that $\tilde{q}(a+b)=\tilde{q}(a)+\tilde{q}(b)+2a\cdot b$. A Pin$^-$ structure gives rise to such a quadratic enhancement. Since every closed $2$-manifold admits a Pin$^-$ structure \cite{Kirby_Taylor_1991} but only orientable surfaces admit a Spin structure, this provides a generalisation of the quadratic refinement above. On spin surfaces, the quadratic enhancement $\tilde{q}$ is related to the quadratic refinement $q$ as
\begin{equation}
    \tilde{q}=2q\mod 4.
\end{equation}
We omit further details of this construction and refer the interested reader to \cite{Kirby_Taylor_1991}.

Given a $2$-manifold $\Xi$ with Pin$^-$ structure $s$, inducing a quadratic enhancement $\tilde{q}$, we can now define its Arf-Brown-Kervaire invariant $\beta(\Xi,s)\in \mathbb{Z}_8$ via a Gauss sum,
\begin{equation}
    e^{\frac{\pi i \beta(\Xi,s)}{4}}=\frac{1}{\sqrt{|H_1(\Sigma^2;\mathbb{Z}_2)|}}\sum_{a\in H_1(\Sigma^2;\mathbb{Z}_2)}e^{\frac{\pi i \tilde{q}(a)}{2}} \, .
\end{equation}
Crucially, on an orientable manifold, a Spin structure defines a
Pin$^-$ structure. In this case, the ABK invariant is given in terms
of the Arf invariant as
\begin{equation}\label{eq:ABKtoArf}
    \beta(\Xi,s)=4\cdot\text{Arf}(q_{\rho_{\Sigma^2}})\mod 8\,.
\end{equation}
We can now use this invariant to define an invertible $3d$ spin TQFT. On a closed spin $3$-manifold $\mathcal{M}^3$ with Spin structure $\rho$ and a $\mathbb{Z}_2$ background $x$, its action is
\begin{equation}
    \mathcal{A}=\frac{\pi i}{4}\,\beta(\text{PD}(x),s)\,.
\end{equation}
This theory has been discussed in the physics literature, see e.g.~\cite{Guo:2018vij,Grigoletto:2021zyv}. The Poincar\'e dual of $x$ can be represented by a closed surface embedded in $\mathcal{M}^3$. Since every closed surface is pin$^-$, the ambient Spin structure $\rho$ induces a Pin$^-$ structure $s$ on PD$(x)$. Consider our generator of $\Omega^{\text{Spin}}_3(B\mathbb{Z}_2)$. The Poincar\'e dual of the non-trivial element $x\in H^1(\mathbb{RP}^3;\mathbb{Z}_2)$ is an embedded $\mathbb{RP}^2$. Depending on the choice of Pin$^-$ structure, $\beta(\mathbb{RP}^2,s)=\pm1\in\mathbb{Z}_8$. Therefore, the partition function of the theory $\mathcal{A}$ agrees with the exponentiated $\eta$-invariant on the generator of $\Omega^{\text{Spin}}_3(B\mathbb{Z}_2)$. Since both quantities are bordism invariants, we deduce that they agree on all closed spin $3$-manifolds with $\mathbb{Z}_2$ bundle. That is, $\mathcal{A}$ is a presentation of the anomaly theory of our Majorana fermion. 

We are finally ready to apply our general approach to this system. We
wish to evaluate $\mathcal{A}$ on the manifold $\Sigma^2\times
S^1$. We take the $\mathbb{Z}_2^{F_L}$ background
$x\in H^1(\Sigma^2\times S^1;\mathbb{Z}_2)$ to be the pullback of the
unique non-zero element in $H^1( S^1;\mathbb{Z}_2)$ under the
projection map. Concretely, this means that the $\mathbb{Z}_2^{F_L}$
flux has a leg on the $S^1$ but not the $\Sigma^2$. The Poincar\'e
dual of $x$ in $\Sigma^2\times S^1$ is $\Sigma^2$, but by
(\ref{eq:ABKtoArf}), the anomalous phase depends only on the Arf
invariant. By our general argument, the partition function obeys
\begin{equation}
    \mathcal{Z}(\Sigma^2,\rho_{\Sigma^2})=(-1)^{\text{Arf}(q_{\rho_{\Sigma^2}})}\mathcal{Z}(\Sigma^2,\rho_{\Sigma^2})\,.
\end{equation}
This agrees with eq.~(2.2) in \cite{Karch_2019} at $m=0$. We finally deduce that $\mathcal{Z}(\Sigma^2,\rho_{\Sigma^2})$ vanishes  if Arf$(q_{\rho_{\Sigma^2}})=1$, as anticipated.


\section{Chern-Simons theories}\label{Chern-Simons_section}

For chiral $p$-form fields in $d=2p+2$ dimensions, the anomaly of the chiral gauge theory is a Chern-Simons theory associated with a background gauge field $\underline{\check{C}} \in \check{H}^{p+2}(Y^{d+1})$. Roughly speaking, the anomaly theory takes the form,
\begin{equation}
    \mathcal{A}[C_{p+1}] \approx \frac{\kappa}{2} \int_{Y^{d+1}} C_{p+1} \wedge dC_{p+1} \, ,
\end{equation}
for some integer level $\kappa$. The dimension of the Hilbert space of the chiral gauge theory scales as some power of $|\kappa|$ \cite{Hsieh:2020jpj}. If we assume that the partition function is a section of a line bundle, then we shall hereafter restrict our attention to invertible theories where $\kappa = \pm 1$, corresponding respectively to self-dual and anti-self-dual gauge fields.

Note that such an anomaly is not well-defined as an element of $\mathbb{R}/\mathbb{Z}$ when $\kappa$ is odd, so a precise formulation of it requires the introduction of a quadratic refinement $q(\underline{\check{C}})$ \cite{Hopkins:2002rd},\footnote{Unlike the quadratic refinements we used in Section \ref{finite_symmetry_sec} which are defined for intersection pairings, those introduced here are for torsion pairings.} defined to be such that
\begin{equation}
    q(\underline{\check{C}} + \underline{\check{C}}') - q(\underline{\check{C}}) - q(\underline{\check{C}}') + q(0) = \int_{Y^{d+1}} \Big(\underline{C}_{p+1} \cup \underline{G}_{p+2} + (-1)^p \underline{\mathsf{G}}_{p+2} \cup \underline{C}_{p+1} + Q(\underline{G}_{p+2},\underline{G}_{p+2}')\Big) \, ,
\end{equation}
where $q(0)$ can be regarded as some (generally non-vanishing) ``constant'' that depends on $Y^{d+1}$ but not on $\underline{\check{C}}$. Equivalently, if $Y^{d+1}$ is closed and there exists some $(d+2)$-manifold $Z^{d+2}$ with $\partial Z^{d+2} = Y^{d+1}$, then we can rewrite the holonomy of $\underline{\check{C}} \star \underline{\check{C}}'$ above more compactly as
\begin{equation}
    q(\underline{\check{C}} + \underline{\check{C}}') - q(\underline{\check{C}}) - q(\underline{\check{C}}') + q(0) = \int_{Z^{d+2}} \underline{G}_{p+2} \wedge \underline{G}_{p+2}' \, .\label{quadratic_refinement}
\end{equation}

To simplify notation, it is customary to define another quadratic refinement $\tilde{q}(\underline{\check{C}}) \coloneqq q(\underline{\check{C}}) - q(0)$. Following \cite{Hsieh:2020jpj}, the anomaly can be expressed as
\begin{equation}
    \mathcal{A}[\underline{\check{C}}] = -\kappa(\tilde{q}(\underline{\check{C}}) - \mathcal{A}_\text{grav}) \, ,
\end{equation}
where $\mathcal{A}_\text{grav}$ is the gravitational anomaly of the theory.\footnote{The gravitational anomaly is given by $\mathcal{A}_\text{grav} = \eta(\mathcal{D}_{Y_3}^\text{Dirac})$ for $d=2$, and $\mathcal{A}_\text{grav} = 28\eta(\mathcal{D}_{Y_7}^\text{Dirac})$ for $d=6$, where $\mathcal{D}_{Y_{d+1}}^\text{Dirac}$ is the Dirac operator on $Y^{d+1}$, and $\eta$ is the corresponding eta-invariant \cite{Hsieh:2020jpj}.} For $d=2,6$, if $Z^{d+2}$ admits a Spin structure, then one can further pick the convention $q(0)=-\mathcal{A}_\text{grav}$ such that $\mathcal{A}[\underline{\check{C}}] = -\kappa q(\underline{\check{C}})$, with
\begin{equation}
    q(\underline{\check{C}}) = \begin{cases} \displaystyle \int_{Z^{d+2}} \bigg(\frac{1}{2} \, \underline{G}_{p+2} \wedge \underline{G}_{p+2} + \hat{A}_1(\underline{R}_2)\bigg) & d=2 \, ,\\[2ex] \displaystyle \int_{Z^{d+2}} \bigg(\frac{1}{2} \, \underline{G}_{p+2} \wedge \underline{G}_{p+2} - \frac{1}{4} \, \underline{G}_{p+2} \wedge p_1(\underline{R}_2) + 28\hat{A}_2(\underline{R}_2)\bigg) & d=6 \, . \end{cases}\label{2d_6d_quadratic_refinements}
\end{equation}
It is straightforward to check that the choices of quadratic refinements above satisfy the defining relation \eqref{quadratic_refinement}. The first and second terms of the $\hat{A}$ genus are given respectively by
\begin{equation}
    \hat{A}_1(\underline{R}_2) = -\frac{1}{24} \, p_1(\underline{R}_2) \, , \qquad \hat{A}_2(\underline{R}_2) = \frac{1}{5760} \, (7p_1^2(\underline{R}_2) - 4p_2(\underline{R}_2)) \, ,
\end{equation}
where $p_1(\underline{R}_2), p_2(\underline{R}_2)$ are the first and second Pontryagin classes of $TZ^{d+2}$. For our purposes, we will neglect terms that are purely gravitational in our subsequent discussion.

\subsection{Integral lifts of Wu classes}\label{Wu_class_section}

More generally, without any prior assumption on the tangential structure of $Z^{d+2}$, a suitable definition of the quadratic refinement is, up to a choice of $q(0)$,
\begin{equation}
    \tilde{q}(\underline{\check{C}}) = \frac{1}{2} \int_{Z^{d+2}} \underline{G}_{p+2} \wedge (\underline{G}_{p+2} - \underline{\Lambda}_{p+2}) \, ,\label{generic_quadratic_refinement}
\end{equation}
where $\underline{\Lambda}_{p+2}$ is defined such that its characteristic class $[\underline{\Lambda}_{p+2}]_\mathbb{Z} \in H^{p+2}(Z^{d+2};\mathbb{Z})$ is an integral lift of the $(p+2)$-th {\it Wu class} $v_{p+2} \in H^{p+2}(Z^{d+2};\mathbb{Z}_2)$ \cite{Witten:1996md,Hopkins:2002rd}. To see why, note that on an $n$-manifold $M^n$, the Wu class $v_i$ is defined to be a class representing the {\it Steenrod square} operation $\text{Sq}^i: H^\ast(M^n;\mathbb{Z}_2) \to H^{\ast + i}(M^n;\mathbb{Z}_2)$ \cite{10.2307/j.ctt1b7x52h.1}, such that
\begin{equation}
    v_i \cup x_{n-i} = \text{Sq}^i(x_{n-i})\label{Wu_classes_definition}
\end{equation}
for any $x_{n-i} \in H^{n-i}(M^n;\mathbb{Z}_2)$ \cite{5895d0ff-dbfd-39a7-9ffe-3ef746a10f70}. We also have that, by definition, $\text{Sq}^j(x_j) = x_j \cup x_j$ for any $j$. For $n$ even, if we take $i=n/2$, then
\begin{equation}
    (x_{n/2} - v_{n/2}) \cup x_{n/2} = 0 \mod 2 \, .
\end{equation}
If both $x_{n/2}$ and $v_{n/2}$ moreover admit integral lifts, then the combination above constitutes an even integral cohomology class. This makes \eqref{generic_quadratic_refinement} well-defined as a quadratic refinement, assuming the existence of the integral lift $\underline{\Lambda}_{p+2}$.

To determine whether $\underline{\Lambda}_{p+2}$ exists, it is instructive for us to express the Wu classes in terms of Stiefel-Whitney classes via the relation $w=\text{Sq}(v)$. Expanding the relation for the first few orders gives
\begin{equation}
    v_1 = w_1 \, , \qquad v_2 = w_2 + w_1^2 \, , \qquad v_3 = w_2 w_1 \, , \qquad v_4 = w_4 + w_3 w_1 + w_2^2 + w_1^4 \, .
\end{equation}
Recall that the short exact sequence $0 \to \mathbb{Z} \to \mathbb{Z} \to \mathbb{Z}_2 \to 0$ induces a long exact sequence in cohomology,
\begin{equation}
    \cdots \to H^i(-;\mathbb{Z}) \xrightarrow{\text{mod} \, 2} H^i(-;\mathbb{Z}_2) \xrightarrow{\beta} H^{i+1}(-;\mathbb{Z}) \xrightarrow{\times 2} H^{i+1}(-;\mathbb{Z}) \to \cdots \, ,
\end{equation}
so by exactness, an element $x_i \in H^i(-;\mathbb{Z}_2)$ admits an integral lift in $H^i(-;\mathbb{Z})$ if and only if $\beta(x_i) = 0$.

Since $\text{Sq}^1 = (\text{mod} \ 2) \circ \beta$, the question of whether $v_i$ can be lifted essentially boils down to computing $\text{Sq}^1(v_i)$. Here we can make use of the Cartan formula,
\begin{equation}
    \text{Sq}^k(x \cup y) = \sum_{i+j=k} \text{Sq}^i(x) \cup \text{Sq}^j(y) \, ,
\end{equation}
and the Wu formula \cite{may1999concise},
\begin{equation}
    \text{Sq}^i(w_j) = \sum_{t=0}^i \begin{pmatrix} j+t-i-1 \\ t \end{pmatrix} w_{i-t} \, w_{j+t} \, ,
\end{equation}
to find that
\begin{equation}
    \begin{gathered}
        \text{Sq}^1(v_1) = w_1^2 \, , \qquad \text{Sq}^1(v_2) = w_1 w_2 + w_3 \, ,\\
        \text{Sq}^1(v_3) = w_1 w_3 \, , \qquad \text{Sq}^1(v_4) = w_1 w_4 + w_5 \, .
    \end{gathered}
\end{equation}
In general, $\text{Sq}^1(v_{2i-1}) = w_1 w_{2i-1}$ and $\text{Sq}^1(v_{2i}) = w_1 w_{2i} + w_{2i+1}$. They actually coincide (mod 2) with the integral Stiefel-Whitney classes, i.e.
\begin{equation}
    V_{i+1} = \beta(v_i) = \beta(w_i) = W_{i+1} \mod 2 \, .\label{integral_SW_classes}
\end{equation}
We thus conclude that $v_i$ admits some integral lift if and only if $V_{i+1} = 0$.

On spin manifolds, $w_1 = w_2 = 0$, so the second Wu class $v_2$ vanishes identically. Its integral lift can be chosen to be trivial, as we saw in \eqref{2d_6d_quadratic_refinements} for the case of $d=2$ and $p=0$. Similar remarks apply to the case of $d=4$ and $p=1$. When $d=6$ and $p=2$, the fourth Wu class reduces to $w_4$, then the relation \cite{3dcb6f87-ee89-3c37-b8a3-6317f6f6a26b}
\begin{equation}
    \mathfrak{B}(w_2) = p_1 + 2(w_1 \cup \text{Sq}^1(w_2) + w_4) \mod 4 \, ,
\end{equation}
where $\mathfrak{B}: H^i(-;\mathbb{Z}_2) \to H^{2i}(-;\mathbb{Z}_4)$ denotes the Pontryagin square operation, tells us that
\begin{equation}
    v_4 = w_4 = \frac{1}{2} \, p_1 \mod 2 \, ,
\end{equation}
thus reproducing \eqref{2d_6d_quadratic_refinements}.

\subsection{2d chiral boson}\label{chiral_boson_section}

Let us start with the chiral boson in $d=2$. We view this as the gapless edge mode of the $\text{U}(1)_1$ Spin-CS theory, following \cite{Witten:1996hc}. Note that the chirality constraint $*d\phi\stackrel{!}{=}d\phi$ requires a choice of volume form and therefore only makes sense on orientable manifolds. However, as explained in \cite{Delmastro:2019vnj}, the bulk theory has a time-reversal symmetry, such that it can be defined as well on non-orientable manifolds. For the purpose of this work, it suffices for us to analyze the bulk anomaly theory in its own right, assuming that it admits a gapless edge mode (as in \cite{Hsieh:2020jpj}) which we loosely refer to as the ``chiral boson.'' In fact, by invoking considerations of integral Wu structures, our goal is to precisely characterise the conditions under which such a boundary theory may become well-defined, and particularly, admit a non-vanishing partition function. We do not attempt to provide any Lagrangian description of the resultant theory.

Due to degree reasons, any 2-manifold $X^2$ must have $V_3=W_3=0$, i.e.~it admits a $\text{Pin}^c$ structure.\footnote{As a remark, a $\text{Pin}^+$ structure requires $w_2=0$, a $\text{Pin}^-$ structure requires $w_2 + w_1^2 = 0$, and a $\text{Spin}$ structure requires $w_1=w_2=0$. Similarly, a $\text{Pin}^c \coloneqq \text{Pin}^\pm \times_{\mathbb{Z}_2} \U(1)$ structure requires $W_3=0$, while a $\text{Spin}^c \coloneqq \text{Spin} \times_{\mathbb{Z}_2} \U(1)$ structure requires both $w_1=0$ and $W_3=0$.} The integral lift of $v_2 = w_2 + w_1^2$ is given by
\begin{equation}
    \Lambda_2 = F_2 + W_2 \, ,\label{v2_integral_lift}
\end{equation}
where $W_2$ is the second integral Stiefel-Whitney class, and $F_2$ is the first Chern class of the $\text{Pin}^c$ bundle. The relation between $w_2$ and $F_2$ can be depicted by the homotopy pullback diagram below.\footnote{The square on the right is (part of) the homotopy fiber sequence $\cdots \to B^{i-1}\mathbb{Z}_2 \to B^i\mathbb{Z} \to B^i\mathbb{Z} \to B^i\mathbb{Z}_2 \to B^{i+1}\mathbb{Z} \to \cdots$ induced from the short exact sequence $0 \to \mathbb{Z} \to \mathbb{Z} \to \mathbb{Z}_2 \to 0$.}
\begin{equation}
    \begin{tikzcd}
        & B\text{Pin}^c(d) \arrow[r,"F_2"] \arrow[d] & B^2\mathbb{Z} \arrow[r] \arrow[d,"\text{mod}\,2"] & \ast \arrow[d] \\
        X^d \arrow[ru,dashed] \arrow[r] & B\text{O}(d) \arrow[r,"w_2"] \arrow[rr,bend right,"W_3"] & B^2\mathbb{Z}_2 \arrow[r,"\beta"] & B^3\mathbb{Z}
    \end{tikzcd}\label{Pinc_diagram}
\end{equation}
The diagram should be read as follows. If the orthogonal structure on $X^d$ can be lifted to a $\text{Pin}^c$ structure, then the dashed arrow becomes a solid arrow. In this case, the diagram commutes if and only if $W_3 = \beta(w_2) = 0$ on $X^d$, and equivalently, $w_2 = F_2$ mod 2.

As far as the mapping torus is concerned, let us assume $Z^4$, where $\partial Z^4 = X^2 \times S^1$, to also admit a $\text{Pin}^c$ structure, such that the integral lift \eqref{v2_integral_lift} exists. At the level of differential cocycles, the quadratic refinement on the mapping torus can be expressed as
\begin{align}
    \tilde{q}(\underline{\check{C}}) & = \frac{1}{2} \int_{X^2 \times S^1} \Big(\underline{C}_1 \cup \underline{G}_2 + \underline{\mathsf{G}}_2 \cup \underline{C}_1 + Q(\underline{G}_2,\underline{G}_2) - \underline{C}_1 \cup \underline{\Lambda}_2 - \underline{\mathsf{G}}_2 \cup \underline{\lambda}_1 - Q(\underline{G}_2,\underline{\Lambda}_2)\Big) \, ,
\end{align}
where $\underline{\check{C}} = \check{C} + \check{c} \star \check{s}$ and $\underline{\check{\lambda}} = \check{\lambda} = (\mathsf{F}_2 + \mathsf{W}_2,A_1 + w_1,F_2)$. Note that $\check{\lambda}$ is a ``background gauge field'' associated with the tangential structure of $X^2$, which we do not attempt to gauge, so a term like $\check{\ell} \star \check{s}$ should not be included in $\underline{\check{\lambda}}$. Further reducing over $S^1$ then yields the basepoint anomaly for the chiral boson,
\begin{equation}
    \widetilde{\mathcal{A}}[\check{C},\check{c}] = -\kappa \int_{X^2} [c_0]_{\mathbb{R}/\mathbb{Z}} \cup \bigg([\mathsf{G}_2]_\mathbb{Z} - \frac{1}{2} \, \Big([\mathsf{F}_2]_\mathbb{Z} + [\mathsf{W}_2]_\mathbb{Z}\Big)\bigg) \, .
\end{equation}
Since $\kappa = \pm 1$, the phase $e^{2\pi i \widetilde{\mathcal{A}}[\check{C},\check{c}]}$ in
\begin{equation}
    \mathcal{Z}[\check{C}] = e^{2\pi i (\widetilde{\mathcal{A}}[\check{C},\check{c}] + \kappa \mathcal{A}_\text{grav})} \mathcal{Z}[\check{C}]
\end{equation}
is an arbitrary element in $\mathbb{R}/\mathbb{Z}$, so the partition function for the chiral boson is non-vanishing only if
\begin{equation}
    [\mathsf{G}_2]_\mathbb{Z} = \frac{1}{2} \, \Big([\mathsf{F}_2]_\mathbb{Z} + [\mathsf{W}_2]_\mathbb{Z}\Big) \, .\label{chiral_boson_FW_anomaly}
\end{equation}

The condition above illustrates the importance of taking into account the quadratic refinement for the anomaly theory. Without doing so, it would be substituted by $[\mathsf{G}_2]_\mathbb{Z}=0$, so one might naïvely conclude that the background gauge field $\check{C}$ cannot be topologically non-trivial. This would be true if we consider only 2-manifolds equipped with a Spin structure. However, the fact that all 2-manifolds necessarily admit a $\text{Pin}^c$ structure means that we can couple the chiral boson not only to $\check{C}$, but also to the gauge field $\check{A}$ of the $\text{Pin}^c$ bundle, which comes for free, so as to relax the constraint on the integral class $[\mathsf{G}_2]_\mathbb{Z}$.

On top of that, there is a contribution from the second integral Stiefel-Whitney class $W_2$, whose mod 2 reduction is $w_1^2$. Let us provide a physical interpretation of the case when such a class contributes non-trivially. Similarly to the $\text{Pin}^c$ structure whose obstruction is measured by $W_3$, we will refer to an $\text{SO}^c$ structure as the tangential structure whose obstruction is measured by $W_2$. This can be seen via a diagram analogous to \eqref{Pinc_diagram} as follows.
\begin{equation}
    \begin{tikzcd}
        & B\text{SO}^c(d) \arrow[r,"f_1"] \arrow[d] & B\mathbb{Z} \arrow[r] \arrow[d,"\text{mod}\,2"] & \ast \arrow[d] \\
        X^d \arrow[ru,dashed] \arrow[r] & B\text{O}(d) \arrow[r,"w_1"] \arrow[rr,bend right,"W_2"] & B\mathbb{Z}_2 \arrow[r,"\beta"] & B^2\mathbb{Z}
    \end{tikzcd}
\end{equation}
As we can infer from above, the orthogonal structure of a manifold can be lifted to an $\text{SO}^c$ structure if and only if $W_2=0$, in which case commutativity of the diagram implies $w_1 = f_1$ mod 2, where $f_1$ can be identified as the characteristic class of the axionic 1-form field strength of the compact scalar. This is an analogue of our earlier Maxwell example in Section \ref{FW_anomaly_section}, wherein the $\text{Spin}^c$ structure implies that $w_2$ coincides mod 2 with $F_2$, which is the field strength of the dynamical gauge field.

As an aside, we can similarly interpret the vanishing of $V_{i+1} = \beta(v_i)$ as giving rise to a $\text{Wu}_i^c$ structure \cite{Sati:2011rw}, which is depicted by the diagram below.
\begin{equation}
    \begin{tikzcd}
        & B\text{Wu}_i^c(d) \arrow[r,"\Lambda_i"] \arrow[d] & B^i\mathbb{Z} \arrow[r] \arrow[d,"\text{mod}\,2"] & \ast \arrow[d] \\
        X^d \arrow[ru,dashed] \arrow[r] & B\text{O}(d) \arrow[r,"v_i"] \arrow[rr,bend right,"V_{i+1}"] & B^i\mathbb{Z}_2 \arrow[r,"\beta"] & B^{i+1}\mathbb{Z}
    \end{tikzcd}
\end{equation}
Equipping $X^d$ with a $\text{Wu}_i^c$ structure is therefore equivalent to picking an integral lift $v_i = \Lambda_i$ mod 2. We also observe that $\text{Wu}_1^c = \text{SO}^c$ and $\text{Wu}_2^c = \text{Pin}^c$.

A class of examples of manifolds which does not admit an $\text{SO}^c$ structure, i.e.~$W_2 \neq 0$, is those admitting a $\text{Pin}^-$ structure but not a $\text{Pin}^+$ structure. Recall that all 2-manifolds admit a $\text{Pin}^-$ structure but not necessarily a $\text{Pin}^+$ structure \cite{Kirby_Taylor_1991}. This implies all orientable 2-manifolds admit a Spin structure (and of course an $\text{SO}^c$ structure). For non-orientable 2-manifolds, a simple example that does not admit an $\text{SO}^c$ structure is $\mathbb{RP}^2$, which has $w_2 = w_1^2 \neq 0$.

There is a minor subtlety for orientable (and $\text{pin}^c$) manifolds. If we equip $X^2$ with an $\text{SO}^c$ structure, then we claim that the appropriate non-vanishing condition should be
\begin{equation}
    [\mathsf{G}_2]_\mathbb{Z} = \frac{1}{2} \, \Big([\mathsf{F}_2]_\mathbb{Z} + [\mathsf{f}_1]_\mathbb{Z} \cup [\mathsf{f}_1]_\mathbb{Z}\Big) \, .
\end{equation}
For the RHS to be well-defined as an integral class, this amounts to imposing a suitable quantisation condition on the field strength $f_1$ of the compact scalar, otherwise $X^2$ is equipped only with an $\text{SO}$ structure, in which case the non-vanishing condition above becomes
\begin{equation}
    [\mathsf{G}_2]_\mathbb{Z} = \frac{1}{2} \, [\mathsf{F}_2]_\mathbb{Z} \, .
\end{equation}

\subsection{M5-brane worldvolume theory}

For the M5-brane, its anomaly theory can be modeled by a Chern-Simons theory in $d=6$ with the M-theory 3-form $C_3$ \cite{Witten:1996hc}. The 11d M-theory spacetime is often assumed to admit a Spin structure, and typically one considers the wrapping of an M5-brane on a spin manifold, in which case \eqref{2d_6d_quadratic_refinements} does the job. Particularly, $\Lambda_4 = \frac{1}{2} p_1$ is a suitable integral lift of $v_4$. To keep the subsequent discussion general, we do not assume any tangential structure on the worldvolume $X^6$ of the M5-brane.

Note that Wu classes in degrees above half the dimension of the manifold vanish, so $v_4 = 0$ on 6-manifolds and thus $\Lambda_4$ exists.\footnote{This follows from the axiom that $\text{Sq}^i(x_j) = 0$ if $i>j$ \cite{5895d0ff-dbfd-39a7-9ffe-3ef746a10f70}.} In addition, recall from \eqref{integral_SW_classes} that the terms in $v_4$ which do not automatically admit integral lifts are $w_4$ and $w_2^2$, but $V_5 = \beta(v_4) = 0$ on $X^6$ implies that the combination $w_4 + w_2^2$ must be a mod 2 reduction.\footnote{Even though $\text{Sq}^1(w_2^2)=0$, the quantity $\beta(w_2^2)$ is generally a non-vanishing even integral class. In contrast, $\beta(w_1^2) = \beta(W_2 \ \text{mod} \ 2)$ must be vanishing as an integral class by exactness.} Let us denote it as $J_4$. In short, we can parametrise the integral lift of $v_4$ as
\begin{equation}
    \Lambda_4 = J_4 + W_4 + W_2^2 \, ,
\end{equation}
using the facts that $w_3 w_1 = W_4$ mod 2 and $w_1^4 = W_2^2$ mod 2.

Suppose we pick the 8-manifold $Z^8$, where $\partial Z^8 = X^6 \times S^1$, to also be such that $W_5=0$, then the quadratic refinement on the mapping torus can be expressed as
\begin{align}
    \tilde{q}(\underline{\check{C}}) & = \frac{1}{2} \int_{X^6 \times S^1} \Big(\underline{C}_3 \cup \underline{G}_4 + \underline{\mathsf{G}}_4 \cup \underline{C}_3 + Q(\underline{G}_4,\underline{G}_4) - \underline{C}_3 \cup \underline{\Lambda}_4 - \underline{\mathsf{G}}_4 \cup \underline{\lambda}_3 - Q(\underline{G}_4,\underline{\Lambda}_4)\Big) \, ,
\end{align}
where $\underline{\check{C}} = \check{C} + \check{c} \star \check{s}$ and $\underline{\check{\lambda}} = \check{\lambda} = (\mathsf{J}_4 + \mathsf{W}_4 + \mathsf{W}_2^2,j_3 + w_3 + w_1^3,J_4)$. Hence, the basepoint anomaly of the M5-brane is given by
\begin{equation}
    \widetilde{\mathcal{A}}[\check{C},\check{c}] = -\kappa \int_{X^6} [c_2]_{\mathbb{R}/\mathbb{Z}} \cup \bigg([\mathsf{G}_4]_\mathbb{Z} - \frac{1}{2} \, \Big([\mathsf{J}_4]_\mathbb{Z} + [\mathsf{W}_4]_\mathbb{Z} + [\mathsf{W}_2]_\mathbb{Z} \cup [\mathsf{W}_2]_\mathbb{Z}\Big)\bigg) \, ,\label{M5-brane_basepoint_anomaly}
\end{equation}
and so the M5-brane partition function is non-vanishing only if
\begin{equation}
    [\mathsf{G}_4]_\mathbb{Z} = \frac{1}{2} \, \Big([\mathsf{J}_4]_\mathbb{Z} + [\mathsf{W}_4]_\mathbb{Z} + [\mathsf{W}_2]_\mathbb{Z} \cup [\mathsf{W}_2]_\mathbb{Z}\Big) \, .\label{M5-brane_FW_anomaly}
\end{equation}
It was argued by \cite{Diaconescu:2003bm} that the partition function being non-vanishing is necessary for the M5-brane to decouple from the dynamics of the 11d supergravity bulk. Only when satisfied can the M5-brane worldvolume theory be treated as a standalone 6d (2,0) superconformal field theory (SCFT) \cite{Witten:1995em}.

Let us examine what \eqref{M5-brane_FW_anomaly} becomes when we impose various commonly quoted tangential structures. Here it is crucial to distinguish between the {\it property} of a manifold admitting a given structure, and the {\it data} of actually equipping it with such a structure (i.e.~coupling the theory to the relevant gauge fields). If $X^6$ is orientable, then $W_4=W_2=0$, so we obtain
\begin{equation}
    [\mathsf{G}_4]_\mathbb{Z} = \frac{1}{2} \, [\mathsf{J}_4]_\mathbb{Z} \, .
\end{equation}
If $X^6$ is $\text{so}^c$, then $w_1 = f_1$ mod 2 where $f_1$ is the aforementioned axion class, so we obtain
\begin{equation}
    [\mathsf{G}_4]_\mathbb{Z} = \frac{1}{2} \, \Big([\mathsf{J}_4]_\mathbb{Z} + [\mathsf{W}_4]_\mathbb{Z} + [\mathsf{f}_1]_\mathbb{Z} \cup [\mathsf{f}_1]_\mathbb{Z} \cup [\mathsf{f}_1]_\mathbb{Z} \cup [\mathsf{f}_1]_\mathbb{Z}\Big) \, .
\end{equation}
If $X^6$ is $\text{pin}^+$, then $W_4=0$, so we obtain
\begin{equation}
    [\mathsf{G}_4]_\mathbb{Z} = \frac{1}{2} \, \Big([\mathsf{J}_4]_\mathbb{Z} + [\mathsf{W}_2]_\mathbb{Z} \cup [\mathsf{W}_2]_\mathbb{Z}\Big) \, .
\end{equation}
If $X^6$ is $\text{pin}^-$, then $w_2^2 + w_1^4 = 0$, so we obtain
\begin{equation}
    [\mathsf{G}_4]_\mathbb{Z} = \frac{1}{2} \, \Big([\mathsf{J}_4]_\mathbb{Z} + [\mathsf{W}_4]_\mathbb{Z}\Big) \, .
\end{equation}
If $X^6$ is $\text{pin}^c$, then $w_3 w_1 = F_2 W_2$ mod 2 and $I_4 = F_2^2$ where $F_2$ is the first Chern class, so we obtain
\begin{equation}
    [\mathsf{G}_4]_\mathbb{Z} = \frac{1}{2} \, \Big([\mathsf{J}_4]_\mathbb{Z} + [\mathsf{F}_2]_\mathbb{Z} \cup [\mathsf{W}_2]_\mathbb{Z} + [\mathsf{F}_2]_\mathbb{Z} \cup [\mathsf{F}_2]_\mathbb{Z} + [\mathsf{W}_2]_\mathbb{Z} \cup [\mathsf{W}_2]_\mathbb{Z}\Big) \, .
\end{equation}
If $X^6$ is $\text{spin}^c$, then $W_4 = W_2 = 0$ and $I_4 = F_2^2$, so we obtain
\begin{equation}
    [\mathsf{G}_4]_\mathbb{Z} = \frac{1}{2} \, \Big([\mathsf{J}_4]_\mathbb{Z} + [\mathsf{F}_2]_\mathbb{Z} \cup [\mathsf{F}_2]_\mathbb{Z}\Big) \, .
\end{equation}
Lastly, if $X^6$ is spin, then $w_4 = \frac{1}{2} p_1$ mod 2 and $W_4=w_2=W_2=0$, so we obtain\footnote{As was shown in \cite{Hsieh:2020jpj}, $\frac{1}{4} p_1$ is indeed always an integral cohomology class on spin manifolds of dimension less than or equal to 7, due to the fact that $v_4=w_3=w_2=w_1=0$. Similar arguments apply to the previous cases.}
\begin{equation}
    [\mathsf{G}_4]_\mathbb{Z} = \frac{1}{4} \, [\mathsf{p}_1]_\mathbb{Z} \, .\label{spin_M5-brane_FW_anomaly}
\end{equation}
Our result for the spin case is indeed compatible with the well-known shifted quantisation of the $G_4$ flux in M-theory \cite{Witten:1996md}, where
\begin{equation}
    \int_{\mathcal{C}^4} \bigg(G_4 - \frac{1}{4} \, p_1(R_2)\bigg) \in \mathbb{Z}
\end{equation}
for any (spin) 4-cycle $\mathcal{C}^4$ of the 11d spacetime. From our perspective, such a ``charge'' being non-zero is an obstruction to have a non-vanishing partition function for the M5-brane.

The Freed-Witten anomaly cancellation condition \eqref{Freed-Witten_anomaly_cancellation_condition} can be interpreted as the D3-brane admitting a twisted $\text{Spin}^c$ structure \cite{Freed:1999vc}, with the twist given by $[\mathsf{H}_3]_\mathbb{Z}$ or $[\mathsf{G}_3]_\mathbb{Z}$. There is a similar interpretation for \eqref{spin_M5-brane_FW_anomaly}. Note that the quantity $\frac{1}{2} p_1$ is also known as the first fractional Pontryagin class. It measures the obstruction to lift a Spin structure to a String structure \cite{Killingback:1986rd,Stolz_Teichner_2004}, as depicted by the diagram below.
\begin{equation}
    \begin{tikzcd}
        & B\text{String}(d) \arrow[r] \arrow[d] & \ast \arrow[d] \\
        X^d \arrow[ru,dashed] \arrow[r] & B\text{Spin}(d) \arrow[r,"\frac{1}{2} p_1"] & B^4\mathbb{Z}
    \end{tikzcd}
\end{equation}
Therefore, we can interpret \eqref{spin_M5-brane_FW_anomaly} as the M5-brane admitting a twisted String structure \cite{Sati:2009ic,Sati:2010dc}, with the twist given by $[\mathsf{G}_4 + \frac{1}{4} \mathsf{p}_1]_\mathbb{Z}$.

On a related note, based on the arguments in \cite{Witten:1998xy,Freed:1999vc} leading to \eqref{Freed-Witten_anomaly_cancellation_condition}, it was conjectured by \cite{Witten:1999vg} that the analogue for the M5-brane should take the form,
\begin{equation}
    [\mathsf{G}_4]_\mathbb{Z} \stackrel{?}{=} \beta([\theta_3]_{\mathbb{R}/\mathbb{Z}}) \, ,
\end{equation}
for some $[\theta_3]_{\mathbb{R}/\mathbb{Z}} \in H^3(X^6;\mathbb{R}/\mathbb{Z})$. Note that our general result can be expressed compactly as $[\mathsf{G}_4]_\mathbb{Z} = \frac{1}{2} [\Lambda_4]_\mathbb{Z}$. Its relation to $v_4$ can be understood in terms of the following long exact sequence.
\begin{equation}
    \begin{tikzcd}[row sep=tiny,column sep=scriptsize]
        \cdots \arrow[r] & H^3(X^6;\mathbb{Z}_2) \arrow[r,"\beta"] & H^4(X^6;\mathbb{Z}) \arrow[r,"\times 2"] & H^4(X^6;\mathbb{Z}) \arrow[r,"\text{mod} \, 2"] &[1em] H^4(X^6;\mathbb{Z}_2) \arrow[r] & \cdots \\
        & \theta_3 \arrow[r,mapsto,"?"] & \frac{1}{2} \Lambda_4 \arrow[r,mapsto] & \Lambda_4 \arrow[r,mapsto] & v_4 &
    \end{tikzcd}
\end{equation}
By exactness, if such a $\theta_3$ exists, then $\beta(\theta_3)$ has to vanish as an integral cohomology class upon multiplication by 2, i.e.~$\Lambda_4=0$, or equivalently, $\frac{1}{2} \Lambda_4$ is 2-torsion. Given that $v_4=0$ on 6-manifolds, $\Lambda_4=0$ is trivially an integral lift, but we also want $\frac{1}{2} \Lambda_4$ to be non-trivial. This is not always guaranteed. Nonetheless, a possible scenario is when $w_4 + w_2^2 = 0$, such that $\Lambda_4 = W_4 + W_2^2 = \beta(w_3 + w_1^3)$, in which case we have $\theta_3 = \frac{1}{2} (w_3 + w_1^3)$. If, furthermore, $w_2=0$, then $\theta_3 = \frac{1}{2} w_1^3$.

For completeness, we expect that \eqref{M5-brane_FW_anomaly} should be modified in the presence of sources, analogously to \eqref{D3-brane_FW_anomaly_with_sources}. By M/F-theory duality \cite{Johnson:1997wf}, the Wilson operators are now M2-branes ending on a 2-submanifold of $X^6$. Schematically, the non-vanishing condition for the resulting M5-brane partition function becomes
\begin{equation}
    [\mathsf{G}_4]_\mathbb{Z} + \text{PD}([\text{M2}]) = \frac{1}{2} \, [\Lambda_4]_\mathbb{Z} \, .
\end{equation}
As alluded to earlier, this relation should be interpreted with care since the M5-brane can no longer be regarded as a decoupled theory with respect to the M-theory background \cite{Diaconescu:2003bm}.


\section{D3-brane in F-theory backgrounds}\label{S-folds_section}

We would like to study the D3-brane again, but now viewed as a dimensional reduction of the M5-brane. Specifically, we take the worldvolume of the M5-brane to be an elliptic fibration
\begin{equation}
    T^2 \hookrightarrow X^6 \to X^4 \, ,
\end{equation}
with $X^4$ closed and path-connected. For simplicity, we assume that the fiber is non-singular. In general, this is a non-trivial fiber bundle with a system of {\it local coefficients} specified by a representation,
\begin{equation}
    \rho: \pi_1(X^4) \to \text{Aut}_\mathbb{Z}(H_\ast(T^2;\mathbb{Z})) \, ,
\end{equation}
which is equivalent to $H_\ast(T^2;\mathbb{Z})$ regarded as a $\mathbb{Z}[\pi_1(X^4)]$-module \cite{b435f69a-14cc-3b23-b856-e21562c71558,MR1867354}.

Notably, in the middle degree we have $\text{Aut}_\mathbb{Z}(\mathbb{Z} \oplus \mathbb{Z}) = \text{GL}(2,\mathbb{Z})$, i.e.~the mapping class group of the torus. For our purposes, it suffices to restrict to the orientation-preserving subgroup $\text{SL}(2,\mathbb{Z})$. As a result, we interpret the resultant D3-brane to be placed in a Type IIB string theory background admitting a non-trivial $\text{SL}(2,\mathbb{Z})$-action. One familiar example of such is an orientifold background, e.g.~$X^4 = S^1 \times \mathbb{RP}^3 \subset \text{AdS}_5 \times \mathbb{RP}^5$, where $\rho$ induces a flip in sign of the $\text{SL}(2,\mathbb{Z})$-doublet $([\mathsf{H}_3],[\mathsf{G}_3]) \in H^3(\mathbb{RP}^3;(\mathbb{Z} \oplus \mathbb{Z})_\rho)$ as we go around a non-contractible loop \cite{Witten:1998xy}.\footnote{Whenever the context is clear, we will abbreviate $(\mathbb{Z} \oplus \mathbb{Z})_\rho$ as $\mathbb{Z}^2_\rho$.} Using the M5-brane basepoint anomaly \eqref{M5-brane_basepoint_anomaly} as the starting point, our goal is to derive the Freed-Witten anomaly cancellation condition for the D3-brane in generic F-theory backgrounds by reducing over the fiber $T^2$.\footnote{As a reminder, this torus $T^2$ is independent from the mapping torus $X^d \times S^1$.}

\subsection{Trivial fibration}

Let us begin with the simplest scenario where the elliptic fibration is a product manifold
\begin{equation}
    X^6 = X^4 \times T^2 \, ,
\end{equation}
and the $\text{SL}(2,\mathbb{Z})$-representation $\rho$ is trivial. Using the Künneth formula, we can decompose
\begin{align}
    [c_2]_{\mathbb{R}/\mathbb{Z}} & = [\varepsilon_2]_{\mathbb{R}/\mathbb{Z}} + [b_1]_{\mathbb{R}/\mathbb{Z}} \cup [\mathsf{s}_1]_\mathbb{Z} + [c_1]_{\mathbb{R}/\mathbb{Z}} \cup [\mathsf{s}_1']_\mathbb{Z} + [\kappa_0]_{\mathbb{R}/\mathbb{Z}} \cup [\omega_2]_\mathbb{Z} \, ,\label{c2_ansatz}\\
    [\mathsf{G}_4]_\mathbb{Z} & = [\mathsf{E}_4]_\mathbb{Z} + [\mathsf{H}_3]_\mathbb{Z} \cup [\mathsf{s}_1]_\mathbb{Z} + [\mathsf{G}_3]_\mathbb{Z} \cup [\mathsf{s}_1']_\mathbb{Z} + [\mathsf{K}_2]_\mathbb{Z} \cup [\omega_2]_\mathbb{Z} \, ,\label{G4_ansatz}
\end{align}
where $[\mathsf{s}_1]_\mathbb{Z},[\mathsf{s}_1']_\mathbb{Z} \in H^1(T^2;\mathbb{Z})$ are Poincaré-dual to the two 1-cycles of $T^2$, satisfying
\begin{equation}
    \int_{T^2} [\mathsf{s}_1]_\mathbb{Z} \cup [\mathsf{s}_1']_\mathbb{Z} = \int_{T^2} [\omega_2]_\mathbb{Z} = 1 \, ,
\end{equation}
and the rest are elements of $H^\ast(X^4;\mathbb{Z})$. The pullbacks in the ansatzes above are implicitly understood. Note that terms like $[b_1]_{\mathbb{R}/\mathbb{Z}} \cup [\mathsf{s}_1]_\mathbb{Z}$ in \eqref{c2_ansatz} are well-defined as the cup product $\cup: H^i(-;\mathbb{R}/\mathbb{Z}) \times H^j(-;\mathbb{Z}) \to H^{i+j}(-;\mathbb{R}/\mathbb{Z})$.

Meanwhile, the fourth Wu class $v_4$ decomposes under the Whitney sum formula as
\begin{equation}
    v_4(X^6) = v_4(X^4) + v_3(X^4) \cup v_1(T^2) + v_2(X^4) \cup v_2(T^2) \, .\label{Wu_class_Whitney_sum}
\end{equation}
What we actually need in the quadratic refinement, though, is its integral lift $\Lambda_4$. Let us first look at the terms $v_1(T^2) = w_1(T^2)$ and $v_2(T^2) = w_2(T^2) + w_1^2(T^2)$, both of which vanish as mod 2 classes. Combined with the facts that $\beta(v_1(T^2))=0$ and $\beta(v_2(T^2))=0$, they admit integral lifts which are necessarily even. Recall that on $T^2$, we have the identities
\begin{equation}
    w_1 \cup x_1 = v_1 \cup x_1 = \text{Sq}^1(x_1) = x_1 \cup x_1
\end{equation}
for all $x_1 \in H^1(T^2;\mathbb{Z}_2)$, so one can check that
\begin{equation}
    2\big([\mathsf{s}_1]_\mathbb{Z} + [\mathsf{s}_1']_\mathbb{Z}\big) \in H^1(T^2;\mathbb{Z})\label{non-canonical_w1_lift}
\end{equation}
is a suitable integral lift of $v_1(T^2)$ whose mod 2 reduction satisfies the defining relations above. Of course, one could have chosen the integral lift to simply be zero, but this would kill the term $v_3(X^4)$ in \eqref{Wu_class_Whitney_sum} after dimensionally reducing the M5-brane, thus leaving us with a less general constraint. Alternatively, one could also choose any linear combination of $[\mathsf{s}_1]_\mathbb{Z}$ and $[\mathsf{s}_1']_\mathbb{Z}$ with non-vanishing even integer coefficients, but as we will see in a moment, $v_3(X^4)$ can be replaced by $W_3(X^4)$. The latter is 2-torsion, so the ``minimal'' non-vanishing integral lift \eqref{non-canonical_w1_lift} can indeed be chosen without loss of generality. In our terminology, this amounts to picking an $\text{SO}^c$ structure on $T^2$, defined by the choice of an integral lift $f_1$ such that $w_1 = f_1$ mod 2.

The situation for $v_2(T^2) = w_2(T^2) + w_1^2(T^2)$ is similar. Note that the cup product of \eqref{non-canonical_w1_lift} with itself is zero, which is compatible with the fact that $\text{Sq}^1(w_1(T^2)) = w_1^2(T^2) = 0$. The canonical integral lift of the top-degree Stiefel-Whitney class $w_2(T^2)$ is the Euler class $e_2(T^2)$, which vanishes for the torus. Just like before, after dimensional reduction this would give rise to a constraint that might be too restrictive. We will thereby choose $2[\omega_2]_\mathbb{Z} = 2[\mathsf{s}_1]_\mathbb{Z} \cup [\mathsf{s}_1']_\mathbb{Z} \in H^2(T^2;\mathbb{Z})$ as the integral lift of $v_2(T^2)$ in \eqref{Wu_class_Whitney_sum}.

We now turn our focus to the Wu classes of $X^4$, hereafter suppressing the explicit dependence on it to simplify notation. Since $V_5 = \beta(v_4) = 0$ by degree reasons, $v_4 = w_4 + w_3 w_1 + w_2^2 + w_1^4$ must admit an integral lift. In fact, each term in $v_4$ can be lifted, and we can express
\begin{equation}
    \Lambda_4 = e_4 + W_4 + I_4 + W_2^2 \, ,
\end{equation}
where $e_4$ is the Euler class of $X^4$, and we denote the integral lift of $w_2^2$ as $I_4$, suggesting that it is essentially the instanton class.\footnote{As an aside, on (smooth and closed) orientable 4-manifolds, it follows from Poincaré duality and Rokhlin's theorem that $w_4 \neq 0$ implies $w_2 \neq 0$, but not vice versa.}

For $v_3 = w_2 w_1$, we again utilise the fact that Wu classes vanish at degrees above half the dimension of the manifold, so $V_4 = \beta(v_3) = 0$ implies $v_3$ admits an integral lift. We would like to argue that, in this particular context, one can use $W_3$ as an effective integral lift of $v_3$. To see how, after dimensionally reducing \eqref{M5-brane_basepoint_anomaly} over $T^2$ to a 4d action, we will find terms of the form $x_1 \cup w_2 \cup w_1$, where $x_1 \in H^1(X^4;\mathbb{Z}_2)$ roughly corresponds either to $[b_1]_{\mathbb{R}/\mathbb{Z}}$ or $[c_1]_{\mathbb{R}/\mathbb{Z}}$ in \eqref{c2_ansatz}. Recall that from the perspective of anomaly inflow, the anomaly theory of the D3-brane is defined on a 6-manifold $Z^6$ with $\partial Z^6 = X^4 \times S^1$. In this viewpoint, $x_1$ originates from a 3-cocycle $x_3 \in H^3(Z^6;\mathbb{Z}_2)$.
It follows from the Cartan formula that
\begin{equation}
    \text{Sq}^1(x_3 \cup w_2) = \text{Sq}^1(x_3) \cup w_2 + x_3 \cup \text{Sq}^1(w_2) \, .
\end{equation}
By definition, we can rewrite the LHS as $\text{Sq}^1(x_3 \cup w_2) = v_1 \cup x_3 \cup w_2 = w_1 \cup x_3 \cup w_2$. We have $\text{Sq}^1(x_3)=0$ as well since $x_3$ admits an integral lift by construction. Together we obtain
\begin{equation}
    x_3 \cup w_2 \cup w_1 = -x_3 \cup W_3 \mod 2 \, ,
\end{equation}
where $W_3 = \beta(w_2)$. At the level of the 4d action, this amounts to substituting the integral lift of $v_3$ with $-W_3$, with the sign merely being a choice of convention.

We should stress that $W_3$ is not literally the integral lift of $v_3$. On 4-manifolds, given that $v_3 = w_2 w_1 = 0$, the Wu formula $\text{Sq}^1(w_2) = w_1 w_2 + w_3 = w_3$ tells us that the mod 2 reduction of $W_3$ is actually $w_3$. To understand the relation between all these quantities, we observe that on any $n$-manifold,
\begin{equation}
    v_3 \cup x_{n-3} = \text{Sq}^3(x_{n-3}) = \text{Sq}^1 \circ \text{Sq}^2(x_{n-3}) \, ,
\end{equation}
where the Adem relations are used in the second equality. The integral uplift of $\text{Sq}^3: H^\ast(-;\mathbb{Z}_2) \to H^{\ast+3}(-;\mathbb{Z}_2)$ is precisely the differential $d_3 = \beta \circ \text{Sq}^2 \circ (\text{mod} \ 2)$ on the third page of the {\it Atiyah-Hirzebruch spectral sequence} for (complex) K-theory. Hence, $\text{Sq}^3(x_{n-3})=0$ corresponds to the condition for $x_{n-3}$ to admit a K-theory lift, whereas $W_3=0$ (plus $w_1=0$) is the condition for a manifold to be orientable in K-theory (see, e.g.~\cite{Diaconescu:2000wy,Freed:2000ta,Yonekura:2024bvh}).

The case for $v_2 = w_2 + w_1^2$ is somewhat different, which does not automatically vanish on 4-manifolds like $v_3$ does. Its vanishing requires $X^4$ to admit a $\text{Pin}^c$ structure, i.e.~$W_3=0$. When satisfied, the integral lift of $v_2$ can be taken to be $-(F_2 + W_2)$ as we saw in Section \ref{chiral_boson_section}, with the minus sign being a choice again. On the other hand, if $W_3 \neq 0$, we will simply take the integral lift of $v_2(T^2)$ in \eqref{Wu_class_Whitney_sum} to be (canonically) zero, such that one needs not worry about lifting $v_2(X^4)$.

Collecting our results, we find that reducing \eqref{M5-brane_basepoint_anomaly} over $T^2$ gives rise to the following basepoint anomaly for the D3-brane,
\begin{align}
    \widetilde{\mathcal{A}} & = -\kappa \int_{X^4} \bigg([\kappa_0]_{\mathbb{R}/\mathbb{Z}} \cup \bigg([\mathsf{E}_4]_\mathbb{Z} - \frac{1}{2} \, \Big([\mathsf{e}_4]_\mathbb{Z} + [\mathsf{W}_4]_\mathbb{Z} + [\mathsf{I}_4]_\mathbb{Z} + [\mathsf{W}_2]_\mathbb{Z} \cup [\mathsf{W}_2]_\mathbb{Z}\Big)\bigg)\nonumber\\
    & \phantom{=\ } + [c_1]_{\mathbb{R}/\mathbb{Z}} \cup \Big([\mathsf{H}_3]_\mathbb{Z} + [\mathsf{W}_3]_\mathbb{Z}\Big) - [b_1]_{\mathbb{R}/\mathbb{Z}} \cup \Big([\mathsf{G}_3]_\mathbb{Z} + [\mathsf{W}_3]_\mathbb{Z}\Big)\nonumber\\
    & \phantom{=\ } + [\varepsilon_2]_{\mathbb{R}/\mathbb{Z}} \cup \Big([\mathsf{K}_2]_\mathbb{Z} + [\mathsf{F}_2]_\mathbb{Z} + [\mathsf{W}_2]_\mathbb{Z}\Big)\bigg) \, ,\label{D3-brane_basepoint_anomaly}
\end{align}
where the terms $[\mathsf{F}_2]_\mathbb{Z} + [\mathsf{W}_2]_\mathbb{Z}$ are understood to be absent if $[\mathsf{W}_3]_\mathbb{Z} \neq 0$. In other words, the D3-brane partition function is non-vanishing only if
\begin{equation}
    \begin{gathered}
        [\mathsf{E}_4]_\mathbb{Z} = \frac{1}{2} \, \Big([\mathsf{e}_4]_\mathbb{Z} + [\mathsf{W}_4]_\mathbb{Z} + [\mathsf{I}_4]_\mathbb{Z} + [\mathsf{W}_2]_\mathbb{Z} \cup [\mathsf{W}_2]_\mathbb{Z}\Big) \, ,\\
        [\mathsf{H}_3]_\mathbb{Z} + [\mathsf{W}_3]_\mathbb{Z} = 0 \, , \qquad [\mathsf{G}_3]_\mathbb{Z} + [\mathsf{W}_3]_\mathbb{Z} = 0 \, , \qquad [\mathsf{K}_2]_\mathbb{Z} + [\mathsf{F}_2]_\mathbb{Z} + [\mathsf{W}_2]_\mathbb{Z} = 0 \, .
    \end{gathered}\label{D3-brane_FW_anomaly}
\end{equation}
We have thus ``rederived'' the Freed-Witten anomaly cancellation condition \eqref{Freed-Witten_anomaly_cancellation_condition} for the D3-brane by a torus compactification of the M5-brane, particularly providing a 6d origin for the shifts $\check{B} \to \check{B} + \check{w}$ and $\check{C} \to \check{C} + \check{w}$ in the 4d effective theory.

On top of that, we have obtained two additional constraints that were not included earlier when we modeled the D3-brane with Maxwell theory. It is obvious that the conditions $[\mathsf{H}_3]_\mathbb{Z} + [\mathsf{W}_3]_\mathbb{Z} = [\mathsf{G}_3]_\mathbb{Z} + [\mathsf{W}_3]_\mathbb{Z} = 0$ arises from the Maxwell term $2\pi i \int_{X^4} C_2 \wedge (B_2 + F_2)$ in the Wess-Zumino action for the D3-brane, whereas the condition $[\mathsf{E}_4]_\mathbb{Z} = \frac{1}{2} \big([\mathsf{e}_4]_\mathbb{Z} + [\mathsf{W}_4]_\mathbb{Z} + [\mathsf{I}_4]_\mathbb{Z} + [\mathsf{W}_2]_\mathbb{Z} \cup [\mathsf{W}_2]_\mathbb{Z}\big)$ is evidently concerned with the theta term
\begin{equation}
    2\pi i \int_{X^4} \frac{1}{2} \, C_0 \wedge (B_2 + F_2)^2 \, .
\end{equation}
In particular, the Euler class $e_4$ describes the effect of coupling the theory to an Euler counterterm, and the instanton class $I_4$ (i.e.~two times the second Chern character) describes the correction from coupling the Chan-Paton gauge field to the worldvolume fermions.

The origins of $W_4$ and $W_2^2$ are less transparent. To understand better, let us consider the case that $X^4$ is not (necessarily) orientable but is equipped with an $\text{SO}^c$ structure, then $w_1 = f_1$ mod 2 where $f_1$ can be identified with the field strength of the axioic mode coupled to the worldvolume fermions, in analogy to $\text{spin}^c$ fermions on orientable (but not spin) manifolds coupling to a $\U(1)$ gauge field.\footnote{An example of a 4-manifold with $w_1 \neq 0$, $w_2 \neq 0$, $w_1^2 \neq 0$ is $\mathbb{RP}^2 \times \mathbb{RP}^2$, whereas a counterexample with $w_1 \neq 0$, $w_2 \neq 0$, $w_1^2 = 0$ is $(K \times S^2) \# \mathbb{CP}^2$, where $K$ denotes the Klein bottle and $\#$ denotes the connected sum (credit to the contributors of this Mathematics Stack Exchange \href{https://math.stackexchange.com/questions/2071685/4-manifold-with-w-1-neq-0-w-12-0-w-2-neq-0}{post}).} In this case, one finds $W_4 \to W_3 f_1$ and $W_2^2 \to f_1^4$, so the former arises from the interaction between the Chan-Paton gauge field and the axion, while the latter can be interpreted as the self-interaction of the axion.

\subsection{Non-trivial fibration with constant local system}

We shall now consider a non-trivial elliptic fibration $T^2 \xrightarrow{\iota} X^6 \xrightarrow{\pi} X^4$ which is not a product manifold. The base $X^4$ can be allowed to be not simply-connected, i.e.~$\pi_1(X^4) \neq 0$, but we demand that it acts trivially on the cohomology of the fiber $T^2$, such that the system of local coefficients appearing in the spectral sequence below is constant. We will treat the most general case in the next subsection. To obtain the basepoint anomaly of the D3-brane on $X^4$, starting from that of the M5-brane on $X^6$, we need to perform a {\it fiber integration} over $T^2$. Formally, the mathematical tool relating the cohomology groups between the total space, the base, and the fiber is the {\it Leray-Serre spectral sequence} \cite{1b68bd4a-2ddc-338a-95d3-30d18474da26}. The primary input data for the spectral sequence are entries on the second page,
\begin{equation}
    E_2^{p,q} = H^p(X^4;H^q(T^2;G)) \, ,
\end{equation}
where $G$ can be $\mathbb{Z}$ or $\mathbb{R}/\mathbb{Z}$ for our purposes, then one can algorithmically compute the cohomology of $X^6$. A review of the essential technical details can be found in Appendix \ref{Serre_spectral_sequence_appendix}.

To summarise, cohomology classes of the D3-brane worldvolume $X^4$ are related to those of the M5-brane worldvolume $X^6$ by the pullback map
\begin{equation}
    \pi^\ast: H^i(X^4;G) \to H^i(X^6;G) \, .\label{cohomology_pullback}
\end{equation}
Similarly, we can relate cohomology classes of $X^6$ to those of the fiber $T^2$ via
\begin{equation}
    \iota^\ast: H^i(X^6;G) \to H^i(T^2;G) \, .
\end{equation}
We also have a dual pushforward map
\begin{equation}
    \iota_\ast: H_i(T^2;G) \to H_i(X^6;G)\label{homology_pushforward}
\end{equation}
sending homology classes of $T^2$ to those of $X^6$. Suppose we modify the parametrisation of the ansatz \eqref{G4_ansatz} for $[\mathsf{G}_4]_\mathbb{Z}$ as
\begin{equation}
    [\mathsf{G}_4]_\mathbb{Z} = \pi^\ast\big([\mathsf{E}_4]_\mathbb{Z}\big) + \pi^\ast\big([\mathsf{H}_3]_\mathbb{Z}\big) \cup [\mathsf{s}_1]_\mathbb{Z} + \pi^\ast\big([\mathsf{G}_3]_\mathbb{Z}\big) \cup [\mathsf{s}_1']_\mathbb{Z} + \pi^\ast\big([\mathsf{K}_2]_\mathbb{Z}\big) \cup [\omega_2]_\mathbb{Z} \, ,
\end{equation}
such that
\begin{equation}
    \int_{T^2} \iota^\ast\big([\mathsf{s}_1]_\mathbb{Z} \cup [\mathsf{s}_1']_\mathbb{Z}\big) = \int_{T^2} \iota^\ast\big([\omega_2]_\mathbb{Z}\big) = 1 \, ,
\end{equation}
and likewise for $[c_2]_{\mathbb{R}/\mathbb{Z}}$ in \eqref{c2_ansatz}. Denoting each summand in the product $[c_2]_{\mathbb{R}/\mathbb{Z}} \cup [\mathsf{G}_4]_\mathbb{Z}$ schematically as $x = \pi^\ast(b) \cup f$, we define the fiber integration over $T^2$ as a map
\begin{equation}
    \pi_!: H^i(X^6;\mathbb{R}/\mathbb{Z}) \to H^{i-2}(X^4;\mathbb{R}/\mathbb{Z}) \, , \qquad x \mapsto b \cup \big(f \cap \iota_\ast([T^2]_\mathbb{Z})\big) \, ,\label{fiber_integration_definition}
\end{equation}
where $[T^2]_\mathbb{Z} \in H_2(T^2;\mathbb{Z})$ is the fundamental class of the fiber. Note that the cap product $\cap: H^2(X^6;\mathbb{Z}) \times H_2(X^6;\mathbb{Z}) \to \mathbb{Z}$ outputs an integer, so $b \cup \big(f \cap \iota_\ast([T^2]_\mathbb{Z})\big)$ is indeed an element of $H^{i-2}(X^4;\mathbb{R}/\mathbb{Z})$.

Such a definition is compatible with the decomposition of the Stiefel-Whitney classes (and hence their integral lifts). Specifically, the tangent bundle of the total space of a fiber bundle $F \xrightarrow{\iota} X \xrightarrow{\pi} B$ decomposes as the direct sum,
\begin{equation}
    TX \cong \pi^\ast(TB) \oplus T_\pi X \, ,
\end{equation}
where $T_\pi X = \text{ker}(d\pi)$ denotes the vertical tangent bundle, i.e.~the bundle of vectors in $TX$ tangent to the fibers \cite{2a7f5c42-423c-3a9f-b9ae-23d812c6429c}, so the Stiefel-Whitney classes of the total space decompose as
\begin{align}
    w(TX) & = \pi^\ast w(TB) \cup w(T_\pi X)\nonumber\\
    & \supset 1 + \pi^\ast w_1(TB) + w_1(T_\pi X) + \pi^\ast w_2(TB) + \pi^\ast w_1(TB) \cup w_1(T_\pi X) + w_2(T_\pi X) \, .
\end{align}
Consequently, the fiber integration of the M5-brane basepoint anomaly \eqref{M5-brane_basepoint_anomaly} essentially yields the same result for the D3-brane as in \eqref{D3-brane_basepoint_anomaly}, such that the non-vanishing condition for its partition function is also given by \eqref{D3-brane_FW_anomaly}.

This is to perhaps not too surprising. As long as there is no non-trivial $\text{SL}(2,\mathbb{Z})$-action on the fluxes supported on the base $X^4$, which we are going to examine next, one should not be able to distinguish between a trivial and a non-trivial fibration from the bottom-up perspective of the D3-brane as an effective theory.

\subsection{Non-trivial fibration with non-trivial local system}

Interesting complications arise when we consider a fibration $T^2 \xrightarrow{\iota} X^6 \xrightarrow{\pi} X^4$ where $\pi_1(X^4)$ acts non-trivially on the cohomology of $T^2$. In this case, the second page of the Leray-Serre spectral sequence has entries
\begin{equation}
    E_2^{p,q} = H^p(X^4;H^q(T^2;G)_\rho)
\end{equation}
where $H^\ast(T^2;G)_\rho$ denotes the cohomology groups of $T^2$ regarded as a system of local coefficients specified by the representation $\rho: \pi_1(X^4) \to \text{Aut}_\mathbb{Z}(H^\ast(T^2;G))$. As usual, there exists a dual construction in homology. To be concrete, suppose $G=\mathbb{Z}$, then
\begin{equation}
    \text{Aut}_\mathbb{Z}(H^0(T^2;\mathbb{Z})) = \text{Aut}_\mathbb{Z}(H^2(T^2;\mathbb{Z})) = \mathbb{Z}_2 \, , \qquad \text{Aut}_\mathbb{Z}(H^1(T^2;\mathbb{Z})) = \text{GL}(2,\mathbb{Z}) \, .
\end{equation}
For simplicity, we focus on orientation-preserving automorphisms, in which case $\rho$ is trivial in degrees 0 and 2, and it is an $\text{SL}(2,\mathbb{Z})$-representation in degree 1.

Of particular interest is the doublet $(B_2,C_2)$ in Type IIB string theory which transforms non-trivially under $\text{SL}(2,\mathbb{Z})$. When restricted to the worldvolume of the D3-brane, its characteristic class $[\mathcal{H}_3] = ([\mathsf{H}_3],[\mathsf{G}_3])$ is an element of $H^3(X^4;\mathbb{Z}^2_\rho)$. One might intuitively expect that performing a fiber integration of \eqref{M5-brane_basepoint_anomaly} over $T^2$ would result in a basepoint anomaly in terms of $[\mathcal{H}_3]_{\mathbb{Z}^2_\rho} \in H^3(X^4;\mathbb{Z}^2_\rho)$. This is unfortunately not true. Even when working with the Leray-Serre spectral sequence with local coefficients, we can unpack the definitions and check that the standard fiber integration always maps between the ordinary cohomology of the total space and that of the base.

Heuristically, the main reason why the codomain of the fiber integration is ordinary cohomology (of the base), rather than cohomology with local coefficients, is because the construction involves taking the cap product with the fundamental class of the fiber as in \eqref{fiber_integration_definition}. This in turns sends us to cohomology with coefficients given by $H_2(T^2;\mathbb{Z})_\rho = \mathbb{Z}$. Morally, we want to instead take the cap product with 1-cycles of the fiber, so as to land in cohomology with coefficients given by $H_1(T^2;\mathbb{Z})_\rho = \mathbb{Z}^2_\rho$.

We propose in Appendix \ref{twisted_fiber_integration_appendix} that under suitable conditions, one can construct a notion of ``twisted fiber integration'' whose codomain is cohomology with local coefficients (aka twisted cohomology). Let us briefly describe the construction. The analogue of \eqref{cohomology_pullback} is a pullback map
\begin{equation}
    \tilde{\pi}^\ast: H^i(X^4;\mathbb{Z}^2_\rho) \to H^{i+1}(X^6;\mathbb{Z}) \, ,
\end{equation}
which shifts the degree by 1. This shift is necessary for the pairing with 1-cycles of $T^2$. Meanwhile, the analogue of \eqref{homology_pushforward} is a ``pushforward'' map
\begin{equation}
    \iota_\ast: H_0(X^4;H_1(T^2;\mathbb{Z})_\rho) = H_0(X^4;\mathbb{Z}^2_\rho) \to H_1(X^6;\mathbb{Z}) \, .
\end{equation}
The domain $H_0(X^4;H_1(T^2;\mathbb{Z})_\rho)$ can be understood as 1-cycles of $T^2$ that transform via Dehn twists as one goes around a non-contractible loop in the base $X^4$. Given any class $x \in H^6(X^6;\mathbb{Z})$ which factorises as $x = \tilde{\pi}^\ast(b) \cup f$ for some $b \in H^4(X^4;\mathbb{Z}^2_\rho)$, we define the twisted fiber integration as a map
\begin{equation}
    \tilde{\pi}_!: H^6(X^6;\mathbb{Z}) \to H^4(X^4;\mathbb{Z}^2_\rho) \, , \qquad x \mapsto b \cup \big(f \cap \iota_\ast([\mathcal{F}]_{\mathbb{Z}^2_\rho})\big) \, ,\label{twisted_fiber_integration_definition}
\end{equation}
where $[\mathcal{F}]_{\mathbb{Z}^2_\rho} \in H_0(X^4;\mathbb{Z}^2_\rho)$ is taken to be a sum of the independent generators of $H_0(X^4;\mathbb{Z}^2_\rho)$, up to a preferred choice of normalisation.\footnote{A more precise definition of $[\mathcal{F}]_{\mathbb{Z}^2_\rho}$ is provided in Appendix \ref{twisted_fiber_integration_appendix}. It is indeed not always guaranteed that such a class can be constructed and satisfies the assumptions therein, but we have checked that it is possible to do so in the examples considered in this paper.} Note that the cap product is taken with respect to the ordinary (co)homology of the total space $X^6$, so it simply outputs an element of $\mathbb{Z}$ without twists.

In light of the prescription above, we will employ the following ansatz for the cup product $[c_2]_{\mathbb{R}/\mathbb{Z}} \cup [\mathsf{G}_4]_\mathbb{Z} \in H^6(X^6;\mathbb{R}/\mathbb{Z})$ in \eqref{M5-brane_basepoint_anomaly},
\begin{equation}
    [c_2]_{\mathbb{R}/\mathbb{Z}} \cup [\mathsf{G}_4]_\mathbb{Z} \supset \frac{\vartheta}{2\pi} \, \tilde{\pi}^\ast\big([\mathfrak{b}_1]_{\mathbb{Z}^2_\rho} \cup [\mathcal{H}_3]_{\mathbb{Z}^2_\rho}\big) \cup [\mathfrak{s}_1]_\mathbb{Z} \, ,\label{c2G4_twisted_ansatz}
\end{equation}
such that $[\mathfrak{s}_1]_\mathbb{Z} \cap \iota_\ast([\mathcal{F}]_{\mathbb{Z}^2_\rho}) = 1$, and $\vartheta \in [0,2\pi)$. Here we have neglected terms that are $\text{SL}(2,\mathbb{Z})$-singlets, which were already addressed in the previous subsection. The next question is what the analogue for $[c_2]_{\mathbb{R}/\mathbb{Z}} \cup \frac{1}{2} [\Lambda_4]_\mathbb{Z}$ is, and in particular, whether there is some $[\mathcal{W}_3]_{\mathbb{Z}^2_\rho} \in H^3(X^4;\mathbb{Z}^2_\rho)$ playing the same role as the third integral Stiefel-Whitney class in the ordinary case. To be self-consistent, a suitable candidate must give rise to a basepoint anomaly for the D3-brane which agrees with that when the representation $\rho$ is trivial.

Taking inspiration from \cite{10.1215/kjm/1250517912,Greenblatt2006Homology}, we postulate a generalisation of the Stiefel-Whitney classes,
\begin{equation}
    \mathfrak{w}_i \in H^i(B\text{O}(n);(\mathbb{Z}_2 \oplus \mathbb{Z}_2)_\rho) \, ,
\end{equation}
defined as characteristic classes of the classifying space of $\text{O}(n)$ with local system $\rho$.\footnote{We assume the action of $\pi_1(X^4)$ on $\text{Aut}_\mathbb{Z}(\mathbb{Z}_2 \oplus \mathbb{Z}_2) = \text{GL}(2,\mathbb{Z}_2)$ is a mod 2 reduction of that on $\text{Aut}_\mathbb{Z}(\mathbb{Z} \oplus \mathbb{Z}) = \text{GL}(2,\mathbb{Z})$, so the local system is denoted as $\rho$ in both cases.} They can be viewed as obstructions to construct linearly independent sections of a real vector bundle (e.g.~the tangent bundle) in a local system of coefficients. The short exact sequence of $\mathbb{Z}[\pi_1]$-modules,
\begin{equation}
    0 \to (\mathbb{Z} \oplus \mathbb{Z})_\rho \xrightarrow{(\times 2,\times 2)} (\mathbb{Z} \oplus \mathbb{Z})_\rho \xrightarrow{(\text{mod} \, 2,\text{mod} \, 2)} (\mathbb{Z}_2 \oplus \mathbb{Z}_2)_\rho \to 0 \, ,
\end{equation}
induces a long exact sequence in cohomology with local coefficients,
\begin{equation}
    \cdots \to H^i(-;(\mathbb{Z} \oplus \mathbb{Z})_\rho) \to H^i(-;(\mathbb{Z}_2 \oplus \mathbb{Z}_2)_\rho) \xrightarrow{\beta} H^{i+1}(-;(\mathbb{Z} \oplus \mathbb{Z})_\rho) \to \cdots \, .\label{twisted_les}
\end{equation}
We can then define $\mathcal{W}_{i+1} = \beta(\mathfrak{w}_i) \in H^{i+1}(B\text{O}(n);(\mathbb{Z} \oplus \mathbb{Z})_\rho)$ using the Bockstein homomorphism above.

To motivate our proposal, consider the case where $\rho$ acts on $\mathbb{Z} \oplus \mathbb{Z}$ simply by a sign flip on both copies. Such a local system factorises as $(\mathbb{Z} \oplus \mathbb{Z})_\rho \cong \widetilde{\mathbb{Z}} \oplus \widetilde{\mathbb{Z}}$, where $\widetilde{\mathbb{Z}} \coloneqq \mathbb{Z}_{w_1}$ is the orientation module corresponding to a non-trivial first Stiefel-Whitney class $w_1$. Meanwhile, $\rho$ acts trivially on $\mathbb{Z}_2$, i.e.~$(\mathbb{Z}_2 \oplus \mathbb{Z}_2)_\rho \cong \mathbb{Z}_2 \oplus \mathbb{Z}_2$, so we recover a doublet $\mathfrak{w}_i = (w_i,w_i)$ with $w_i \in H^i(B\text{O}(n);\mathbb{Z}_2)$ being the standard $i$-th Stiefel-Whitney class. The cohomology ring $H^\ast(B\text{O}(n);\mathbb{Z}_{w_1})$ is indeed non-trivial \cite{10.1215/kjm/1250517912}, such that $\mathcal{W}_{i+1} = \beta(\mathfrak{w}_i)$ measures the obstruction to find twisted integral lifts of $\mathfrak{w}_i$.

We may also define ``twisted Wu classes'' via relations analogous to \eqref{Wu_classes_definition}, i.e.
\begin{equation}
    \mathfrak{v}_i \cup \mathfrak{a}_{n-i} = \mathfrak{Sq}^i(\mathfrak{a}_{n-i})\label{twisted_Wu_classes_definition}
\end{equation}
for any $\mathfrak{a}_{n-i} \in H^{n-i}(M^n;(\mathbb{Z}_2 \oplus \mathbb{Z}_2)_\rho)$, where $\mathfrak{Sq}^i: H^\ast(M^n;(\mathbb{Z}_2 \oplus \mathbb{Z}_2)_\rho) \to H^{\ast+i}(M^n;(\mathbb{Z}_2 \oplus \mathbb{Z}_2)_\rho)$ are regarded as cohomology operations with local coefficients \cite{1c7ecebe-252f-3c02-8b70-bd46894bf791} satisfying axioms analogous to those of the ordinary version. Furthermore, assuming they are related to the twisted Stiefel-Whitney classes by the Wu formula
\begin{equation}
    \mathfrak{w} = \mathfrak{Sq}(\mathfrak{v}) \, ,\label{twisted_Wu_formula}
\end{equation}
we can use the ansatz
\begin{equation}
    [c_2]_{\mathbb{R}/\mathbb{Z}} \cup \frac{1}{2} \, [\Lambda_4]_\mathbb{Z} \supset \frac{\vartheta}{2\pi} \, \tilde{\pi}^\ast\big([\mathfrak{b}_1]_{\mathbb{Z}^2_\rho} \cup [\mathcal{W}_3]_{\mathbb{Z}^2_\rho}\big) \cup [\mathfrak{s}_1]_\mathbb{Z}
\end{equation}
in combination with \eqref{c2G4_twisted_ansatz}. Finally, applying the twisted fiber integration \eqref{twisted_fiber_integration_definition} on \eqref{M5-brane_basepoint_anomaly} gives us the following basepoint anomaly for the D3-brane,
\begin{equation}
    \widetilde{\mathcal{A}} \supset -\kappa \, \frac{\vartheta}{2\pi} \int_{X^4} [\mathfrak{b}_1]_{\mathbb{Z}^2_\rho} \cup \Big([\mathcal{H}_3]_{\mathbb{Z}^2_\rho} + [\mathcal{W}_3]_{\mathbb{Z}^2_\rho}\Big) \, ,
\end{equation}
so the D3-brane partition function is non-vanishing only if
\begin{equation}
    [\mathcal{H}_3]_{\mathbb{Z}^2_\rho} + [\mathcal{W}_3]_{\mathbb{Z}^2_\rho} = 0 \, .\label{D3-brane_F-theory_FW_anomaly}
\end{equation}
As a sanity check, this does reduce to $[\mathsf{H}_3]_\mathbb{Z} + [\mathsf{W}_3]_\mathbb{Z} = [\mathsf{G}_3]_\mathbb{Z} + [\mathsf{W}_3]_\mathbb{Z} = 0$ when the $\text{SL}(2,\mathbb{Z})$-action is trivial.

\subsection{D3-brane on S-folds}

\label{sec:D3s-on-S-folds}

As a concrete application, we would like to study the behavior of the D3-brane partition function in a class of non-trivial F-theory backgrounds known as {\it S-folds}. These are generalisations of orientifolds in Type IIB string theory \cite{Witten:1998xy}. The latter are holographically dual to 4d $\mathcal{N}=4$ SCFTs, while the former are dual to 4d $\mathcal{N}=3$ SCFTs \cite{Garcia-Etxebarria:2015wns,Aharony:2016kai}.\footnote{See also \cite{Apruzzi:2020pmv,Giacomelli:2020gee} for a construction of $\mathcal{N}=2$ S-folds.}

The 10d background geometry of an S-fold is $\text{AdS}_5 \times S^5/\mathbb{Z}_k$, over which there is a non-trivial $\text{SL}(2,\mathbb{Z})$ bundle acting on the doublet $(B_2,C_2)$, characterised by
\begin{equation}
    \rho: \pi_1(S^5/\mathbb{Z}_k) \to \text{SL}(2,\mathbb{Z}) \, , \qquad \begin{pmatrix} B_2 \\ C_2 \end{pmatrix} \xmapsto{\rho = \begin{pmatrix} a & b \\ c & d \end{pmatrix}} \begin{pmatrix} a B_2 + b C_2 \\ c B_2 + d C_2 \end{pmatrix} \, ,
\end{equation}
as one goes around a non-contractible loop in $S^5/\mathbb{Z}_k$. S-folds arise precisely from the non-trivial finite subgroups, namely, $\mathbb{Z}_2, \mathbb{Z}_3, \mathbb{Z}_4, \mathbb{Z}_6$, of $\text{SL}(2,\mathbb{Z})$, with the matrix $\rho_k$ given by
\begin{equation}
    \rho_2 = \begin{pmatrix} -1 & 0 \\ 0 & -1 \end{pmatrix} \, , \quad \rho_3 = \begin{pmatrix} -1 & -1 \\ 1 & 0 \end{pmatrix} \, , \quad \rho_4 = \begin{pmatrix} 0 & -1 \\ 1 & 0 \end{pmatrix} \, , \quad \rho_6 = \begin{pmatrix} 0 & -1 \\ 1 & 1 \end{pmatrix} \, .\label{rho_matrices}
\end{equation}
The $k=2$ case corresponds to the orientifold, where $(B_2,C_2)$ acquires a sign flip but the components do not mix. Note that all the matrices in \eqref{rho_matrices} can be regarded as elements of $\text{SL}(2,\mathbb{Z}_2)$ as well.

When restricted to $S^5/\mathbb{Z}_k$, the flux $\mathcal{H}_3 = (H_3,G_3)$ is classified by $H^3(S^5/\mathbb{Z}_k;(\mathbb{Z} \oplus \mathbb{Z})_{\rho_k})$. Let us briefly review how the twisted cohomology groups of lens spaces $S^{2n+1}/\mathbb{Z}_k$, with $n \geq 1$, can be computed. For $A$ some $\mathbb{Z}[\pi_1]$-module where $\pi_1(S^{2n+1}/\mathbb{Z}_k) = \mathbb{Z}_k$, we can construct a cochain complex
\begin{equation}
    0 \to C^0 \xrightarrow{1-t} C^1 \xrightarrow{1+t+\cdots+t^{k-1}} C^2 \xrightarrow{1-t} \cdots \xrightarrow{1+t+\cdots+t^{k-1}} C^{2n} \xrightarrow{1-t} C^{2n+1}\to 0 \, ,
\end{equation}
where $C^i \cong A$ and $t$ is a generator of $\mathbb{Z}_k$ \cite{Davis2001LectureNI,MR1867354,Aharony:2016kai,Heckman:2022muc,Etheredge:2023ler}. By construction, $1+t+\cdots+t^{k-1}=0$, and $d \circ d = (1-t)(1+t+\cdots+t^{k-1}) = 1-t^k = 0$, i.e.~the differential is nilpotent, as desired.

Particularly, when $A = \mathbb{Z} \oplus \mathbb{Z}$, it follows that
\begin{equation}
    \begin{aligned}
        H^\text{even}(S^{2n+1}/\mathbb{Z}_k,(\mathbb{Z} \oplus \mathbb{Z})_{\rho_k}) & = 0 \, ,\\[1ex]
        H^\text{odd}(S^{2n+1}/\mathbb{Z}_k,(\mathbb{Z} \oplus \mathbb{Z})_{\rho_k}) & = \begin{cases} \mathbb{Z}_2 \oplus \mathbb{Z}_2 & k=2 \, ,\\ \mathbb{Z}_3 & k=3 \, ,\\ \mathbb{Z}_2 & k=4 \, ,\\ 0 & k=6 \, . \end{cases}
    \end{aligned}
\end{equation}
On the other hand, when $A = \mathbb{Z}_2 \oplus \mathbb{Z}_2$, we find that for all $0 \leq i \leq 2n+1$,
\begin{equation}
    \begin{aligned}
        H^i(S^{2n+1}/\mathbb{Z}_k,(\mathbb{Z}_2 \oplus \mathbb{Z}_2)_{\rho_k}) & = \begin{cases} \mathbb{Z}_2 \oplus \mathbb{Z}_2 & k=2 \, ,\\ 0 & k=3 \, ,\\ \mathbb{Z}_2 & k=4 \, ,\\ 0 & k=6 \, , \end{cases}
    \end{aligned}
\end{equation}
or more generally, when $A = \mathbb{Z}_m \oplus \mathbb{Z}_m$ for any $m \in \mathbb{Z}^+$,
\begin{equation}
    \begin{aligned}
        H^i(S^{2n+1}/\mathbb{Z}_k,(\mathbb{Z}_m \oplus \mathbb{Z}_m)_{\rho_k}) & = \begin{cases} \mathbb{Z}_{\gcd(m,2)} \oplus \mathbb{Z}_{\gcd(m,2)} & k=2 \, ,\\ \mathbb{Z}_{\gcd(m,3)} & k=3 \, ,\\ \mathbb{Z}_{\gcd(m,2)} & k=4 \, ,\\ 0 & k=6 \, . \end{cases}
    \end{aligned}
\end{equation}
For comparison, the ordinary cohomology groups of lens spaces are given by
\begin{equation}
    \begin{aligned}
        H^i(S^{2n+1}/\mathbb{Z}_k;\mathbb{Z}) & = \begin{cases} \mathbb{Z} & i=0, 2n+1 \, ,\\ 0 & i = 2j+1 < 2n+1 \, ,\\ \mathbb{Z}_k & i = 2j > 0 \, , \end{cases}\\[1ex]
        H^i(S^{2n+1}/\mathbb{Z}_k;\mathbb{Z}_m) & = \begin{cases} \mathbb{Z}_m & i = 0, 2n+1 \, ,\\ \mathbb{Z}_{\gcd(m,k)} & \text{otherwise} \, , \end{cases}
    \end{aligned}
\end{equation}
where the latter can be derived using the universal coefficient theorem.

\subsubsection{Discrete torsion and twisted Stiefel-Whitney classes}

Apart from the integer $k$, an S-fold is also characterised by a choice of {\it discrete torsion} $\Theta = (\Theta_\text{NS},\Theta_\text{RR}) \in H^3(S^5/\mathbb{Z}_k;(\mathbb{Z} \oplus \mathbb{Z})_{\rho_k})$ \cite{Aharony:2016kai}, which determines the cohomology class of $\mathcal{H}_3 = (H_3,G_3)$ in the supergravity background. For example, when $k=2$, the gauge group of the 4d SCFT dual to the supergravity background is $\text{Sp}(M)$ or $\text{SO}(M)$ for some $M \in \mathbb{Z}^+$, depending on whether $\Theta_\text{NS}$ is non-trivial. When $\Theta_\text{NS}$ is trivial, the rank of $\text{SO}(M)$ depends on whether $\Theta_\text{RR}$ is non-trivial \cite{Witten:1998xy}. Suppose we wrap a D3-brane on, say, $S^1 \times S^3/\mathbb{Z}_k \subset \text{AdS}_5 \times S^5/\mathbb{Z}_k$, our question is whether its partition function is vanishing or not in the presence of a non-trivial discrete torsion. Equivalently, we know from \eqref{D3-brane_F-theory_FW_anomaly} that we need the 4-manifold $S^1 \times S^3/\mathbb{Z}_k$ to have a non-trivial $\mathcal{W}_3$ that cancels $\mathcal{H}_3$.\footnote{For the twisted fiber integration \eqref{twisted_fiber_integration_definition} to work, we require $E_\infty^{5,0}=0$ and and $H_0(X^4;\mathbb{Z}^2_\rho)$ to be non-trivial, as explained in Appendix \ref{twisted_fiber_integration_appendix}. The former is always satisfied because of degree reasons, while the latter indeed holds for our lens spaces, except for the trivial case of $k=6$.}

For $k=6$, since $H^3(S^3/\mathbb{Z}_6,(\mathbb{Z} \oplus \mathbb{Z})_{\rho_6})=0$, i.e.~there cannot be any non-trivial discrete torsion, the question is redundant. For $k=3$, the fact that $H^2(S^3/\mathbb{Z}_3,(\mathbb{Z}_2 \oplus \mathbb{Z}_2)_{\rho_3})=0$ implies $\mathcal{W}_3 = \beta(\mathfrak{w}_2)$ is necessarily zero, so the D3-brane partition function must be vanishing when the discrete torsion $\Theta \in \mathbb{Z}_3$ is non-trivial.

For $k=2$, we observe that $\rho_2$ corresponds to the identity matrix in $\text{SL}(2,\mathbb{Z}_2)$, because a sign flip mod 2 is not meaningful. The local system is then trivial when considering $\mathbb{Z}_2 \oplus \mathbb{Z}_2$ coefficients. In this case, $\mathfrak{w}_2 = (w_2,w_2) \in H^2(S^3/\mathbb{Z}_2;\mathbb{Z}_2 \oplus \mathbb{Z}_2)$ is simply two identical copies of the second Stiefel-Whitney class. Provided that $S^3/\mathbb{Z}_2 \cong \mathbb{RP}^3$ is spin, and so $\mathcal{W}_3 = \beta(\mathfrak{w}_2) = 0$, we again conclude the D3-brane partition function must be vanishing when the discrete torsion $\Theta \in \mathbb{Z}_2 \oplus \mathbb{Z}_2$ is non-trivial, which matches with the analysis of the orientifold in \cite{Witten:1998xy}.

The only case left to consider is when $k=4$. With some minimal assumptions, we will argue that $\mathfrak{w}_2=0$. Recall that, when $k$ is even, the ordinary cohomology groups $H^i(S^{2n+1}/\mathbb{Z}_k;\mathbb{Z}_2)$ is generated either by a single generator $a_1$ with $|a_1|=1$, or by a pair of generators $(a_1,b_2)$ with $|b_2|=2$. See Appendix \ref{Lens_spaces_SW_classes_appendix} for a review of how the (ordinary) Stiefel-Whitney classes of lens spaces can be computed. We assume the same for the twisted cohomology groups $H^i(S^{2n+1}/\mathbb{Z}_4;(\mathbb{Z}_2 \oplus \mathbb{Z}_2)_{\rho_4})$.

Suppose $H^i(S^{2n+1}/\mathbb{Z}_4;(\mathbb{Z}_2 \oplus \mathbb{Z}_2)_{\rho_4}) = \mathbb{Z}_2$ is generated only by $\mathfrak{a}_1$, i.e.~its non-trivial element is $\mathfrak{a}_1^i$. Using \eqref{twisted_Wu_classes_definition}, we have
\begin{equation}
    \mathfrak{v}_i \cup \mathfrak{a}_1^{2n+1-i} = \mathfrak{Sq}^i(\mathfrak{a}_1^{2n+1-i}) = \begin{pmatrix} 2n+1-i \\ i \end{pmatrix} \mathfrak{a}_1^{2n+1-i} \, ,
\end{equation}
where the second equality follows from the Cartan formula, and specifically,
\begin{equation}
    \mathfrak{v}_1 \cup \mathfrak{a}_1^{2n} = 2n \mathfrak{a}_1^{2n} = 0 \mod 2 \, .
\end{equation}
This implies $\mathfrak{w}_1 = \mathfrak{v}_1 = 0$ by \eqref{twisted_Wu_formula}. Similarly,
\begin{equation}
    \mathfrak{v}_2 \cup \mathfrak{a}_1^{2n-1} = (n-1)(2n-1) \mathfrak{a}_1^{2n+1} \, ,
\end{equation}
so $\mathfrak{w}_2 = \mathfrak{v}_2 - \mathfrak{w}_1^2 = \mathfrak{v}_2$ is vanishing if and only if $n$ is odd.

Another possible scenario is that $H^i(S^{2n+1}/\mathbb{Z}_4;(\mathbb{Z}_2 \oplus \mathbb{Z}_2)_{\rho_4})$ is generated by $\mathfrak{a}_1$ and $\mathfrak{b}_2$ with $\mathfrak{a}_1^2=0$, so the even-degree elements are $\mathfrak{b}_2^j$ and the odd-degree elements are $\mathfrak{a}_1 \cup \mathfrak{b}_2^j$ for some $j$. Consider the relation,
\begin{equation}
    \mathfrak{v}_1 \cup \mathfrak{b}_2^n = \mathfrak{Sq}^1(\mathfrak{b}_2^n) = (\text{mod} \ 2) \circ \beta(\mathfrak{b}_2^n) \, .
\end{equation}
Note that $H^{2n}(S^{2n+1}/\mathbb{Z}_4;(\mathbb{Z} \oplus \mathbb{Z})_{\rho_4}) = 0$ means $\mathfrak{b}_2^n$ cannot be a mod 2 reduction of some integral lift, the exactness of \eqref{twisted_les} then asserts $\beta(\mathfrak{b}_2^n)$ must be non-trivial. Since $H^{2n+1}(S^{2n+1}/\mathbb{Z}_4;(\mathbb{Z} \oplus \mathbb{Z})_{\rho_4}) = \mathbb{Z}_2$, we can simply take $\beta(\mathfrak{b}_2^n)$ to be odd, thus $(\text{mod} \ 2) \circ \beta(\mathfrak{b}_2^n) = \mathfrak{a}_1 \cup \mathfrak{b}_2^n$. In other words, $\mathfrak{w}_1 = \mathfrak{v}_1 = \mathfrak{a}_1$ is always non-vanishing. We now proceed to evaluate
\begin{align}
    \mathfrak{v}_2 \cup \mathfrak{a}_1 \cup \mathfrak{b}_2^{n-1} & = \mathfrak{Sq}^2(\mathfrak{a}_1 \cup \mathfrak{b}_2^{n-1})\nonumber\\
    & = \mathfrak{Sq}^1(\mathfrak{a}_1) \cup \mathfrak{Sq}^1(\mathfrak{b}_2^{n-1}) + \mathfrak{a}_1 \cup \mathfrak{Sq}^2(\mathfrak{b}_2^{n-1})\nonumber\\
    & = \mathfrak{a}_1^2 \cup (\mathfrak{a}_1 \cup \mathfrak{b}_2^{n-1}) + \mathfrak{a}_1 \cup \begin{pmatrix} n-1 \\ 1 \end{pmatrix} \mathfrak{b}_2^n + \mathfrak{a}_1 \cup \begin{pmatrix} n-1 \\ 2 \end{pmatrix} \mathfrak{a}_1^2 \cup \mathfrak{b}_2^{n-1}\nonumber\\
    & = \bigg(\bigg(\frac{(n-1)(n-2)}{2} + 1\bigg) \mathfrak{a}_1^2 + (n-1) \mathfrak{b}_2\bigg) \cup \mathfrak{a}_1 \cup \mathfrak{b}_2^{n-1} \, ,
\end{align}
which gives
\begin{equation}
    \mathfrak{w}_2 = \mathfrak{v}_2 - \mathfrak{w}_1^2 = \frac{(n-1)(n-2)}{2} \, \mathfrak{a}_1^2 + (n-1) \mathfrak{b}_2 \, .
\end{equation}
For $\mathfrak{w}_2$ to vanish, we need $(n-1)/2$ to be an even integer.

To summarise, we see that when $n=1$, without needing to explicitly determine whether $H^\ast(S^3/\mathbb{Z}_4;(\mathbb{Z}_2 \oplus \mathbb{Z}_2)_{\rho_4})$ is generated by one or two generators, the second twisted Stiefel-Whitney class $\mathfrak{w}_2$ necessarily vanishes. Consequently, $\mathcal{W}_3 = \beta(\mathfrak{w}_2) = 0$, and so the D3-brane partition function must be vanishing when the discrete torsion $\Theta \in \mathbb{Z}_2$ is non-trivial.

\subsubsection{Non-Abelian corrections}

In the presence of discrete torsion, although we cannot wrap a single D3-brane on $S^1 \times S^3/\mathbb{Z}_k$ without trivialising its partition function, one can ask whether the vanishing can be circumvented by wrapping multiple coincident D3-branes instead.\footnote{Technically, this goes beyond the realm of our assumption where the original M5-brane anomaly theory is invertible, but we expect our arguments to still hold in general.} Such a possibility was demonstrated in Section \ref{coincident_branes_section} in the context of a trivial F-theory fibration, by allowing the Chan-Paton bundle of the stack of D3-branes to admit a non-trivial gauge group $\text{SU}(N) \times_{\mathbb{Z}_m} \U(1)$, characterised by the flat background connections $\zeta_2^B,\zeta_2^C \in H^2(X^4;\mathbb{Z}_m)$.\footnote{We have replaced $k$ with $m$ here to avoid clashing of notation.} This construction can be readily generalised to the case of S-folds as follows.

Similarly to \eqref{twisted_les}, the short exact sequence of $\mathbb{Z}[\pi_1]$-modules,
\begin{equation}
    0 \to (\mathbb{Z} \oplus \mathbb{Z})_\rho \xrightarrow{(\times 2,\times 2)} (\mathbb{Z} \oplus \mathbb{Z})_\rho \xrightarrow{(\text{mod} \, m,\text{mod} \, m)} (\mathbb{Z}_m \oplus \mathbb{Z}_m)_\rho \to 0 \, ,
\end{equation}
induces a long exact sequence in cohomology with local coefficients,
\begin{equation}
    \cdots \to H^i(-;(\mathbb{Z} \oplus \mathbb{Z})_\rho) \to H^i(-;(\mathbb{Z}_m \oplus \mathbb{Z}_m)_\rho) \xrightarrow{\beta} H^{i+1}(-;(\mathbb{Z} \oplus \mathbb{Z})_\rho) \to \cdots \, .
\end{equation}
Since $H^2(S^3/\mathbb{Z}_k;(\mathbb{Z} \oplus \mathbb{Z})_{\rho_k})=0$, any non-trivial element $\xi_2 = (\zeta_2^B,\zeta_2^C) \in H^2(S^3/\mathbb{Z}_k;(\mathbb{Z}_2 \oplus \mathbb{Z}_2)_{\rho_k})$ cannot be a mod $m$ reduction, which implies $\beta(\xi_2) \neq 0$ by exactness. The corresponding stack of D3-branes then has a non-vanishing partition function only if
\begin{equation}
    [\mathcal{H}_3] + \beta([\xi_2]) = 0 \, ,
\end{equation}
where we implicitly used the previous result that $[\mathcal{W}_3]=0$.

For $k=2$, we have $H^2(S^3/\mathbb{Z}_2;(\mathbb{Z}_m \oplus \mathbb{Z}_m)_{\rho_2}) = \mathbb{Z}_{\gcd(m,2)} \oplus \mathbb{Z}_{\gcd(m,2)}$, so there exists some non-trivial $\xi_2$ as long as $m$ is even. This means that we can take a stack of $N$ coincident D3-branes with $N$ even, and pick the gauge group of the Chan-Paton bundle to be $\text{SU}(N) \times_{\mathbb{Z}_m} \U(1)$ for some even divisor $m$ of $N$. More explicitly, suppose the discrete torsion is $\Theta = (1,0) \in \mathbb{Z}_2 \oplus \mathbb{Z}_2$, then we couple the dynamical Chan-Paton gauge field to a background connection $\zeta_2^B \in H^2(X^4;\widetilde{\mathbb{Z}}_m) = \mathbb{Z}_2$ for some even $m$. Upon going around a non-contractible loop in $X^4$, its characteristic class $\beta(\zeta_2^B) \in H^3(X^4;\widetilde{\mathbb{Z}})$ acquires a sign flip, which precisely counteracts the effect of the pullback of a non-trivial $\mathsf{H}_3$ from the supergravity background, such that the overall partition function is non-vanishing. Similar remarks apply when $\Theta = (0,1)$, in which case we simply replace $\zeta_2^B$ with $\zeta_2^C$, or more generally, we need both when $\Theta = (1,1)$.

For $k=3$, we have $H^2(S^3/\mathbb{Z}_3;(\mathbb{Z}_m \oplus \mathbb{Z}_m)_{\rho_3}) = \mathbb{Z}_{\gcd(m,3)}$. Accordingly, we can take a stack of D3-branes with any $N \in 3\mathbb{Z}$ and $m \in 3\mathbb{Z}$ some divisor of $N$, such that the characteristic class $\beta(\xi_2)$ cancels the effect of the pullback of $\mathcal{H}_3$. There are three inequivalent non-Abelian Chan-Paton structures (one being trivial), determined by $\xi_2$, that are in one-to-one correspondence with the discrete torsion $\Theta \in \mathbb{Z}_3$.

By the same token, for $k=4$, we have $H^2(S^3/\mathbb{Z}_4;(\mathbb{Z}_m \oplus \mathbb{Z}_m)_{\rho_4}) = \mathbb{Z}_{\gcd(m,2)}$, so we can take any $N \in 2\mathbb{Z}$ and $m \in 2\mathbb{Z}$ a divisor of $N$ to match the discrete torsion $\Theta \in \mathbb{Z}_2$. This generalises the specific $N=4$ example studied in \cite{Aharony:2016kai} to an infinite family of candidates. The trivial case of $k=6$ is uninteresting as usual. All in all, from the perspective of obstruction theory, we have constructed non-Abelian D3-brane configurations on S-folds which have non-vanishing partition functions in the presence of discrete torsion.

\subsection{Class $\mathcal{S}$ theories}

We briefly outline a generalisation that is applicable to the compactification of the M5-brane worldvolume theory, viewed as a 6d $(2,0)$ SCFT, over a generic Riemann surface $\Sigma^2$, which results in what are referred to as 4d {\it Class $\mathcal{S}$ theories} \cite{Gaiotto:2009hg} (see also \cite{Lawrie:2018jut}). These constructions typically preserve $\mathcal{N}=2$ supersymmetry, and often admit no known Lagrangian descriptions \cite{Moore:2012yp,Tachikawa:2013kta,Akhond:2021xio}. For the sake of illustration, let us assume below that the Riemann surface is compact, orientable, and has no punctures, so $\Sigma^2$ is characterised only by its genus $g$.

Suppose the fibration is trivial, i.e.~$X^6 = X^4 \times \Sigma^2$, then there are $2g$ 2-form Kaluza-Klein zero modes associated with 1-cycles of $\Sigma^2$. For example, the M-theory 4-form flux decomposes similarly to \eqref{G4_ansatz} as
\begin{equation}
    [\mathsf{G}_4]_\mathbb{Z} = [\mathsf{E}_4]_\mathbb{Z} + \sum_{i=1}^g \Big([\mathsf{H}_{3,i}]_\mathbb{Z} \cup [\mathsf{s}_{1,i}]_\mathbb{Z} + [\mathsf{G}_{3,i}]_\mathbb{Z} \cup [\mathsf{s}_{1,i}']_\mathbb{Z}\Big) + [\mathsf{K}_2]_\mathbb{Z} \cup [\omega_2]_\mathbb{Z} \, ,
\end{equation}
with the intersection pairing $([\mathsf{s}_{1,i}]_\mathbb{Z},[\mathsf{s}_{1,j}']_\mathbb{Z}) = \delta_{ij}$. A parallel computation then tells us that the non-vanishing conditions for the partition function is essentially the same as \eqref{D3-brane_FW_anomaly}, promoting $[\mathsf{H}_3]_\mathbb{Z} \to [\mathsf{H}_{3,i}]_\mathbb{Z}$ and $[\mathsf{G}_3]_\mathbb{Z} \to [\mathsf{G}_{3,i}]_\mathbb{Z}$ for $i=1,\dots,g$.

More generally, the fibration $\Sigma^2 \xrightarrow{\iota} X^6 \xrightarrow{\pi} X^4$ can be non-trivial, and $\pi_1(X^4)$ can act on $H^1(\Sigma^2;\mathbb{Z})=\mathbb{Z}^{2g}$. A local system is, broadly speaking, specified by a representation $\rho: \pi_1(X^4) \to \text{Aut}_\mathbb{Z}(\mathbb{Z}^{2g}) = \text{GL}(2g,\mathbb{Z})$. On the other hand, the (oriented) mapping class group of $\Sigma^2$ is given by the extension,
\begin{equation}
    0 \to \text{T}(\Sigma^2) \to \text{MCG}(\Sigma^2) \to \text{Sp}(2g,\mathbb{Z}) \to 0 \, ,
\end{equation}
where $\text{Sp}(2g,\mathbb{Z})$ is defined to be the intersection $\text{Sp}(2g,\mathbb{R}) \cap \text{GL}(2g,\mathbb{Z})$, while the Torelli group $\text{T}(\Sigma^2)$ denotes the group that acts trivially on $H_1(\Sigma^2;\mathbb{Z})$. When $g=1$, we recover $\text{MCG}(T^2) = \text{Sp}(2,\mathbb{Z}) \cong \text{SL}(2,\mathbb{Z})$. Therefore, we shall focus on 4d theories where the 3-form fluxes $\mathcal{H}_3 = \{H_{3,i},G_{3,i}\}$ form a multiplet under an $\text{Sp}(2g,\mathbb{Z})$-action described by
\begin{equation}
    \rho: \pi_1(X^4) \to \text{Sp}(2g,\mathbb{Z}) \, ,
\end{equation}
such that it preserves the symplectic structure of the Riemann surface.

One can check that our construction of the twisted fiber integration in Appendix \ref{twisted_fiber_integration_appendix} can be readily modified to account for fibrations where the fiber is $\Sigma^2$. If we further assume the existence of the twisted Stiefel-Whitney classes
\begin{equation}
    \mathfrak{w}_i \in H^i(B\text{O}(n);(\mathbb{Z}_2^{2g})_\rho)
\end{equation}
and twisted Wu classes $\mathfrak{v}_i$ which satisfy axioms analogous to the ordinary version, then we conjecture that the necessary condition for such a Class $\mathcal{S}$ theory to have a non-vanishing partition function is
\begin{equation}
    [\mathcal{H}_3] + [\mathcal{W}_3] = 0 \, ,
\end{equation}
where $[\mathcal{H}_3],[\mathcal{W}_3] \in H^3(X^4;\mathbb{Z}^{2g}_\rho)$ are understood to live in cohomology with local coefficients.


\acknowledgments

We thank T.~Daniel Brennan, Victor Carmona, Dan Freed, Andrea Grigoletto, Patrick Jefferson, Nitu Kitchloo, Ho Tat Lam, Daniel Roggenkamp, Raffaele Savelli, Sebastian Schulz, and Thomas Waddleton for useful discussion and correspondence. EL is grateful to the Department of Mathematical Sciences at Durham University for hospitality during a visit contributing to this work. The work of IGE is supported by the ``Global Categorical Symmetries'' Simons Collaboration grant (award number 888990) and the STFC ST/X000591/1 consolidated grant. 


\appendix

\addtocontents{toc}{\protect\setcounter{tocdepth}{1}}


\section{Leray-Serre spectral sequence}\label{Serre_spectral_sequence_appendix}

Consider a Serre fibration $F \xrightarrow{\iota} X \xrightarrow{\pi} B$ where $\pi_1(B)$ acts trivially on $H^\ast(F;M)$, with $M$ an $R$-module and $R$ a commutative ring. The second page $E_2$ of the Leray-Serre spectral sequence has entries
\begin{equation}
    E_2^{p,q} = H^p(B;H^q(F;M)) \, .
\end{equation}
For each entry on the $r$-th page, there is a differential defined as a homomorphism
\begin{equation}
    d_r^{p,q}: E_r^{p,q} \to E_r^{p+r,q-r+1} \, ,
\end{equation}
such that the entries on the $(r+1)$-th page are defined as
\begin{equation}
    E_{r+1}^{p,q} = \frac{\text{ker}(d_r^{p,q})}{\text{im}(d_r^{p-r,q+r-1})} \, .
\end{equation}
The entries eventually stabilise to some $E_\infty^{p,q}$, and the associated graded group of $H^n(X;M)$ is given by
\begin{equation}
    \text{Gr} H^n(X;M) = \bigoplus_p E_\infty^{p,n-p} = \bigoplus_p \frac{F^n_p}{F^n_{p+1}} \, .
\end{equation}
In other words, we have the following filtration,
\begin{equation}
    \begin{tikzcd}[row sep=scriptsize,column sep=small]
        E_\infty^{n,0} \arrow[r,phantom,"="] & F^n_n \arrow[r,hookrightarrow] & F^n_{n-1} \arrow[d,twoheadrightarrow] \arrow[r,hookrightarrow] & F^n_{n-2} \arrow[d,twoheadrightarrow] \arrow[r,hookrightarrow] & \cdots \arrow[r,hookrightarrow] & F^n_0  \arrow[d,twoheadrightarrow] \arrow[r,phantom,"="] & H^n(X;M) \, .\\
        & & E_\infty^{n-1,1} & E_\infty^{n-2,2} & & E_\infty^{0,n}
    \end{tikzcd}\label{Serre_spectral_sequence_filtration}
\end{equation}

For simplicity, we will hereafter assume that $M$ is an Abelian group $G$, i.e.~it is a $\mathbb{Z}$-module. Suppose the fiber $F$ is path-connected, then $H^0(F;G) \cong G$, so the bottom row of the second page has entries
\begin{equation}
    E_2^{p,0} = H^p(B;G) \, .
\end{equation}
Since the bottom row necessarily has trivial outgoing differentials on all pages, i.e.~$E_{r+1}^{p,0} = \text{coker}(d_r^{p-r,r-1})$, we have a sequence of surjections, $E_2^{p,0} \twoheadrightarrow E_3^{p,0} \twoheadrightarrow \cdots \twoheadrightarrow E_\infty^{p,0}$. Composing the maps yields the {\it horizontal edge homomorphism},
\begin{equation}
    H^p(B;G) = E_2^{p,0} \xrightarrowdbl{\pi^\ast} E_\infty^{p,0} \subseteq H^p(X;G) \, ,
\end{equation}
which can be identified as the pullback of the projection $\pi: X \to B$. Similarly, if the base $B$ is path-connected, then the left column of the second page has entries
\begin{equation}
    E_2^{0,q} = H^q(F;G) \, .
\end{equation}
On all pages, the left column has trivial incoming differentials, i.e.~$E_{r+1}^{0,q} = \text{ker}(d_r^{0,q})$, and so we have a sequence of inclusions, $E_\infty^{0,q} \subseteq \cdots \subseteq E_3^{0,q} \subseteq E_2^{0,q}$. Composing the maps yields the {\it vertical edge homomorphism},
\begin{equation}
    H^q(X;G) \xrightarrowdbl{\iota^\ast} E_\infty^{0,q} \subseteq E_2^{0,q} = H^q(F;G) \, ,
\end{equation}
which can be identified as the pullback of the inclusion $F \xrightarrow{\iota} X$.

There is a dual spectral sequence in homology with
\begin{equation}
    E^2_{p,q} = H_p(B;H_q(F;G)) \Rightarrow H_{p+q}(X;G) \, ,
\end{equation}
and differentials defined in the opposite direction as homomorphisms
\begin{equation}
    d^r_{p,q}: E^r_{p,q} \to E^r_{p-r,q+r-1} \, .
\end{equation}
Consequently, we have pushforward maps given respectively by
\begin{equation}
    H_p(X;G) \xrightarrowdbl{\pi_\ast} E^\infty_{p,0} \subseteq E^2_{p,0} = H_p(B;G) \, ,
\end{equation}
and also
\begin{equation}
    H_q(F;G) = E^2_{0,q} \xrightarrowdbl{\iota_\ast} E^\infty_{0,q} \subseteq H_q(X;G) \, .
\end{equation}

We may make use of the maps constructed above to construct the fiber integration of a cohomology class $x \in H^{p+q}(X;G)$. Suppose it factorises as
\begin{equation}
    x = f \cup \pi^\ast(b)\label{total_space_cocycle_factorization}
\end{equation}
for some $b \in H^p(B;G)$ and $f \in \text{coker}(\pi^\ast) \subset H^q(X;G)$, then we define a ``wrong-way'' homomorphism (also known as the {\it umkehr map})
\begin{equation}
    \pi_!: H^{p+q}(X;G) \to H^{q-\text{dim}(F)}(X;G) \times H^p(B;G) \, , \qquad x \mapsto (f \cap \iota_\ast([F])) \cup b \, ,\label{wrong-way_homomorphism}
\end{equation}
where $[F] \in H_{\text{dim}(F)}(F;G)$ is the fundamental class of the fiber (assuming it is closed), and $\cap: H^q(X;G) \times H_{\text{dim}(F)}(X;G) \to H^{q-\text{dim}(F)}(X;G)$ is the cap product on the total space.\footnote{The prescription works similarly if we have a more general decomposition $x = \sum_i f_i \cup \pi^\ast(b_i)$.} Naturality of the cap product implies that
\begin{equation}
    \iota^\ast(f \cap \iota_\ast([F])) = \iota^\ast(f) \cap [F] \, ,
\end{equation}
where the cap product on the RHS is understood to be that on the fiber. For $q=\text{dim}(F)$, we always have
\begin{equation}
    H^0(X;G) = E_\infty^{0,0} = E_2^{0,0} = H^0(F;G) = H^0(B;G) = G \, ,
\end{equation}
so $(f \cap \iota_\ast([F])) \cup b$ can indeed be regarded as the cup product
\begin{equation}
    \cup: H^0(B;G) \times H^p(B;G) \to H^p(B;G) \, .
\end{equation}
Hence, this gives us a notion of fiber integration
\begin{equation}
    \int_F: H^{p+\text{dim}(F)}(X;G) \to H^p(B;G) \, .
\end{equation}
A limitation of this approach is the assumption of the factorisation \eqref{total_space_cocycle_factorization}, which does not a priori exist for an arbitrary $x \in H^{p+q}(X;G)$.

More formally, such a wrong-way map can be defined analogously to before using an edge homomorphism of the spectral sequence, without assuming the aforementioned factorisation. Note that all the entries $E_r^{p,q}$ with $q>\text{dim}(F)$ are identically zero on all the pages due to degree reasons, so for $q=\text{dim}(F)$ all the incoming differentials are trivial, i.e.~$E_{r+1}^{p,\text{dim}(F)} = \text{ker}(d_r^{p,\text{dim}(F)})$, which gives rise to a sequence of inclusions, $E_\infty^{p,\text{dim}(F)} \subseteq \cdots \subseteq E_3^{p,\text{dim}(F)} \subseteq E_2^{p,\text{dim}(F)}$. Together with the fact that $E_\infty^{p,q>\text{dim}(F)}=0$ for all $p$, i.e.
\begin{equation}
    \begin{tikzcd}[row sep=scriptsize,column sep=small]
        \cdots \arrow[r,hookrightarrow] & F^{p+\text{dim}(F)}_{p+1} \arrow[d,twoheadrightarrow] \arrow[r,hookrightarrow] & F^{p+\text{dim}(F)}_p \arrow[d,twoheadrightarrow] \arrow[r,phantom,"="] & \cdots \arrow[r,phantom,"="] & F^{p+\text{dim}(F)}_0 \arrow[r,phantom,"="] & H^{p+\text{dim}(F)}(X;G) \, ,\\
        & E_\infty^{p+1,\text{dim}(F)-1} & E_\infty^{p,\text{dim}(F)}
    \end{tikzcd}\label{top_edge_homomorphism_filtration}
\end{equation}
we can compose the previous inclusion maps to obtain the {\it top edge homomorphism},
\begin{equation}
    H^{p+\text{dim}(F)}(X;G) = F^{p+\text{dim}(F)}_p \xrightarrowdbl{\pi_!} E_\infty^{p,\text{dim}(F)} \subseteq E_2^{p,\text{dim}(F)} = H^p(B;G) \, ,
\end{equation}
where we used the assumption that the fiber $F$ is closed, orientable, and path-connected to conclude $E_2^{p,\text{dim}(F)} = H^p(B;H^{\text{dim}(F)}(F;G)) = H^p(B;G)$. If $F$ is non-orientable instead, then one should replace the coefficients in $H^p(B;G)$ accordingly.


\section{Twisted fiber integration}\label{twisted_fiber_integration_appendix}

If the fundamental group $\pi_1(B)$ acts non-trivially on $H^\ast(F;G)$, the corresponding Leray-Serre spectral sequence becomes
\begin{equation}
	E_2^{p,q} = H^p(B;H^q(F;G)_\rho) \Rightarrow H^{p+q}(X;G) \, ,
\end{equation}
where the representation $\rho: \pi_1(B) \to \text{Aut}_\mathbb{Z}(H^\ast(F;G))$ is regarded as a $\mathbb{Z}[\pi_1(B)]$-module, and so $H^\ast(B;H^\ast(F;G)_\rho)$ is taken to be {\it cohomology with local coefficients} \cite{MR1867354}. Importantly, the spectral sequence abuts to {\it ordinary cohomology} of the total space $X$.

For simplicity, let us continue to assume that the fiber $F$ and the base $B$ are both closed and path-connected (and $F$ is also orientable), in which case we have
\begin{equation}
    E_2^{p,0} = E_2^{p,\text{dim}(F)} = H^p(B;G) \, , \qquad E_2^{0,q} = H^0(B;H^q(F;G)_\rho) \, .
\end{equation}
Here we used the fact that the action of $\pi_1(B)$ on $H^0(F;G)$ must be trivial when $F$ is path-connected. Repeating the exercise in Appendix \ref{Serre_spectral_sequence_appendix} with edge homomorphisms gives us the following pullback maps in cohomology, along with a wrong-way map,
\begin{gather}
    H^p(B;G) = E_2^{p,0} \xrightarrowdbl{\pi^\ast} E_\infty^{p,0} \subseteq H^p(X;G) \, ,\\
    H^q(X;G) \xrightarrowdbl{\iota^\ast} E_\infty^{0,q} \subseteq E_2^{0,q} = H^0(B;H^q(F;G)_\rho) \, ,\\
    H^{p+\text{dim}(F)}(X;G) = F^{p+\text{dim}(F)}_p \xrightarrowdbl{\pi_!} E_\infty^{p,\text{dim}(F)} \subseteq E_2^{p,\text{dim}(F)} = H^p(B;G_\rho) \, .
\end{gather}
In many cases, $H^{\text{dim}(F)}(F;G) \cong G$ is also invariant under the action of $\pi_1(B)$, so the codomain of the wrong-way map reduces simply to $H^p(B;G)$. We will hereafter assume that such a condition holds. Likewise, one obtains the following pushforward maps in homology,
\begin{gather}
    H_p(X;G) \xrightarrowdbl{\pi_\ast} E^\infty_{p,0} \subseteq E^2_{p,0} = H_p(B;G) \, ,\\
    H_0(B;H_q(F;G)_\rho) = E^2_{0,q} \xrightarrowdbl{\iota_\ast} E^\infty_{0,q} \subseteq H_q(X;G) \, .
\end{gather}
Note particularly that the second line above becomes $H_{\text{dim}(F)}(F;G) \xrightarrowdbl{\iota_\ast} E^\infty_{0,\text{dim}(F)} \subseteq H_{\text{dim}(F)}(X;G)$ when $q = \text{dim}(F)$. Similarly to before, for a cohomology class $x \in H^{p+\text{dim}(F)}$ which factorises as
\begin{equation}
    x = f \cup \pi^\ast(b)
\end{equation}
for some $b \in H^p(B;G)$ and $f \in \text{coker}(\pi^\ast) \subset H^{\text{dim}(F)}(X;G)$, we define a wrong-way homomorphism
\begin{equation}
    \pi_!: H^{p+\text{dim}(F)}(X;G) \to H^p(B;G) \, , \qquad x \mapsto (f \cap \iota_\ast([F])) \cup b
\end{equation}
where $[F] \in H_{\text{dim}(F)}(F;G)$ is the fundamental class of the fiber. We thus see that the standard definition of fiber integration always takes us to ordinary cohomology of the base, but not cohomology with local coefficients as desired.

To achieve our goal, we propose the following construction. Consider a torus fibration $T^2 \xrightarrow{\iota} X \xrightarrow{\pi} B$. As before, we have a ``pushforward'' map
\begin{equation}
    H_0(B;\mathbb{Z}^2_\rho) = H_0(B;H_1(T^2;\mathbb{Z})_\rho) = E^2_{0,1} \xrightarrowdbl{\iota_\ast} E^\infty_{0,1} \subseteq H_1(X;\mathbb{Z}) \, .
\end{equation}
If $E_\infty^{p+1,0}=0$, then the filtration \eqref{Serre_spectral_sequence_filtration} for $n=p+1$ becomes
\begin{equation}
    \begin{tikzcd}[row sep=scriptsize,column sep=small]
        E_\infty^{p,1} \arrow[r,phantom,"="] & F^{p+1}_p \arrow[r,hookrightarrow] & F^{p+1}_{p-1} \arrow[d,twoheadrightarrow] \arrow[r,phantom,"="] & \cdots \arrow[r,phantom,"="] & F^{p+1}_0  \arrow[r,phantom,"="] & H^{p+1}(X;\mathbb{Z}) \, . \\
        & & E_\infty^{p-1,2} & &
    \end{tikzcd}
\end{equation}
By construction, $E_\infty^{p,1} = E_3^{p,1} = \text{ker}(d_2^{p,1})/\text{im}(d_2^{p-2,2})$, which can be regarded as a map $E_2^{p,1} \to E_\infty^{p,1}$. For example, this map is surjective if $E_\infty^{p,1} = \text{coker}(d_2^{p-2,2})$. Composing the maps above yields a pullback map from cohomology with local coefficients to ordinary cohomology\footnote{The notation $\tilde{\pi}^\ast$ is used to distinguish it from the standard pullback $\pi^\ast$ as introduced earlier.}
\begin{equation}
    H^p(B;\mathbb{Z}^2_\rho) = E_2^{p,1} \xrightarrow{\tilde{\pi}^\ast} E_\infty^{p,1} \subseteq H^{p+1}(X;\mathbb{Z}) \, .\label{twisted_pullback}
\end{equation}
Importantly, note that the pullback shifts the degree of a cocycle by 1. Suppose we have a factorisation for a class $x \in H^{p+2}(X;\mathbb{Z})$ as
\begin{equation}
    x = f \cup \tilde{\pi}^\ast(b)
\end{equation}
for some $b \in H^p(B;\mathbb{Z}^2_\rho)$ and $f \in \text{coker}(\tilde{\pi}^\ast) \subset H^1(X;\mathbb{Z})$, then one can define a notion of ``twisted fiber integration'' with respect to 1-cycles of the fiber $T^2$,
\begin{equation}
    \tilde{\pi}_!: H^{p+2}(X;\mathbb{Z}) \to H^p(B;\mathbb{Z}^2_\rho) \, , \qquad x \mapsto (f \cap \iota_\ast([\mathcal{F}])) \cup b \, ,\label{twisted_fiber_integration}
\end{equation}
where $[\mathcal{F}] \in H_0(B;\mathbb{Z}^2_\rho)$, assuming such a class exists. Contrary to the fundamental class, the choice of $[\mathcal{F}]$ is generally not unique, especially if $H_0(B;\mathbb{Z}^2_\rho)$ has multiple factors. In general, we may define $[\mathcal{F}]$ to be a sum of the independent generator(s) for each factor therein. Loosely speaking, this choice corresponds to a set of fundamental classes for all 1-cycles in the fiber, but also collectively twisted by the local system.

Similarly to before, one may try to define the notion of twisted fiber integration directly using something akin to an edge homomorphism. If $E_\infty^{p-1,2}=0$, then the filtration \eqref{top_edge_homomorphism_filtration} becomes
\begin{equation}
    \begin{tikzcd}[row sep=scriptsize,column sep=small]
        \cdots \arrow[r,hookrightarrow] & F^p_{p+1} \arrow[d,twoheadrightarrow] \arrow[r,phantom,"="] & F^{p+1}_{p-1} \arrow[r,phantom,"="] & \cdots \arrow[r,phantom,"="] & F^{p+1}_1 \arrow[r,phantom,"="] & F^{p+1}_0 \arrow[r,phantom,"="] & H^{p+1}(X;\mathbb{Z}) \, ,\\
        & E_\infty^{p,1}
    \end{tikzcd}
\end{equation}
and if $E_\infty^{p,1} = \text{ker}(d_2^{p,1})$, we can construct a map
\begin{equation}
    H^{p+1}(X;\mathbb{Z}) \xrightarrowdbl{\hat{\pi}_!} E_\infty^{p,1} \subseteq E_2^{p,1} = H^p(B;\mathbb{Z}^2_\rho) \, .
\end{equation}
Evidently, the domain of such a map is different from the previous map \eqref{twisted_fiber_integration}. These two notions of twisted fiber integration are a priori not equivalent, and the existence of each of them respectively requires suitable conditions to hold, as we described above.\footnote{These constructions can be generalised for fibers other than $T^2$, but we refrain from a discussion in full generality.} For our purposes, we will adopt the former definition.


\section{Stiefel-Whitney classes of lens spaces}\label{Lens_spaces_SW_classes_appendix}

Consider the lens space $S^{2n+1}/\mathbb{Z}_k$ with $n \geq 1$. When $k$ is odd, $H^i(S^{2n+1}/\mathbb{Z}_k;\mathbb{Z}_2)$ is trivial unless $i=0,2n+1$, so the Stiefel-Whitney classes $w_i$ automatically vanish for $1 \leq i \leq 2n$. Particularly, $S^{2n+1}/\mathbb{Z}_k$ is orientable and spin. In fact, $w_{2n+1}$ vanishes as well because it is the mod 2 reduction of the Euler class, which is trivial for odd-dimensional, compact, oriented manifolds.

When $k=2m$ is even, $H^i(S^{2n+1}/\mathbb{Z}_k;\mathbb{Z}_2) = \mathbb{Z}_2$ and $H^i(S^{2n+1}/\mathbb{Z}_k;\mathbb{Z}_k) = \mathbb{Z}_k$ for all $0 \leq i \leq 2n+1$. If we denote the generators $\hat{a}_1 \in H^1(S^{2n+1}/\mathbb{Z}_k;\mathbb{Z}_k)$ and $\hat{b}_2 \in H^2(S^{2n+1}/\mathbb{Z}_k;\mathbb{Z}_k)$, then it follows from the simplicial complex of the lens space that (cf.~Examples 3.9 and 3.41 in \cite{MR1867354})
\begin{equation}
    \hat{a}_1^2 = m\hat{b}_2 \, .\label{lens_space_cup_product_structure}
\end{equation}
The mod 2 reduction $\mathbb{Z}_k \xrightarrow{\text{mod} \, 2} \mathbb{Z}_2$ induces a ring homomorphism $H^\ast(S^{2n+1}/\mathbb{Z}_k;\mathbb{Z}_k) \to H^\ast(S^{2n+1}/\mathbb{Z}_k;\mathbb{Z}_2)$. Let us denote the generators of the latter as $a_1 \in H^1(S^{2n+1}/\mathbb{Z}_2;\mathbb{Z}_k)$ and $b_2 \in H^2(S^{2n+1}/\mathbb{Z}_k;\mathbb{Z}_2)$, then applying mod 2 reduction to \eqref{lens_space_cup_product_structure} yields
\begin{equation}
    a_1^2 = mb_2 \, .
\end{equation}

If $m$ is odd, $a_1^2=b_2$, then $H^\ast(S^{2n+1}/\mathbb{Z}_k;\mathbb{Z}_2)$ is generated only by $a_1$. The Cartan formula tells us that
\begin{equation}
    v_i \cup a_1^{2n+1-i} = \text{Sq}^i(a_1^{2n+1-i}) = \begin{pmatrix} 2n+1-i \\ i \end{pmatrix} a_1^{2n+1} \, .
\end{equation}
In particular, we have
\begin{equation}
    v_1 \cup a_1^{2n} = \text{Sq}^1(a_1^{2n}) = \begin{pmatrix} 2n \\ 1 \end{pmatrix} a_1^{2n+1} = 0 \, ,
\end{equation}
and also
\begin{equation}
    v_2 \cup a_1^{2n-1} = \text{Sq}^2(a_1^{2n-1}) = \begin{pmatrix} 2n-1 \\ 2 \end{pmatrix} a_1^{2n+1} = (n-1)(2n-1) a_1^{2n+1} \, ,
\end{equation}
where we used the fact that $\text{Sq}^2(a_1)=0$ due to degree reasons. Using the Wu formula, $w=\text{Sq}(v)$, we find
\begin{equation}
    w_1 = v_1 = 0 \, , \qquad w_2 = v_2 - w_1^2 = (n-1) a_1^2 \, .
\end{equation}
The higher Stiefel-Whitney classes can be computed inductively like so.

On the other hand, if $m$ is even, $a_1^2=0$, then $H^\ast(S^{2n+1}/\mathbb{Z}_k;\mathbb{Z}_2)$ is generated by $a_1$ and $b_2$. In this case,
\begin{equation}
    v_1 \cup b_2^n = \text{Sq}^1(b_2^n) = (\text{mod} \ 2) \circ \beta(b_2^n) \, .
\end{equation}
By exactness, $\beta(b_2^n)$ is 2-torsion, but since $H^{2n+1}(S^{2n+1}/\mathbb{Z}_k;\mathbb{Z})=\mathbb{Z}$ is torsion-free, it must be trivial. Similarly,
\begin{align}
    v_2 \cup a_1 \cup b_2^{n-1} & = \text{Sq}^2(a_1 \cup b_2^{n-1})\nonumber\\
    & = \text{Sq}^1(a_1) \cup \text{Sq}^1(b_2^{n-1}) + a_1 \cup \text{Sq}^2(b_2^{n-1})\nonumber\\
    & = a_1^2 \cup \big((\text{mod} \ 2) \circ \beta(b_2^{n-1})\big) + a_1 \cup \begin{pmatrix} n-1 \\ 1 \end{pmatrix} b_2^n\nonumber\\
    & \phantom{=\ } + a_1 \cup \begin{pmatrix} n-1 \\ 2 \end{pmatrix} \big((\text{mod} \ 2) \circ \beta(b_2)\big)^2 \cup b_2^{n-3}\nonumber\\
    & = (n-1) \, a_1 \cup b_2^n \, ,
\end{align}
where $\beta(b_2)=\beta(b_2^{n-1})=0$ since $H^{2i+1}(S^{2n+1}/\mathbb{Z}_k;\mathbb{Z})=0$ for all $i<n$. As a result, we obtain
\begin{equation}
    w_1 = v_1 = 0 \, , \qquad w_2 = v_2 - w_1^2 = (n-1) b_2 \, .
\end{equation}
To conclude, when $k$ is even, $S^{2n+1}/\mathbb{Z}_k$ is always orientable, and it is spin if and only if $n$ is odd.


\bibliographystyle{./JHEP}
\bibliography{./refs}


\end{document}